\documentclass[printer]{aa}
\usepackage[utf8]{inputenc}
\usepackage{natbib}
\usepackage{graphicx}
\usepackage{amsmath}
\usepackage{gensymb}
\usepackage{subcaption}
\usepackage{dblfloatfix}
\usepackage{makecell}
\usepackage{ragged2e}
\usepackage[colorlinks=true,allcolors=blue]{hyperref}
\usepackage[scientific-notation=true]{siunitx}
\usepackage{xcolor}
\usepackage{mhchem}

\title{Constraining the gas mass of Herbig disks using CO isotopologues}

\author{L. M. Stapper\inst{\ref{inst1}}\fnmsep \and M. R. Hogerheijde\inst{\ref{inst1}, \ref{inst2}} \and E. F. van Dishoeck\inst{\ref{inst1},\ref{inst3}} \and L. Lin\inst{\ref{inst1}} \and A. Ahmadi\inst{\ref{inst1}} \and A. S. Booth\inst{\ref{inst1}} \and S. L. Grant\inst{\ref{inst3}} \and K. Immer\inst{\ref{inst1}} \and M. Leemker\inst{\ref{inst1}} \and A. F. Pérez-Sánchez\inst{\ref{inst1}}}

\institute{Leiden Observatory, Leiden University, PO Box 9513, 2300 RA Leiden, The Netherlands \\e-mail: \texttt{stapper@strw.leidenuniv.nl} \label{inst1} \and Anton Pannekoek Institute for Astronomy, University of Amsterdam, PO Box 94249, 1090 GE, Amsterdam, The Netherlands \label{inst2} \and Max-Planck-Institut für Extraterrestrische Physik, Giessenbachstrasse 1, 85748 Garching, Germany \label{inst3}}

\date{\today}

\abstract
{ %Context 
The total disk mass sets the formation potential for exoplanets. Obtaining the disk mass is however not an easy feat, as one needs to consider the optical thickness, temperature, photodissociation, and freeze-out of potential mass tracers. Carbon-monoxide (CO) has been used as a gas mass tracer in T~Tauri disks, but was found to be less abundant than expected due to freeze-out and chemical conversion of CO on the surfaces of cold dust grains. The disks around more massive intermediate mass pre-main sequence stars called Herbig disks are likely to be warmer, allowing for the possibility of using CO as a more effective total gas mass tracer.
}
{ %Aims
This work aims to obtain the gas mass and size of Herbig disks observed with ALMA and compare these to previous works on T~Tauri disks and debris disks.
}
{ %Methods
Using ALMA archival data and new NOEMA data of \ce{^12CO}, \ce{^13CO}, and \ce{C^18O} transitions of 35 Herbig disks within 450~pc, the masses are determined using the thermo-chemical code Dust And LInes (DALI). A grid of models is run spanning five orders of magnitude in disk mass, for which the model CO line luminosities can be linked to the observed luminosities. Survival analysis is used to obtain cumulative distributions of the resulting disk masses. These are compared with dust masses from previous work to obtain gas-to-dust ratios for each disk. In addition, radii for all three isotopologues are obtained.
}
{ %Results
The majority of Herbig disks for which \ce{^13CO} and \ce{C^18O} are detected are optically thick in both. For these disks, the line flux essentially only traces the disk size and only lower limits to the mass can be obtained. Computing the gas mass using a simple optically thin relation between line flux and column density results in an underestimate of the gas mass of at least an order of magnitude compared to the masses obtained with DALI. The inferred gas masses with DALI are consistent with a gas-to-dust ratio of at least 100. These gas-to-dust ratios are two orders of magnitude higher compared to those found for T~Tauri disks using similar techniques, even over multiple orders of magnitude in dust mass, illustrating the importance of chemical conversion of CO in colder T~Tauri disks. Similar high gas-to-dust ratios are found for Herbig group I and II disks. Since group II disks have dust masses comparable to T~Tauri disks, their higher CO gas masses illustrate the determining role of temperature. Compared to debris disks, Herbig disks have gas masses higher by four orders of magnitude. At least one Herbig disk, HD~163296, has a detected molecular disk wind, but our investigation has not turned up other detections of the CO disk wind in spite of similar sensitivities.

}
{ %Conclusions
Herbig disks are consistent with a gas-to-dust ratio of at least 100 over multiple orders of magnitude in dust mass. This indicates a fundamental difference between CO emission from Herbig disks and T~Tauri disks, which is likely linked to the warmer temperature of the Herbig disks.
}


\begin{document}

\maketitle

\section{Introduction}
\label{sec:introduction}

The formation of exoplanets depends heavily on the amount of material present in a planet-forming disk. Hence, efforts have been made to determine the total disk mass and link this to planet formation occurring inside them. The most common disk mass tracer used is the millimeter dust continuum of the disk. Assuming the canonical gas-to-dust mass ratio of 100 \citep{Bohlin1978}, the total disk mass can be determined after making a few assumptions on the grain properties, most notably the grain emissivity \citep{Beckwith1990}. Many population studies on dust masses have been done with the Atacama Large Millimeter/submillimeter Array (ALMA) \citep[see, e.g.,][]{Ansdell2016, Barenfeld2016, Pascucci2016, Eisner2018, Cazzoletti2019, Anderson2022, vanTerwisga2022}. These studies have shown clear trends regarding disk mass with stellar mass \citep{Ansdell2017, Manara2022}, disk radius \citep{Hendler2020}, and stellar accretion rate \citep{Testi2022}. However, it is yet unknown if the dust continuum indeed traces the total disk mass, i.e., gas and dust, directly. Recent works have shown that the dust can be optically thick, underestimating the total disk mass by an order of magnitude for the most massive disks \citep[e.g.,][]{Liu2022, Kaeufer2023}, and may even be optically thick at 3~mm \citep{Xin2023}. 
% Hence, \citet{stapper2022} suggested that this the seen increase in dust mass for Herbig disks is thus also affected by their relatively larger size compared to T~Tauri disks.

A handle on the total (gas) mass of the disk is therefore much needed (see \citealt{Miotello2022} for a recent overview). The most abundant molecule in a planet-forming disk is \ce{H_2}. However, its faint emission at the typically low temperatures ($\sim20$~K) present in a planet-forming disk do not make it a viable disk mass tracer. Hence, other more indirect tracers are necessary. One promising tracer is the \ce{H_2} isotopologue hydrogen deuteride (HD). Using thermo-chemical models, the HD emission can be used to determine the bulk mass of a disk \citep[see e.g.,][]{Bergin2013, Schwarz2016, Trapman2017, Sturm2023}. HD $J=1-0$ observed with the \textit{Herschel} Space Observatory has been detected in three disks resulting in gas mass measurements \citep{Bergin2013, McClure2016}, and a dozen non-detections in Herbig disks resulting in gas mass upper limits \citep{Kama2016, Kama2020}. However, after the end of the Herschel mission, there are no current facilities to observe the HD $J=1-0$ line, so a different tracer is needed.

\begin{table*}[t]
\caption{Source parameters used for Keplerian masking and computing the gas radii.}
\tiny\centering
\begin{tabular}{lcccccccc}
\hline\hline
\makecell{Name \\ \hspace{1mm}} & \makecell{Distance \\ (pc)} & \makecell{$M_\star$ \\ (M$_\odot$)} & \makecell{Log$_{10}$($L_\star$) \\ (L$_\odot$)} & \makecell{V$_\text{sys}$ \\ (km~s$^{-1}$)} & \makecell{V$_\text{int}$ \\ (km~s$^{-1}$)} & \makecell{PA \\ ($\degree$)} & \makecell{Inc. \\ ($\degree$)}  & \makecell{Ref. \\ \hspace{1mm}} \\ \hline
AB Aur     & 155.0 &  2.4 & 1.7 &   5.8 & 1.5 &  -36 &  23 & \citet{Tang2017}   \\
AK Sco     & 139.2 &  1.7 & 0.8 &   5.4 & 2.5 &  128 & 109 & \citet{Czekala2015}    \\
BH Cep     & 323.9 &  1.5 & 0.8 & --    & --  &  --  & --  &   \\
BO Cep     & 367.9 &  1.4 & 0.6 & -11.6 & 2.0 &  -60 &  30 & This work.    \\
CQ Tau     & 148.6 &  1.5 & 0.8 &   6.2 & 1.5 &  -50 &  35 & \citet{Wolfer2021}    \\
HD 9672    &  57.1 &  1.9 & 1.2 &   2.8 & 2.5 &   73 &  81 & \citet{Hughes2017}   \\
HD 31648   & 155.2 &  1.9 & 1.2 &   5.1 & 1.5 &  148 &  37 & \citet{Liu2019}   \\
HD 34282   & 311.5 &  1.5 & 1.0 &  -2.4 & 1.0 &   65 &  60 & \citet{vanderPlas2017a}   \\
HD 36112   & 155.0 &  1.6 & 0.9 &   6.0 & 1.5 &  -60 &  21 & \citet{Isella2010}   \\
HD 58647   & 302.2 &  4.1 & 2.5 & --    & --  &  --  & --  &    \\
HD 97048   & 184.1 &  2.8 & 1.8 &   4.6 & 1.5 & -176 &  41 & \citet{Walsh2016}   \\
HD 100453  & 103.6 &  1.6 & 0.8 &   5.2 & 1.5 & -145 &  35 & \citet{Rosotti2020}   \\
HD 100546  & 108.0 &  2.1 & 1.3 &   5.6 & 1.5 & -140 &  42 & \citet{Pineda2019}   \\
HD 104237  & 106.5 &  1.9 & 1.3 &   4.8 & 2.0 &  110 &  20 & This work.   \\
HD 135344B & 134.4 &  1.5 & 0.7 &   7.1 & 1.5 &  -62 &  10 & \citet{Cazzoletti2018}   \\
HD 139614  & 133.1 &  1.6 & 0.8 &   6.8 & 1.5 & -100 &  18 & \citet{MuroArena2020}   \\
HD 141569  & 111.1 &  2.1 & 1.4 &   6.4 & 2.0 & -175 &  53 & \citet{White2016}   \\
HD 142527  & 158.5 &  2.2 & 1.4 &   3.6 & 1.5 &   26 &  27 & \citet{Kataoka2016}   \\
HD 142666  & 145.5 &  1.8 & 1.1 &   4.1 & 1.5 &   18 &  62 & \citet{Huang2018}   \\
HD 163296  & 100.6 &  1.9 & 1.2 &   5.9 & 2.0 & -137 &  47 & \citet{Huang2018}   \\
HD 169142  & 114.4 &  1.6 & 0.8 &   6.9 & 0.8 & -175 &  13 & \citet{Fedele2017}   \\
HD 176386  & 154.3 &  2.6 & 1.6 & --    & --  & --   & --  &    \\
HD 200775  & 352.4 &  7.0 & 3.4 & --    & --  & --   & --  &    \\
HD 245185  & 410.4 &  2.2 & 1.5 &  13.1 & 2.0 &  100 & 30  & This work.   \\
HD 290764  & 397.1 &  2.0 & 1.3 &   9.5 & 2.0 &   70 & 30  & \citet{Kraus2017}   \\
HR 5999    & 157.8 &  3.2 & 2.0 & --    & --  & --   & --  &    \\
KK Oph     & 221.1 &  1.5 & 0.7 & --    & --  & --   & --  &    \\
MWC 297    & 407.8 & 20.0 & 4.8 & --    & --  & --   & --  &    \\
SV Cep     & 340.3 &  2.0 & 1.3 & --    & --  & --   & --  &    \\
TY CrA     & 158.3 &  2.6 & 1.6 & --    & --  & --   & --  &    \\
V718 Sco   & 153.9 &  1.7 & 1.1 &   8.4 & 1.5 & --   & --  &    \\
V892 Tau   & 152.5 &  6.0 & 0.1 &   8.4 & 4.0 & -60  & 55  & \citet{Long2021}   \\
VV Ser     & 403.4 &  3.6 & 2.3 & --    & --  & --   & --  &    \\
XY Per     & 419.2 &  3.7 & 2.3 & --    & --  & --   & --  &    \\
Z CMa      & 229.7 &  3.8 & 2.3 & --    & --  & --   & --  &    \\\hline
\end{tabular}
\label{tab:disk_params}\\
\textbf{Notes.} The distances, masses and luminosities are from \citet{GuzmanDiaz2021}, except for KK~Oph, V892~Tau and Z~CMa which are from \citet{Vioque2018}. The distance and luminosity of V892~Tau are also taken from \citet{Vioque2018}, but the stellar mass is from \citet{Long2021}. The PA and inclination references are given in the last column. The PA is defined as the angle in the clockwise direction between north and the blueshifted side of the disk. For three disks the position angle and inclination were determined by eye based on the \ce{^12CO} Keplerian mask. V$_{sys}$ and V$_{int}$ are respectively the system velocity of the disk and the internal velocity which is the range in velocities due to emission from different layers in the disk, as determined by the data.
\end{table*}

Carbon monoxide (CO) has been used extensively as a tracer of both the total gas mass and size of planet-forming disks. Especially its less common isotopologues \ce{^13CO} and \ce{C^18O} \citep[e.g.,][]{Miotello2017, Long2017, Loomis2018}, or even the rarer \ce{C^17O}, \ce{^13C^17O} and \ce{^13C^18O} isotopologues \citep{Booth2019, Booth2020, Loomis2020, Zhang2020, Zhang2021, Temmink2023}, have been used as tracers of the bulk mass. Generally, two main methods are used to determine the gas mass from CO: simple scaling relations assuming an excitation temperature and optically thin emission \citep[e.g.,][]{Loomis2018}, and using thermo-chemical models \citep[e.g.,][]{Miotello2014, Miotello2016}. It has been shown that in T~Tauri disks the CO emission is much weaker than expected, which could be due to carbon and oxygen rich volatiles being locked up as ice resulting in chemical conversion into more complex ices \citep{Bosman2018, Agundez2018} and radial transport \citep{Krijt2018}, both of which lead to low CO fluxes and making CO less ideal to trace the bulk mass \citep[see e.g.,][]{Ansdell2016, Ansdell2017, Pascucci2016, Miotello2017}. Moreover, if CO is not abundant enough it cannot self-shield resulting in a drop in CO abundance due to UV-photons dissociating the molecule \citep{Visser2009}. Alternatively, \citet{Miotello2022} show that the lack of CO emission can also be explained by compact gas disks for those disks that remain unresolved in ALMA observations. Because of the more luminous star, Herbig disks are expected to be warmer and thus less CO conversion should occur \citep{Bosman2018}. Indeed, HD upper limits combined with dust based masses from \citet{Kama2020} imply that this is likely the case. Moreover, using literature values of gas-to-dust ratios \citet{Miotello2022} find that the CO depletion in Herbig disks is lower than for T~Tauri disks. However, a general survey of Herbig gas masses is still lacking.

In addition to the gas mass, CO has also been used to trace the radius of the disk \citep{Ansdell2017, Trapman2019}. The outer disk radius can be used to distinguish between the two main evolutionary scenarios proposed for disks: viscous evolution and wind driven evolution \citep[see for a recent overview][]{Manara2022}. The former results in an increasing outer radius with evolution, while the latter decreases the outer radius of the disk. Modelling has shown that this is also visible in CO emission \citep{Trapman2022}. In the Herbig disk HD~163296, a disk wind has been detected in \ce{^12CO} and \ce{^13CO} \citep{Klaassen2013, Booth2021}, suggesting a possible wind driven scenario operating in that Herbig disk. However, clear conclusions on which of the two scenarios is the main driver of evolution in disks is still missing. A survey is lacking here as well.

This work builds upon the previous work by \citet{stapper2022} who obtained the dust masses of a comprehensive survey of Herbig disks observed with ALMA. We use their sample in addition to five observations done with the Northern Extended Millimeter Array (NOEMA) to obtain information on all Herbig disks with millimeter CO isotopologue observations. These CO observations will be used to determine the gas masses in the same manner as \citet{Miotello2016}, as well as gas disk radii. The obtained masses and radii will be used to answer the following important questions in the field. Is there less CO conversion in Herbig disks compared to T~Tauri disks? Can CO be used as an effective gas mass tracer in Herbig disks? What is the efficiency of radial drift in Herbig disks, indicated by comparison of gas and dust radii? Are there any more wind signatures present in the available Herbig disk ALMA data? What are the mass and gas-to-dust ratio differences compared to T~Tauri disks and debris disks?
 
This work is structured as follows. Section \ref{sec:data_and_reduction} details the selection of the sample and accompanying data reduction. Section \ref{sec:model_setup} describes the \texttt{DALI} model grid parameters. The results are presented in Section \ref{sec:results}, where the integrated-intensity maps of \ce{^12CO}, \ce{^13CO} and \ce{C^18O} are presented in \S\ref{subsec:integrated_maps}, the \ce{^12CO} radii are compared to the dust radii in \S\ref{subsec:12CO_radius}, the luminosities of the observations and models are compared in \S\ref{subsec:13CO_C18O_luminosities} and the resulting gas mass estimates are shown in \S\ref{subsec:obtaining_mass}. We discuss these results in Section \ref{sec:discussion}, where in \S\ref{subsec:rare_isotopologues} we will investigate how to unravel the different masses of the disks using rare isotopologues, and introduce a new technique to compute the full mass of a disk using CO isotopologues in \S\ref{subsec:peeling_the_onion}. In \S\ref{subsec:comparisons} we compare the obtained masses to those of T~Tauri disks in \S\ref{subsec:TTauri}, between the group~I and group~II Herbig disks in \S\ref{subsec:GI_vs_GII}, and to those of debris disks in \S\ref{subsec:debris_disks}. Lastly, in \S\ref{subsec:viscous_vs_wind} we present results into finding disk winds in the data, and discuss the wind driven versus viscously driven evolution of Herbig disks using the obtained disk radii. Section \ref{sec:conclusion} summarizes our conclusions.

\section{Sample and data reduction}
\label{sec:data_and_reduction}
The sample used in this work is compiled by \citet{stapper2022}, who selected all Herbig disks from \citet{Vioque2018} with ALMA data within 450~pc. In this work, we update this sample by using the GAIA~DR3 updated stellar parameters and distances from \citet{GuzmanDiaz2021}. We include HD~34282, KK~Oph, V892~Tau and Z~CMa from \citet{Vioque2018} as well, which are not included in \citet{GuzmanDiaz2021}.

A selection of the available data on the ALMA archive\footnote{\url{https://almascience.eso.org/aq/}} was made which cover the \ce{^12CO}, \ce{^13CO}, and/or \ce{C^18O} lines. In general, the $J=2-1$ transition was chosen, except in cases where only the $J=3-2$ transition was available. The data were chosen such that the largest angular scale is at least as large as the continuum disk size, as to not filter out larger scales. In cases where high resolution data were used (either due to a better sensitivity or no low resolution data were available), a uv-taper was applied, to taper the resolution to $\sim0.1''$ to easier be able to determine the radius and integrated flux.

\begin{figure*}[h!]
    \centering
    \includegraphics[width=\textwidth]{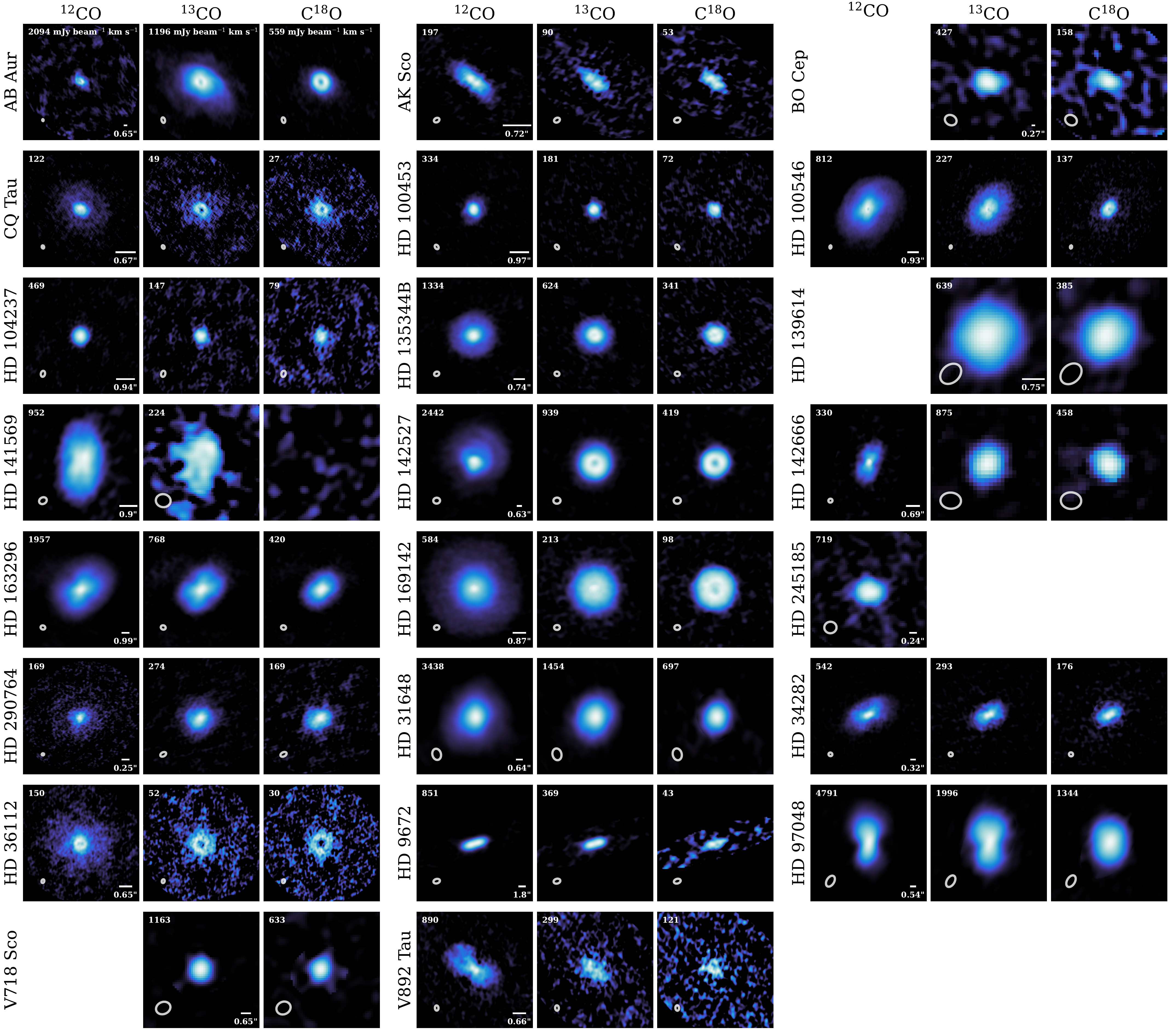}
    \caption{Keplerian masked velocity integrated-intensity maps of the 23 Herbig disks with at least one detection of \ce{^12CO}, \ce{^13CO} or \ce{C^18O}. For each disk \ce{^12CO}, \ce{^13CO} and \ce{C^18O} are given in the first, second and third panel from left to right. If nothing is shown, the molecule is not covered. The bar in every first panel from the left indicates a size of 100 au with the corresponding angular size below. The size of the beam is indicated in the bottom left of each panel. A \textit{sinh} stretch is used to make the fainter parts of the maps better visible. The minimum value is set to zero, and the maximum value in mJy~beam$^{-1}$~km~s$^{-1}$ is indicted on each panel in the top left corner. The disks with no detections in all three isotopologues are not shown here.}
    \label{fig:gallery}
\end{figure*}

These selection criteria resulted in 30 disks for which at least one \ce{^12CO}, \ce{^13CO} or \ce{C^18O} transition is available. R~CrA is not included because it only has ACA (Atacama Compact Array) data available, which is prone to having contamination from nearby sources due to the much lower spatial resolution. Out of the 30 data sets, two (VV~Ser and HD~245185) only have \ce{^12CO} observations available.  

In addition to the ALMA archival data, we present observations made with the Northern Extended Millimeter Array (NOEMA) towards 5 Herbig disks which lie too far north for ALMA. These disks (BH~Cep, BO~Cep, HD~200775, SV~Cep and XY~Per) all lie at distances similar to Orion, ranging from 324~pc to 419~pc \citep{GuzmanDiaz2021}. These data were calibrated using the IRAM facilities. Details on both the NOEMA and ALMA data and their project codes can be found in Table \ref{tab:project_codes}.

The data were imaged using the \texttt{Common Astronomy Software Applications} (CASA) version 5.8.0 \citep{McMullin2007}. The data were first self-calibrated using the continuum, which was made by combining all spectral windows and flagging the channels containing line emission. For the ALMA data, up to six rounds of phase-only self-calibration were computed, starting with a solution interval equal to the duration of a single scan. The solution intervals were shortened by a factor of two each round. For the NOEMA data, only one phase calibration round was used. After phase-only self-calibration, a single round of amplitude self-calibration was performed. Both the phase and amplitude calibrations were only applied if an increase of more than $2\%$ in peak signal-to-noise ratio ($S/N$) was found, for which the rms was determined by excluding the disk emission and using the full field-of-view of the observation. Typically a factor of 1.8 was gained in peak $S/N$. Afterwards, the resulting calibration table was applied to the line spectral windows by using the \texttt{applycal} task in CASA. See \citet{Richards2022} for more information on self-calibration. After self-calibration each data set was continuum subtracted using the \texttt{uvcontsub} task in CASA using a linear fit. For masking during the CLEAN process, hand-drawn masks were aided by the \texttt{auto-multithres}\footnote{The standard values given on \url{https://casaguides.nrao.edu/index.php/Automasking_Guide} were used.} option in \texttt{tclean} to automatically generate masks. For the resolved data, the data were cleaned using the \texttt{multiscale} algorithm in \texttt{tclean}. The used scales were 0 (point source), 1, 2, 5, 10 and 15 times the size of the beam in pixels ($\sim$5 pixels). The last three scales were only used if the disk morphology allowed for it. All unresolved data (which is, in addition to the NOEMA data, some of the ALMA data as well) were cleaned using the \texttt{hogbom} algorithm. For the ALMA data, every data set was imaged using a Briggs robust weighting of 2.0, optimizing the $S/N$ of the resulting image. For the NOEMA data, a Briggs robust weighting of 0.5 was used, due to its lower spatial resolution, without impacting the $S/N$. Finally, the resulting image cubes were primary beam corrected. All resulting image parameters can be found in Appendix \ref{app:data_sets} in Table \ref{tab:project_codes} which also lists the root-mean-square noise of the data. Additionally, the NOEMA continuum data are presented in Appendix \ref{app:noema_data}.

\begin{table*}[]
\caption{Integrated fluxes in Jy km s$^{{-1}}$ and radii in au of \ce{^12CO}, \ce{^13CO} and \ce{C^18O}.}
\tiny\centering
\begin{tabular}{l|ccc|ccc|ccc}
\hline\hline
& \multicolumn{3}{c|}{\ce{^12CO}} & \multicolumn{3}{c|}{\ce{^13CO}} & \multicolumn{3}{c|}{\ce{C^18O}} \\ \hline
\makecell{Name \\ \hspace{1mm}} & \makecell{Flux \\ (Jy km s$^{{-1}}$)} & \makecell{$R_\text{gas, 68\%}$ \\ (au)} & \makecell{$R_\text{gas, 90\%}$ \\ (au)} & \makecell{Flux \\ (Jy km s$^{{-1}}$)} & \makecell{$R_\text{gas, 68\%}$ \\ (au)} & \makecell{$R_\text{gas, 90\%}$ \\ (au)} & \makecell{Flux \\ (Jy km s$^{{-1}}$)} & \makecell{$R_\text{gas, 68\%}$ \\ (au)} & \makecell{$R_\text{gas, 90\%}$ \\ (au)} \\ \hline
AB~Aur	 & 52.35±0.30$^{*}$\hspace{-4.4pt}	 & 232±12	 & 358±12	 & 37.01±0.18	 & 690±35	 & 1118±35	 & 7.50±0.11	 & 370±35	 & 678±40 \\
AK~Sco	 & 2.18±0.05	 & 78±4	 & 114±5	 & 0.68±0.04	 & 66±6	 & 97±18	 & 0.27±0.02	 & 52±5	 & 69±11 \\
BH~Cep	 & --	 & --	 & --	 & <0.07	 & --	 & --	 & <0.07	 & --	 & -- \\
BO~Cep	 & --	 & --	 & --	 & 1.22±0.08	 & <387 	 &<525 	 &0.35±0.07	 & <371 	 &<529 \\
CQ~Tau	 & 3.44±0.04	 & 103±3	 & 146±3	 & 1.12±0.04	 & 84±5	 & 113±6	 & 0.60±0.02	 & 65±4	 & 99±7 \\
HD~9672	 & 3.61±0.08	 & <285 	 &<460 	 &1.37±0.08	 & <285 	 &<439 	 &0.12±0.04	 & <231 	 &<347 \\
HD~31648	 & 19.97±0.08	 & 389±36	 & 597±36	 & 7.03±0.08	 & 313±38	 & 467±38	 & 2.16±0.05	 & 237±38	 & 363±38 \\
HD~34282	 & 9.47±0.09	 & 468±17	 & 673±17	 & 3.46±0.05	 & 340±18	 & 483±18	 & 1.89±0.04	 & 324±18	 & 568±40 \\
HD~36112	 & 7.98±0.14	 & 211±6	 & 305±12	 & 1.87±0.07	 & 121±9	 & 192±18	 & 0.78±0.04	 & 103±6	 & 141±15 \\
HD~58647	 & <0.03	 & --	 & --	 & <0.02	 & --	 & --	 & <0.02	 & --	 & -- \\
HD~97048	 & 27.40±0.14	 & 389±45	 & 596±45	 & 15.93±0.11	 & 433±48	 & 615±48	 & 8.41±0.08	 & 366±48	 & 512±48 \\
HD~100453	 & 2.40±0.03	 & 51±5	 & 78±5	 & 0.91±0.02	 & 38±5	 & 56±5	 & 0.37±0.01	 & 39±5	 & 59±10 \\
HD~100546	 & 63.56±0.12	 & 230±6	 & 322±6	 & 13.13±0.08	 & 187±6	 & 256±6	 & 3.35±0.06	 & 127±9	 & 236±16 \\
HD~104237	 & 2.43±0.06	 & <43 	 &<64 	 &0.64±0.03	 & <39 	 &<59 	 &0.37±0.03	 & <46 	 &<69 \\
HD~135344B	 & 20.07±0.18$^{*}$\hspace{-4.4pt}	 & 158±10	 & 219±10	 & 8.86±0.08$^{*}$\hspace{-4.4pt}	 & 119±9	 & 171±9	 & 3.55±0.08$^{*}$\hspace{-4.4pt}	 & 86±9	 & 125±9 \\
HD~139614	 & --	 & --	 & --	 & 2.82±0.05	 & <130 	 &<179 	 &1.12±0.04	 & <108 	 &<151 \\
HD~141569	 & 18.43±0.12$^{*}$\hspace{-4.4pt}	 & 162±9	 & 220±9	 & 1.27±0.13	 & 150±17	 & 193±54	 & <0.07	 & --	 & -- \\
HD~142527	 & 27.14±0.10	 & 542±30	 & 793±30	 & 11.88±0.07	 & 334±31	 & 491±31	 & 3.91±0.04	 & 271±31	 & 368±31 \\
HD~142666	 & 4.05±0.03	 & 136±6	 & 186±6	 & 1.51±0.06	 & <182 	 &<260 	 &0.64±0.04	 & <164 	 &<236 \\
HD~163296	 & 51.99±0.09	 & 348±13	 & 496±13	 & 16.98±0.07	 & 287±14	 & 400±14	 & 5.96±0.04	 & 228±14	 & 324±14 \\
HD~169142	 & 18.93±0.05	 & 232±8	 & 347±8	 & 6.37±0.04	 & 156±9	 & 230±9	 & 2.93±0.02	 & 129±9	 & 169±9 \\
HD~176386	 & <0.11	 & --	 & --	 & <0.10	 & --	 & --	 & <0.07	 & --	 & -- \\
HD~200775	 & --	 & --	 & --	 & <0.09	 & --	 & --	 & <0.08	 & --	 & -- \\
HD~245185	 & 2.37±0.10	 & <221 	 &<396 	 &--	 & --	 & --	 & --	 & --	 & -- \\
HD~290764	 & 6.55±0.13$^{*}$\hspace{-4.4pt}	 & 271±11	 & 386±14	 & 2.95±0.05$^{*}$\hspace{-4.4pt}	 & 218±17	 & 328±17	 & 1.51±0.05$^{*}$\hspace{-4.4pt}	 & 188±18	 & 277±23 \\
HR~5999	 & <0.08	 & --	 & --	 & <0.08	 & --	 & --	 & <0.06	 & --	 & -- \\
KK~Oph	 & --	 & --	 & --	 & <0.18	 & --	 & --	 & <0.07	 & --	 & -- \\
MWC~297	 & <0.03	 & --	 & --	 & <0.05	 & --	 & --	 & <0.02	 & --	 & -- \\
SV~Cep	 & --	 & --	 & --	 & <0.08	 & --	 & --	 & <0.08	 & --	 & -- \\
TY~CrA	 & <0.10	 & --	 & --	 & <0.10	 & --	 & --	 & <0.07	 & --	 & -- \\
V718~Sco	 & --	 & --	 & --	 & 1.63±0.09	 & <106 	 &<151 	 &0.86±0.07	 & <113 	 &<170 \\
V892~Tau	 & 16.18±1.87	 & 171±30	 & 242±126	 & 3.92±0.20	 & 128±9	 & 191±18	 & 0.86±0.10	 & 90±15	 & 123±43 \\
VV~Ser	 & <0.24	 & --	 & --	 & --	 & --	 & --	 & --	 & --	 & -- \\
XY~Per	 & --	 & --	 & --	 & <0.08	 & --	 & --	 & <0.08	 & --	 & -- \\
Z~CMa	 & <0.11	 & --	 & --	 & <0.06	 & --	 & --	 & <0.03	 & --	 & -- \\\hline
\end{tabular}\\
\label{tab:fluxes_and_radii}
\textbf{Notes.} The fluxes with an asterisk are for the $J=3-2$ transition, all others are for the $J=2-1$ transition. The upper limits are estimated by using three times the noise in a single channel and multiplying this by the width of the channel and the square root of the number of independent measurements, assuming that the emission is coming from a range of 10~km~s$^{-1}$ in velocity and within a single beam. The error on the flux is estimated by multiplying the noise in an empty channel by the width of the channel and the square root of the number of independent measurements in the Keplerian mask and the area over which the integrated fluxes were computed. All integrated fluxes have an additional 10\% calibration error. 
\end{table*}

The velocity integrated (moment 0) maps of the image cubes were obtained by using a Keplerian mask. We use the same implementation as \citet{Trapman2020} (also see \citealt{Salinas2017, Ansdell2018}). The used inclinations, position angles, system velocities and internal velocities (i.e., a range in velocity to take into account that emission is not geometrically thin and comes from different layers in the disk, extending the velocity range over which the mask is generated, see Appendix~A in \citealt{Trapman2020}) can be found in Table \ref{tab:disk_params}. The internal velocity is for the majority of the disks set at 1.5~km~s$^{-1}$, and generally provides an extra factor to improve the coverage of all emission by the Keplerian mask. The distances and stellar masses are taken from \citet{GuzmanDiaz2021} or \citet{Vioque2018}, obtained via evolutionary tracks on the HR-diagram, except for V892~Tau, for which we use the mass estimate from \citet{Long2021}, because both \citet{GuzmanDiaz2021} and \citet{Vioque2018} do not include a mass estimate. We note that other stars have dynamically estimates of their mass as well (e.g., AK~Sco, \citealt{Czekala2015}; HD~169142, \citealt{Yu2021}; HD~163296, \citealt{Teague2021}; HD~34282 \citealt{Law2023}), but we do not include these to keep the stellar masses homogeneously derived for the complete sample. As all emission is included in the Keplerian masking, the stellar masses are not expected to impact our results. The references for the inclinations and position angles as obtained from continuum observations are given in Table \ref{tab:disk_params}. The position angles have been altered by hand by small amounts such that the Keplerian masks fit the channel maps well. By applying the Keplerian mask the moment~0 maps are made, see Figure \ref{fig:gallery}. The integrated fluxes were determined using a curve-of-growth method, see \citet{stapper2022} for more details. The final luminosity of each line was computed by multiplying the integrated flux by a factor $4\pi d^2$ where $d$ is the distance to the source in parsecs.

For the non-detections, the upper limits are estimated by using three times the noise in a single channel and multiplying this by the width of the channel and the square root of the number of independent measurements, assuming that the emission is coming from a range of 10~km~s$^{-1}$ in velocity and within a single beam. We note that this was also done for the binary XY~Per (see Appendix~\ref{app:noema_data}). Similarly, the error on the integrated fluxes was estimated by multiplying the noise in an empty channel by the width of the channel and the square root of the number of independent measurements in the Keplerian mask and the area over which the integrated fluxes were computed.

In addition to the integrated flux, the curve-of-growth method also gives the size of the disk. We compute the radii encircling 68\% and 90\% of the flux. We use a minimum error on the radius of 1/5 the size of the beam (i.e., the pixel size). Similarly, the errors on the dust radius from \citet{stapper2022} are taken as 1/5 the size of the beam. Upper limits on the size of the disk are given if, after fitting a Gaussian to it with the CASA task \texttt{imfit}, the major or minor axes of the Gaussian are smaller than two times the beam. We note that for BO~Cep, HD~104237 and HD~245185, while marginally resolved with only two beams along the minor axis, we estimate an inclination and position angle to cover all emission with the Keplerian mask. See the resulting values in Table \ref{tab:disk_params}. For the other unresolved disks, we either use the inclination and position angle of previous works (HD~9672, HD~142666, HD~139614) for the Keplerian mask or we use no Keplerian mask (V718~Sco). We include the inferred radii for these seven disks as upper limits in Table \ref{tab:fluxes_and_radii}. Due to the many different projects imaged, the sample has a rather inhomogeneous distribution of spatial scales and sensitivities (see Table~\ref{tab:project_codes}). All \ce{^13CO} and \ce{C^18O} observations are from the same data sets, which is also the case for most \ce{^12CO} observations. The exceptions are: AB~Aur, HD~135344B, HD~141569, HD~142666 and HD~290764, which have a different data set for their \ce{^12CO} observations. 

\begin{figure}
    \centering
    \includegraphics[width=0.5\textwidth]{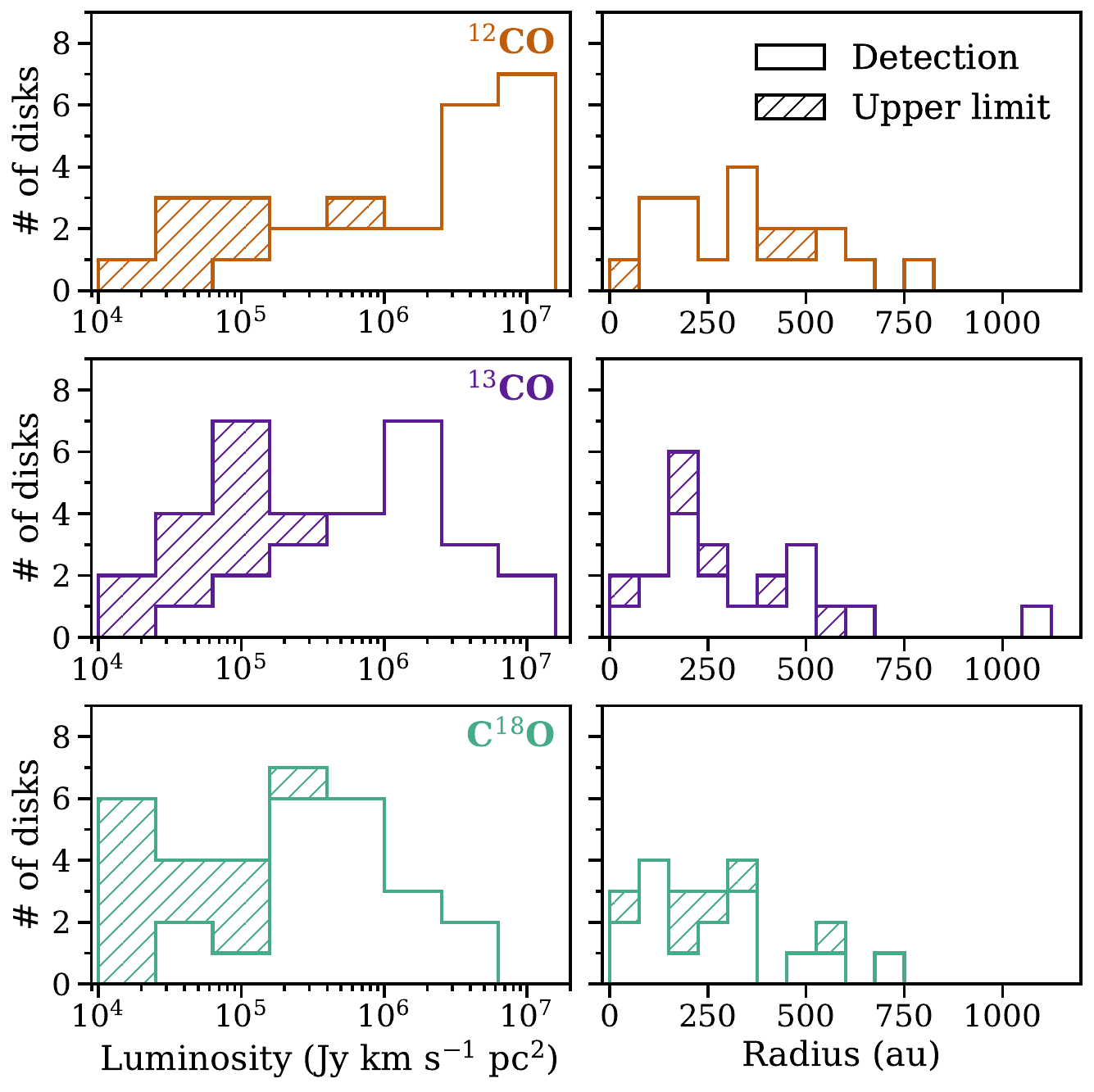}
    \caption{Summary of the luminosities and $90\%$ radii for the \ce{^12CO}, \ce{^13CO} and \ce{C^18O} observations as presented in Table \ref{tab:fluxes_and_radii}.}
    \label{fig:flux_and_radius_hists}
\end{figure}

The fluxes and radii together with their errors are listed in Table~\ref{tab:fluxes_and_radii}. A summary of Table~\ref{tab:fluxes_and_radii} of all luminosities and radii is shown in Figure~\ref{fig:flux_and_radius_hists}.

\begin{table}[t]
\caption{Values of each DALI parameter used in this work.}
\centering
\begin{tabular}{ll}
\hline\hline
Model Parameter                 & Range                                     \\ \hline
                                &                                           \\
\textbf{Chemistry}              &                                           \\
Chemical age                    & 1 Myr                                     \\
Volatile $[$C$]$/$[$H$]$        & $1.35\times10^{-4}$                       \\
Volatile $[$O]/$[$H$]$          & $2.88\times10^{-4}$                       \\
$[$PAH$]$                       & $10^{-2}$ $\times$ ISM                    \\
$\zeta_{cr}$                    & $5.0\times10^{-17}$~s$^{-1}$              \\
                                &                                           \\
\textbf{Physical structure}     &                                           \\
$R_c$                           & 5, 10, 30, 60, 200~au                     \\
$R_\text{subl}$                 & 0.2~au                                    \\
Log$_{10}(M_\text{gas})$        & \{-5, -4.5, ..., -0.5\} M$_\odot$         \\
$\gamma$                        & 0.4, 0.8, 1.5                             \\
$h_c$                           & 0.05, 0.2, 0.4~rad                        \\
$\psi$                          & 0.1, 0.2                                  \\
                                &                                           \\
\textbf{Dust properties}        &                                           \\
Dust populations                & 0.005–1 $\mu$m (small)                    \\
                                & 1–1000 $\mu$m (large)                     \\
$\chi$                          & 0.2                                       \\
$\Delta_{g/d}$                  & 100                                       \\
                                &                                           \\
\textbf{Stellar properties}     &                                           \\
$T_\text{eff}$                  & $10^4$~K                                  \\
$L_\text{bol}$                  & 5, 10, 20, 50~$L_\odot$                   \\
$L_X$                           & $10^{30}$~erg~s$^{-1}$                    \\
$M_\star$                       & 2.5~$M_\odot$                             \\
                                &                                           \\
\textbf{Observational geometry} &                                           \\
$i$                             & 10$\degree$, 40$\degree$, 70$\degree$     \\
$d$                             & 100~pc                                    \\ 
                                &                                           \\\hline
\end{tabular}
\label{tab:model_params}
\end{table}

\section{Model setup}
\label{sec:model_setup}
In this work a grid of \texttt{DALI} models \citep[Dust And LInes,][]{Bruderer2012, Bruderer2013} is run to determine the gas masses of the Herbig disks based on the \ce{^13CO} and \ce{C^18O} luminosities. \texttt{DALI} is a thermo-chemical code which solves for the gas and dust thermal structure of the disk by taking heating, cooling, and chemical processes into account. We use the CO isotopologue chemistry network by \citet{Miotello2016} evolved to 1~Myr, which is a simplified version of the network by \citet{Miotello2014}. This network includes the \ce{^12C}, \ce{^13C}, \ce{^16O}, \ce{^18O}, and \ce{^17O} isotopologues, and processes such as isotope-selective photodissociation, fractionation reactions, self-shielding, and freeze-out.

\texttt{DALI} uses the simple parametric prescription proposed by \citet{Andrews2011} for the density structure. This density structure is motivated by the viscous accretion disk model, for which the solution follows an exponentially tapered power law \citep{LyndenBell1974, Hartmann1998} given by

\begin{equation}
    \Sigma_\text{gas} = \Sigma_c \left( \frac{R}{R_c} \right)^{-\gamma} \exp\left[ -\left( \frac{R}{R_c}\right)^{2-\gamma} \right],
    \label{eq:surface_density}
\end{equation}

\noindent where $\Sigma_c$ and $R_c$ are respectively the surface density and critical radius and $\gamma$ the power law index. The vertical distribution of the gas is given by a Gaussian distribution. The scale height angle of the gas is given by

\begin{equation}
    h = h_c \left( \frac{R}{R_c} \right)^\psi,
    \label{eq:scale_height}
\end{equation}

\noindent where $\psi$ is the flaring index and $h_c$ is the scale height at distance $R_c$. The physical scale height $H$ is then equal to $h R$.

The models include small and large dust populations, for which the small dust follows the gas distribution and the large dust is settled vertically. The dust settling is set by the settling parameter $\chi$. The gas to dust mass ratio in the disk is given by $\Delta_{g/d}$ and is set to the ISM value of 100.

In this work we use a range of parameters to run a grid of \texttt{DALI} models, see Table~\ref{tab:model_params}. All of these parameters combined make a total of 3600 models. Besides varying the gas mass of the models from $10^{-5}$~M$_\odot$ to $10^{-0.5}$~M$_\odot$ in steps of 0.5~dex, we also vary the parameters related to the vertical and radial mass distribution of the gas following eqs. (\ref{eq:surface_density}) and (\ref{eq:scale_height}). We expand the range of values used in \citet{Miotello2016} by also including flat and compact disks, as these are likely present in the group II Herbig disks \citep{stapper2023}, through $h_c$, $\psi$, and $R_c$. Therefore, the critical radius ranges from 5 to 200~au in steps of $\sim0.4$~dex. The power-law index $\gamma$ is set to 0.4, 0.8, and 1.5, which changes the steepness of the turnover in the surface density at large radii. Higher values of $\gamma$ allow for the bulk of the mass to be distributed further out in the disk, effectively changing the size of the disk.

For the vertical distribution of the gas, we alter both the flaring index $\psi$ and the scale height at the critical radius $h_c$. The range in $\psi$ is kept the same as \citet{Miotello2016} used, see Table~\ref{tab:model_params}. For the scale height $h_c$ we use a larger range from a very flat disk with $h_c=0.05$~rad, as very flat disks have been found to exist in the Herbig population \citep{Law2021, Law2022, stapper2023}, up to 0.2~rad and 0.4~rad for thicker disks. Flaring of the disk changes the (vertical) temperature structure of the disk and hence the emitting layer of the CO, which in turn changes the observed luminosity of the line.

To better accommodate for the range in stellar properties, we vary the stellar luminosity from 5~$L_\odot$ to 50~$L_\odot$ in steps of $\sim0.3$~dex to include a similar range in luminosities present in our Herbig sample (not including the most luminous stars), as determined by \citet{GuzmanDiaz2021}. We keep the effective temperature $T_\mathrm{eff}$ at $10^4$~K as this matches well with the effective temperatures of the sample, and we use the sublimation radius $R_\mathrm{subl}$ corresponding to a stellar luminosity of 10~$L_\odot$ (using the scaling relation by \citealt{Dullemond2001}) for all models, as the sublimation radius is much smaller than the scales we probe.

The remaining parameters we keep the same as \citet{Miotello2016}. The PAH abundance is set to 1\% of the ISM, the ISM value being a PAH-to-dust mass ratio of 5\% \citep{Draine2007}. The cosmic-ray ionization rate ($\zeta_{cr}$) is set to $5.0\times10^{-17}$~s$^{-1}$, and the X-ray luminosity ($L_X$) is set to $10^{30}$~erg~s$^{-1}$. As noted in both \citet{Bruderer2012} and \citet{Miotello2016}, $L_X$ only has a minor influence on the CO pure rotational lines, which are the focus in this work.

The models have been run with 40 vertical cells and 42 log-spaced radial cells out to 1000~au. The highest mass models were radially sampled in 62 cells out to 1500~au. This sampling was found to be sufficient to yield line fluxes accurate to $<5\%$. The grid extends vertically to 10 times the scale height, as given by eq.~(\ref{eq:scale_height}).

For the built-in DALI ray-tracer, a distance of 100~pc is used and each model is ray-traced at inclinations of 10$\degree$, 40$\degree$ and 70$\degree$, encompassing all inclinations in our sample (see Table~\ref{tab:disk_params}). A total of six CO isotopologues are ray-traced: \ce{^12CO}, \ce{^13CO}, \ce{C^18O}, \ce{C^17O}, \ce{^13C^18O}, and \ce{^13C^17O}. The $J=2-1$ and $J=3-2$ transitions are ray-traced for each molecule \citep{Yang2010, Schoier2005}.

\begin{figure*}[t]
    \centering
    \includegraphics[width=\textwidth]{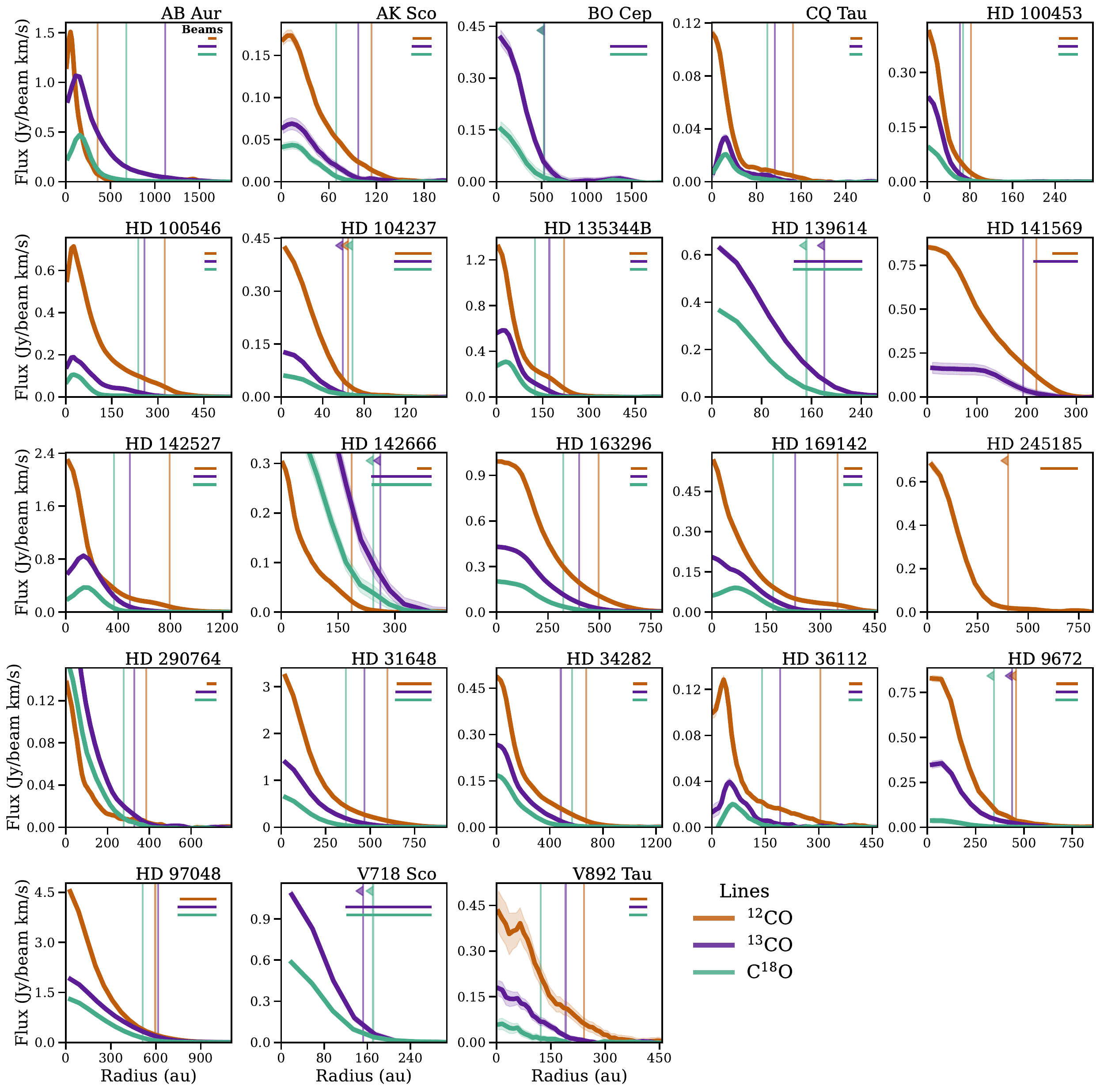}
    \caption{Azimuthally averaged radial profiles of the three CO isotopologues in all 23 Herbig disks in which at least one isotopologue is detected. The $1\sigma$ uncertainty interval is indicated by the shaded region in the same color as the profile. The vertical solid lines indicate the derived 90\% radii for each line. An upper limit on the radius is indicated with a left facing arrow. The beam size is illustrated by the horizontal lines in the top right of each panel.}
    \label{fig:azimuthal_avg}
\end{figure*}

\begin{figure}
    \centering
    \includegraphics[width=0.5\textwidth]{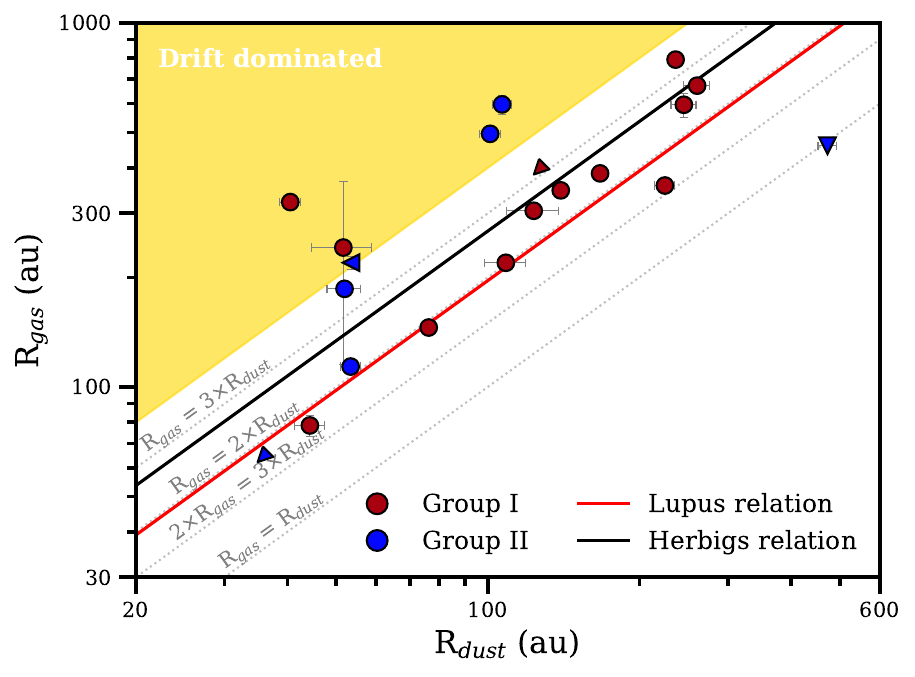}
    \caption{Gas radii versus dust radii of the Herbig disks with detected \ce{^12CO} emission, both $R_{90\%}$. An upper limit on the gas radius or dust radius, or both, is marked with a triangle pointing in the corresponding direction(s). The colors indicate the \citet{Meeus2001} group of the disk: red is group~I, blue group~II. Lines for different $R_\text{dust}$ to $R_\text{gas}$ ratios are shown as well. The black line shows for the resolved disks the relation of the $R_{90\%}$ radii. Using $R_{68\%}$ does not alter the relation significantly. The yellow region indicates the region where the difference between the dust and gas radii cannot solely be explained by optical depth effects, and radial drift is necessary \citep{Trapman2019}. The red line corresponds to the relation found for the disks in Lupus \citep{Ansdell2018}.}
    \label{fig:Rg_vs_Rd}
\end{figure}

\section{Results}
\label{sec:results}
\subsection{Integrated-intensity maps}
\label{subsec:integrated_maps}
Figure \ref{fig:gallery} presents the integrated intensity maps of the disks in which any of the three CO isotopologues is detected: 20 out of 27 for \ce{^12CO}, 22 out of 33 for \ce{^13CO}, and 21 out of 33 for \ce{C^18O} for our 35 sources. If nothing is shown, the molecule is not covered. As \ce{C^18O} is covered but not detected, the integrated-intensity map of HD~141569 is presented as well, which is obtained by integrating over the same velocity range as done for the \ce{^12CO} and \ce{^13CO} emission. All other non-detections are not shown. For the majority of the disks, the transition used is $J=2-1$, except for all lines in the HD~135344B and HD~290764 disks, and \ce{^12CO} in AB~Aur and HD~141569, for which the $J=3-2$ transition is imaged. For the detections, azimuthally averaged radial profiles are presented in Fig.~\ref{fig:azimuthal_avg} which have been made using the inclinations and position angles listed in Table \ref{tab:disk_params}. Fig.~\ref{fig:azimuthal_avg} also presents the radii listed in Table \ref{tab:fluxes_and_radii}.

\begin{figure*}[b]
    \centering
    \includegraphics[width=\textwidth]{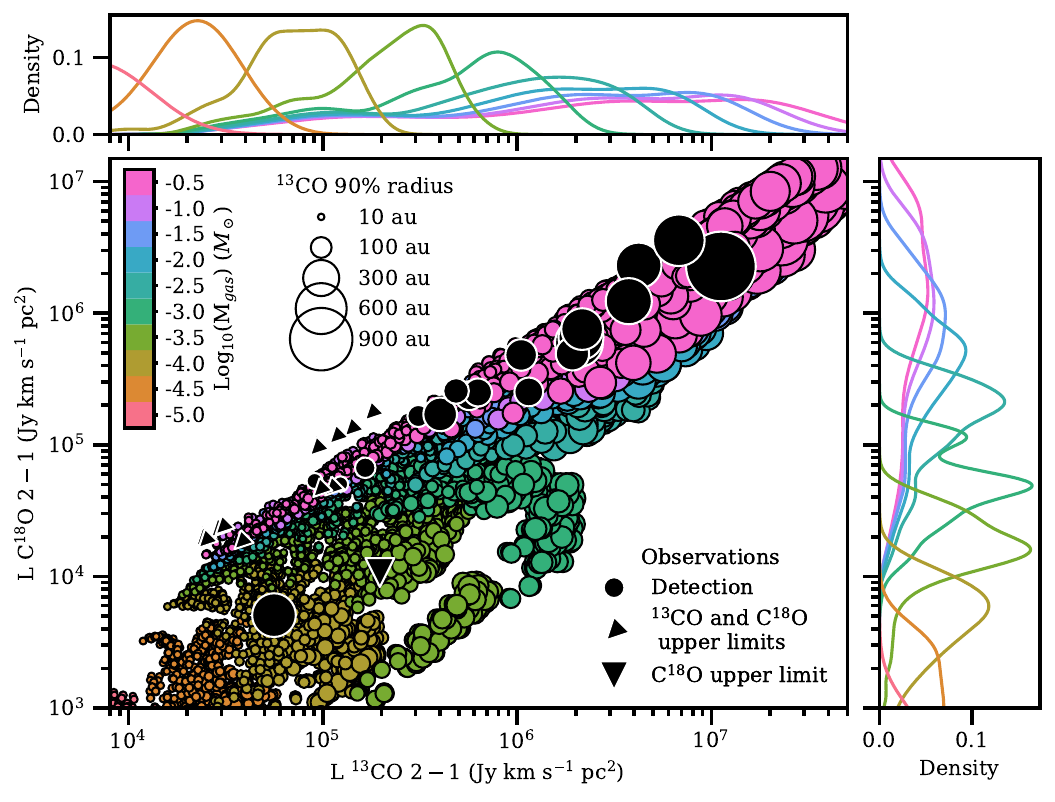}
    \caption{\ce{C^18O} luminosity versus \ce{^13CO} luminosity for the DALI models (colored circles) and the observations (black circles). The colors indicate the gas mass of the disk model. The size of the markers indicate the $90\%$ radius of the \ce{^13CO} emission. Probability density curves of the models for each gas mass are shown along the x-axis in the top panel and along the y-axis in the right panel, i.e., most models of a particular gas mass reside around the peak of each curve.}
    \label{fig:observations_and_models}
\end{figure*}

In general we find that the size of the disk decreases with the abundance of the isotopologue. For the resolved disks the median \ce{^12CO} $R_{68\%}$ ($R_{90\%}$) radius is a factor of 1.3 (1.3) larger compared to \ce{^13CO}, and a factor of 1.8 (1.4) larger compared to \ce{C^18O}. Hence, for \ce{C^18O} the bulk of the emission is generally compact, but it can have more faint extended emission around the compact emission compared to \ce{^12CO} and \ce{^13CO}, resulting in a larger difference between the $R_{68\%}$ and $R_{90\%}$ radii. The decrease with abundance of the isotopologues observed is caused by the smaller column density of the rarer CO isotopologue and isotope selective photodissociation, which lead to optically thin millimeter emission at larger radii reducing the amount of emission. Regarding the luminosity of the disk, we find a clearly decreasing trend with rarity of the isotopologue. Most \ce{^12CO} disks have a luminosity of at least $5\times10^{6}$~Jy~km~s$^{-1}$~pc$^2$, while the \ce{^13CO} and \ce{C^18O} disk luminosities are $2\times10^{6}$~Jy~km~s$^{-1}$~pc$^2$ and $4\times10^{5}$~Jy~km~s$^{-1}$~pc$^2$ respectively. As all three isotopologues are likely optically thick in our sample when detected, this decrease in luminosity is a result of a decrease in size of the detected disk, a reduction in temperature due to the emission from rarer isotopologues originating from deeper in the disk, and/or truncation by photodissociation.

The sizes of the \ce{^12CO} disks are seen to vary significantly. The smallest detected gas disks range from an $R_{90\%}$ of less than 64~au (HD~104237) out to 82~au (HD~100453). The HD~104237 disk is particularly small both in CO and continuum: we measure a continuum disk extent of less than 21~au, which is still unresolved (we use a higher spatial resolution data set than \citealt{stapper2022}, who had an upper limit of 139~au). Most of the \ce{^12CO} disks have a $R_{90\%}$ between 100-400~au (65\%, 13/20; see Fig.~\ref{fig:flux_and_radius_hists}). The median disk size lies close to the middle of this range at 335~au. The $R_{68\%}$ radius has a median of 226~au. The largest disk in \ce{^12CO} is HD~142527 which has a $R_{90\%}$ of 793~au, by far the largest disk in the sample. AB~Aur is likely a very large disk in \ce{^12CO} as well, because of its \ce{^13CO} radius of 1118~au. But due to severe foreground or envelope absorption large parts of the disk are not visible in \ce{^12CO}. A number of disks show large extended emission especially in \ce{^12CO} (but also for \ce{^13CO} and \ce{C^18O}) where there is an initial peak and a much flatter emission `shoulder' towards larger radii. This part contributes to the integrated flux only slightly, and mainly increases the inferred radius of the disk, especially for the 90\% radius. Hence, $R_{90\%}$ is the main tracer of the outer radius, and we will primarily use this over the 68\% radius, except when explicitly stated otherwise.

The \ce{^13CO} and \ce{C^18O} sizes are generally smaller than the \ce{^12CO} sizes. As Fig.~\ref{fig:flux_and_radius_hists} shows, most of the \ce{^13CO} and \ce{C^18O} emission is within 300~au in size. The range of the inferred radii is larger than for \ce{^12CO}, ranging from the smallest resolved \ce{^13CO} disk in the sample of 61~au, out to 1118~au in size for the largest disk. The most stringent upper limit obtained is similar as the smallest resolved disk, at 59~au, for HD~104237. There are also more upper limits on the \ce{^13CO} and \ce{C^18O} radii compared to \ce{^12CO} due to four relatively low resolution data sets used not containing \ce{^12CO}.

Some disks suffer from foreground cloud absorption, such as HD~97048 and V892~Tau, as can be seen in Fig.~\ref{fig:gallery} where emission along the minor axes of these disks seems to be absent. Especially the \ce{^12CO} emission of AB~Aur and some of HD~97048 suffers from this, resulting in a smaller outer radius found for \ce{^12CO} compared to the other isotopologue(s). For \ce{^13CO} and \ce{C^18O} the cloud absorption leaves much less of an imprint on the moment map, and these do not affect the obtained integrated fluxes and the radii significantly. Hence, we will mainly use the \ce{^13CO} radius as a measure of the size of the disk. Several of the non-detections show foreground cloud emission: HD~176386, MWC~297, TY~CrA and Z~CMa. For Z~CMa emission from the disk is present at V$_{\rm LSRK}\sim10$~km~s$^{-1}$, but the foreground cloud emission makes it difficult to extract any information from this, and thus it is considered as an upper limit. For the other disks with foreground cloud emission no signature of a disk is visible in any of the isotopologues, possibly due to the cloud emission.

Due to the different data sets, their spatial resolution is different resulting in unresolved \ce{^13CO} and \ce{C^18O} disks for HD~142666, and different sensitivities resulting in upper limits larger than the \ce{^12CO} disk. This is especially clear from the azimuthal averages in Fig.~\ref{fig:azimuthal_avg}, where one can see that the inferred radii of \ce{^12CO} are indeed smaller than for the rarer isotopologues in these two disks. Lastly, for BO~Cep, HD~104237, HD~142666, and V718~Sco we do find very similar outer radii for \ce{^13CO} and \ce{C^18O} because the disk is unresolved. The other two disks with upper limits on their size, HD~139614 and HD~9672, are marginally resolved, especially along the major axes, resulting in a more significant difference between the \ce{^13CO} and \ce{C^18O} radii.

\begin{figure*}[t]
    \centering
    \includegraphics[width=\textwidth]{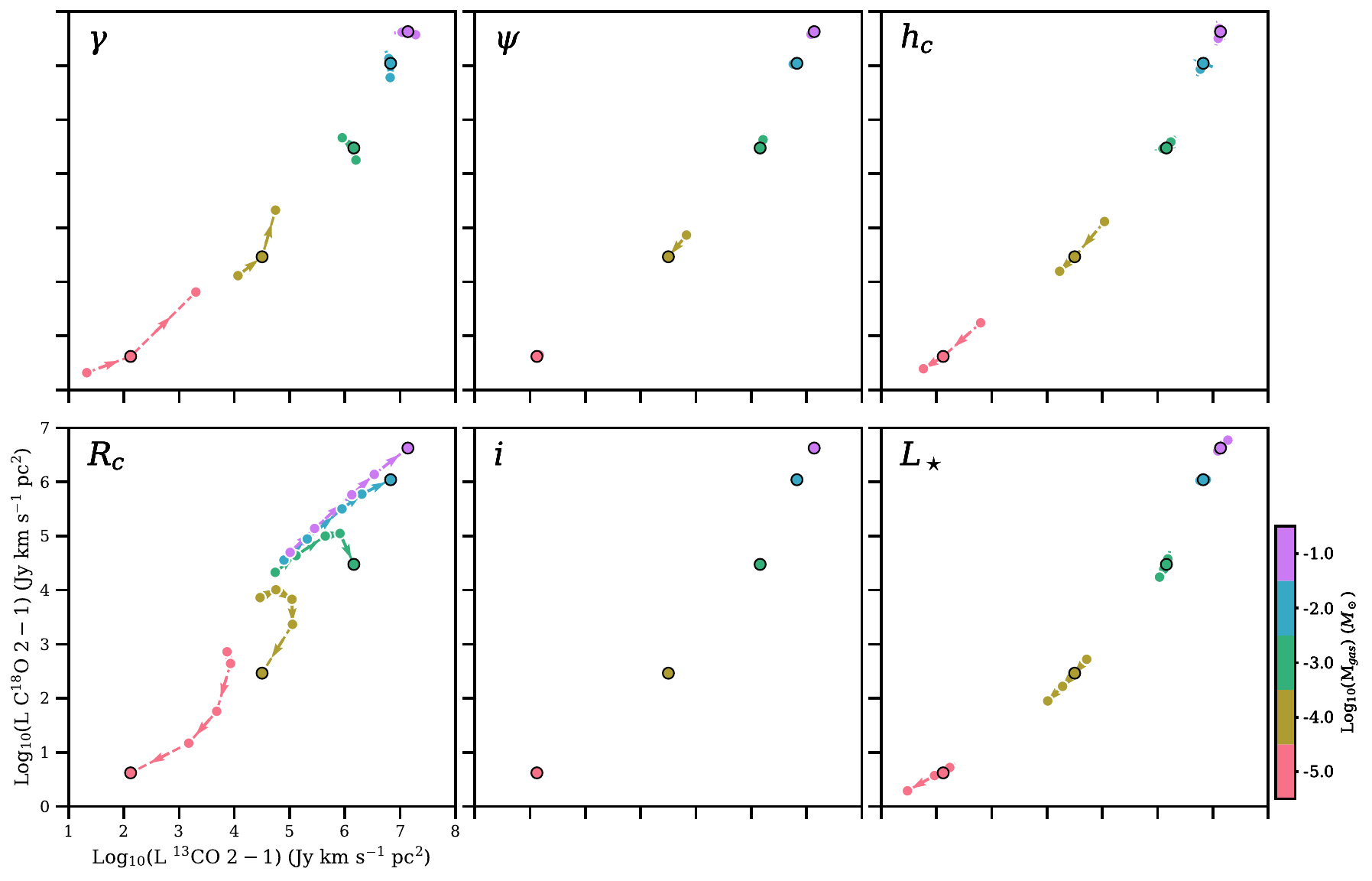}
    \caption{Overview of the different parameters explored in the DALI models and their effect on the luminosity of \ce{^13CO} and \ce{C^18O}. The colors indicate the mass of the model. The arrows indicate in which direction the parameter increases in value (see Table \ref{tab:model_params}). The black outlined circles are the same models for each mass in all panels. The black outlined models have the following parameters: $\gamma=0.8$, $\psi=0.2$, $h_c=0.2$~rad, $R_c=200$~au, $i=10\degree$ and $L_\star=10$~$L_\odot$. In Appendix \ref{app:parameter_overview} the same plot is shown, but with $\gamma=0.4$ and $\gamma=1.5$, and with $R_c=5$~au, $R_c=10$~au, $R_c=30$~au, and $R_c=60$~au, and for the $J=3-2$ transition}.
    \label{fig:parameter_overview}
\end{figure*}

\subsection{\ce{^12CO} radius versus dust radius}
\label{subsec:12CO_radius}
The \ce{^12CO} observations can be used to compare the gas radii of the Herbig disks with the dust radii from \citet{stapper2022}. Figure \ref{fig:Rg_vs_Rd} presents the the dust radius compared to the gas radius. In addition to HD~104237, a higher resolution data set of HD~141569 has been used as well compared to \citet{stapper2022}, which further constrains the dust radius to less than $38$~au for the $68\%$ radius and $54$~au for the $90\%$ radius. In addition to the region in the plot which is radial drift dominated, the relation found for the disks in Lupus is also indicated \citep{Ansdell2018}. If the gas radius is larger, the disk is thought to be radial drift dominated \citep{Trapman2019}. Anything below this value can still be caused by optical depth effects. 

All disks but one (HD~9672) have larger gas radii compared to their dust radii. Most disks lie along the same relation as found for Lupus, where the gas is two times the dust radius. However, a larger spread in ratios is present in the Herbig sample compared to the disks in Lupus. The black line in Fig.~\ref{fig:Rg_vs_Rd} indicates the relation for the resolved disks between the dust and gas radii for the $R_{90\%}$ radii. A simple fitting routine \texttt{curve\_fit} is used from the \texttt{scipy} package to fit the scaling between the dust and gas radii. We find that for the $R_{90\%}$ radii, the ratio between the dust and gas radii is a factor of 2.7. The $R_{68\%}$ are consistent with this value. HD~9672 is the outlier in this sample, showing an equal size in both the dust and gas. The gas radius is slightly unresolved however, and this could make the gas radius even smaller. A comparison with debris disks will be shown in Section \ref{subsec:debris_disks}.

These ratios are high compared to Lupus due to a few disks which are close to or within the regime where radial drift is necessary to explain the differences seen between the dust and gas radii \citep{Trapman2019}. This is the case for HD~100546, HD~142666, HD~163296, HD~31648 (MWC~480) and V892~Tau, all of which have gas disks a factor of more than four larger than their dust radius. This is consistent with other works of for instance MAPS \citep{Zhang2021}. A few disks have gas radii only slightly smaller than four times the dust radius, HD~142527 and HD~245185. We note that HD~100546 has a very faint outer dust ring at 190~au \citep{Walsh2014, Fedele2021}, lowering the ratio to 1.7 instead of almost a factor of 8. Similarly, HD~163296 has a faint outer ring in both the DSHARP and MAPS data as well \citep{Huang2018, Sierra2021}. Hence, Fig.~\ref{fig:Rg_vs_Rd} could indicate around which disks a faint outer ring can be found, possibly linked to giant exoplanet formation happening in these disks.

\subsection{\ce{^13CO} and \ce{C^18O} luminosities}
\label{subsec:13CO_C18O_luminosities}
Figure \ref{fig:observations_and_models} presents the resulting \ce{^13CO} and \ce{C^18O} $J=2-1$ line luminosities of the models in colors and observations in black (the same figure for the $J=3-2$ transition can be found in Appendix \ref{app:other_transitions}). The size of the markers are scaled by the \ce{^13CO} 90\% radius of the models and observations. The disks with upper limits on both \ce{^13CO} and \ce{C^18O} are presented as the diagonal triangles. HD~141569, which has a \ce{C^18O} upper limit, is shown as a downward pointing triangle. The top and right hand-side panels indicate the distribution of the models for the \ce{^13CO} and \ce{C^18O} luminosity respectively.

The two main deciding factors governing how luminous the disk is are the optical depth of the millimeter lines as seen from the observer and the self-shielding capacity of the CO isotopologues against UV \citep[e.g.,][]{Visser2009}. Additionally, these two factors depend on the distribution of the mass and the size of the disk \citep{Trapman2019, Trapman2020}. If the disk is very compact a large range of masses can be `hidden' behind the region where the gas becomes optically thick \citep{Miotello2021}. This results in a large degeneracy in possible luminosities for a single mass, as can be seen in Fig.~\ref{fig:observations_and_models}. The only way to change the observed luminosity is to increase the emitting area as for optically thick emission, the intensity per unit area is constant, and thus increasing the radius of the disk. So a clear trend is present in the models where the smallest disks are at low \ce{^13CO} and \ce{C^18O} luminosities, while the largest disks are on the opposite side with high \ce{^13CO} and \ce{C^18O} luminosities due to an increase in radius. 

Figure \ref{fig:parameter_overview} presents this in another way: for the most massive disks an increase in radius results in an increase in luminosity. In addition, Appendix \ref{app:parameter_overview} shows this for the other values of $R_c$ and $\gamma$. This trend is also evident in the observations, where the largest disks such as AB~Aur and HD~97048 are in the top right of Fig.~\ref{fig:observations_and_models} and the smallest disks such as HD~100453 and HD~104237 are in the bottom left. We can therefore conclude that most disks, apart from HD~9672 and HD~141569, are likely to be optically thick in both \ce{^13CO} and \ce{C^18O} as for these disks both luminosities scale with the size of the disk.

The models with higher mass disks can also make larger disks due to the disk surface density being higher at larger radii. This results in a maximum \ce{^13CO} and \ce{C^18O} luminosity for both the $10^{-1}$~M$_\odot$ and $10^{-2}$~M$_\odot$ models. The distributions presented in the top and right panels of Fig.~\ref{fig:observations_and_models} show that the $10^{-1}$~M$_\odot$ models can have luminosities up to $\sim4\times10^7$~Jy~km~s$^{-1}$~pc$^2$ for \ce{^13CO} and $\sim10^7$~Jy~km~s$^{-1}$~pc$^2$ for \ce{C^18O}. Similarly, the $10^{-2}$~M$_\odot$ models have luminosities up to $\sim2\times10^7$ and $\sim2\times10^6$ Jy~km~s$^{-1}$~pc$^2$ for \ce{^13CO} and \ce{C^18O}, resulting in a region where the mass of the disks needs to be at least $0.1$~M$_\odot$ to explain the observed luminosities.

For the lower mass disks, an interplay between the optical depth and self-shielding of the CO molecules sets the observed luminosities. Due to its larger abundance, \ce{^13CO} becomes optically thin further out in the disk than \ce{C^18O}. For the $10^{-2}$~M$_\odot$ models, first an increase in radius results in an increase in luminosity in both isotopologues, but eventually \ce{C^18O} becomes optically thin. Consequently, an increase in radius does not increase the luminosity of \ce{C^18O}, but it still does so for \ce{^13CO}, and the model moves horizontally with an increase in radius, also see the bottom left panel of Fig.~\ref{fig:parameter_overview}. When the \ce{C^18O} is spread out even more, its self-shielding ability decreases and the \ce{C^18O} starts to photodissociate, decreasing its abundance and thus its luminosity. This results in the model moving down with an increase in radius in Fig.~\ref{fig:parameter_overview}. Consequently, a maximum luminosity for \ce{C^18O} is set by photodissociation for disk masses $\lesssim$ $10^{-3}$~M$_\odot$. \ce{^13CO} is on the other hand still optically thick, and increasing the radius still increases its luminosity, until \ce{^13CO} starts dissociating as well, causing the models to curve towards lower luminosities in both isotopologues with increasing radius for M$_{disk}\lesssim10^{-4}$~M$_\odot$.

Photodissociation of \ce{C^18O} is already the main driver for the luminosity of the lower mass models of almost all sizes, hence almost all models move horizontal or diagonally down to lower \ce{^13CO} luminosities. For the smallest models, while still being optically thin (or marginally optically thick) in \ce{C^18O}, an increase in size can still result in an increase in luminosity if the photodissociation of the \ce{C^18O} does not yet dominate. Also in the M$_{disk}\lesssim10^{-4}$~M$_\odot$ case the \ce{^13CO} will become optically thin for the observer and photodissocation will start to lower its luminosity with an increase in radius. Going to the lowest mass models of $10^{-5}$~M$_\odot$ both \ce{^13CO} and \ce{C^18O} are (marginally) optically thin and photodissociation of both isotopologues reduce their luminosity with increasing radius.

Evidently, the mass and the radius of the disk have the biggest impact on the observables of the disk. The next most impactful parameter affecting the CO luminosity of the disk is $\gamma$ in Eq.~(\ref{eq:surface_density}), which affects the overall distribution of the mass of the disk. A larger $\gamma$ results in a higher surface density at the inner region and decreases the effect of the exponential taper in the outer region, which increases the size of the disk. $\gamma$ especially impacts the lower mass models, see Fig.~\ref{fig:parameter_overview} and the extra Figures in Appendix \ref{app:parameter_overview}. A larger $\gamma$ makes the region where the \ce{C^18O} emission is coming from relatively constant in size and self-shielding occurs less than for smaller $\gamma$ even for the largest $R_c$, while the \ce{^13CO} emission increases with increasing radius because of a larger emitting surface. Hence, the model moves horizontally in the bottom left panel of Fig.~\ref{fig:parameter_overview_gamma_1.5}. For smaller $\gamma$ at \ce{C^18O} can dissociate at large radii due to the lower surface density closer in and the luminosity decreases rapidly, as can be seen as the relatively large jump in luminosity from $R_c=60$~au to 200~au for the $10^{-3}$~M$_\odot$ models in the bottom left panel of Fig.~\ref{fig:parameter_overview}. This jump also causes the bimodal distribution present in the right panel of Fig.~\ref{fig:observations_and_models}. This could have been alleviated by adding a value of $\gamma$ between 0.8 and 1.5. For even lower masses an increase in $\gamma$ results in less change in both \ce{^13CO} and \ce{C^18O} with an increase in radius due to more efficient self-shielding.

All remaining parameters affect the CO luminosity in the same direction as $R_c$, see Figs. \ref{fig:parameter_overview}, and \ref{fig:parameter_overview_Rc_5} to \ref{fig:parameter_overview_gamma_1.5}. Changing $h_c$ does not affect the more massive disks, because these are optically thick already. For the lower mass models increasing $h_c$ has the main affect of increasing the volume of the disk, making the gas optically thin and reducing the abundance of both \ce{^13CO} and \ce{C^18O} due to photodissociation. Increasing the stellar luminosity decreases the abundance of both \ce{^13CO} and \ce{C^18O}, hence the downward left movement of the models in Fig.~\ref{fig:parameter_overview}. For the higher mass disks the CO isotopologues are already shielded and the luminosity barely affects the overall abundance. Lastly, both $\psi$ and the inclination do not affect the observed integrated \ce{^13CO} and \ce{C^18O} luminosities considerably.

\begin{figure*}[t]
    \centering
    \includegraphics[width=\textwidth]{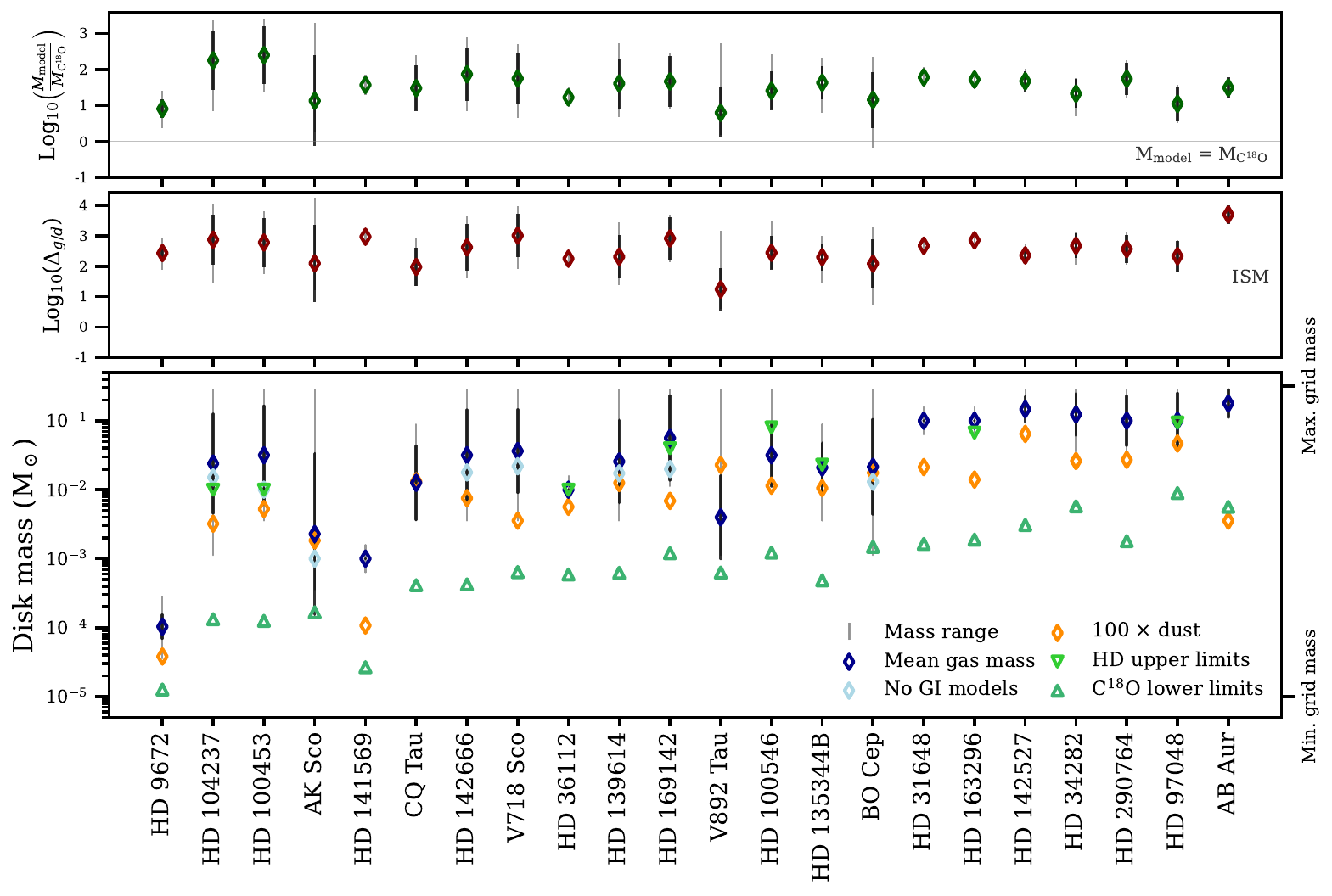}
    \caption{Gas masses of the Herbig disks detected in the CO isotopologues as selected from the models presented in Fig.~\ref{fig:observations_and_models}. The disks are ordered from left to right by increasing \ce{^13CO} luminosity. The gray lines indicate the range in possible disk mass based on the disk parameters listed in Table \ref{tab:disk_params}, while the black lines indicate the standard deviation. The mean mass of these models is given as the dark blue diamond. After removing models which are potentially gravitationally unstable, the mean mass changes into the light blue diamond. The orange diamonds are taken from the dust masses of \citet{stapper2022} multiplied by the canonical gas-to-dust ratio of 100. The HD upper limits from \citet{Kama2020} are shown as green downward pointing triangles. The \ce{C^18O} lower limits are directly obtained from the integrated flux. The middle panel shows the resulting gas-to-dust ratio based on the mass range and mean values, and the top panel shows the factor by which the mass could be underestimated when using \ce{C^18O} assuming optically thin emission.}
    \label{fig:gas_masses}
\end{figure*}

\subsection{Obtaining the mass of the disk}
\label{subsec:obtaining_mass}
As discussed in the previous section, Figure \ref{fig:observations_and_models} shows that most observations are consistent with the highest possible masses based on their \ce{^13CO} and \ce{C^18O} luminosities. However, at least for the observations detecting \ce{^12CO}, \ce{^13CO}, and \ce{C^18O}, the mass range can be further constrained by using the observed size of the disk in the selected isotopologues.

\subsubsection{Masses from detected emission}
\label{subsubsec:detected_obs}
To obtain a measure of the mass of the disk, we select all models within a region around the observations in Fig.~\ref{fig:observations_and_models} based on the confidence intervals of the observations. There are two main sources of uncertainty that we take into account: a systematic uncertainty set by the calibration accuracy, and a statistical uncertainty set by the noise in the moment~0 map. We select all models which fall within $3\sigma$ from a line given by the 10\% systematical uncertainty. For some of the more luminous targets the obtained uncertainty is relatively small and either no or very few models were found within the region confined by the two sources of uncertainty. If less than 200 models were found to fall within the given region of a disk, the closest 200 models were included, this occurred for 11 disks. Typically, the selected models deviated around 10\% from the observed luminosity. For HD~141569 we take all models below the \ce{C^18O} upper limit and within a 10\% systematic uncertainty on the \ce{^13CO} luminosity because it dominates over the noise. 

After obtaining this set of models, selections based on the size of the disk and its inclination, and on the stellar luminosity can be made (see Tables \ref{tab:disk_params} and \ref{tab:fluxes_and_radii}). Given the lack of \ce{^12CO} observations for some of these disks, together with possible cloud contamination, we use the \ce{^13CO} R$_{90\%}$ as a disk size tracer. We select all models within a factor of 1.4 in size compared to the observations, as this is the factor by which the models are on average smaller than the observations in \ce{^13CO}, we discuss this further at the and of this section. Based on the radius, inclination, and stellar luminosity, the constraints on the disk mass can be tightened. Figure \ref{fig:gas_masses} presents the resulting range in possible disk mass values and the (logarithmic) mean mass of the models within that range, as shown by the light gray lines and darkblue diamonds respectively. The spread in mass of the selected models is given by the black lines. For the high-mass disks (on the right), that correspond to bright \ce{^13CO} emission, the masses are well constrained to be above $10^{-1.5}$~M$_\odot$. These stringent lower limits are mainly due to their large and well-known sizes. Our obtained disk masses can be found in Appendix~\ref{app:disk_masses}.

Figure \ref{fig:mgas_vs_lum} shows that the \ce{C^18O} luminosity is sensitive to the gas mass when selecting models with the appropriate size based on our grid of models, each panel showing this for differently sized models in \ce{^12CO}. We note that this is the $R_{90\%}$ radius, which should not be confused with $R_c$. This clearly shows that the size of the disk is an indicator of its mass as well. One cannot make disks larger than $\sim500$~au without the disk mass being higher than $\sim10^{-2}$~M$_\odot$. On the other hand, for the lower mass disks the tenuous \ce{^12CO} gas cannot self-shield in the outer regions anymore for the largest disks, reducing the measured size of the disk. Consequently, one obtains a better constraint on the disk mass of the largest disks in the sample.

For the smaller disks, more masses are compatible with the observed \ce{^13CO} and \ce{C^18O} line luminosity due to the higher optical depth. This results in low lower limits on the gas mass from our models. For example AK~Sco and HD~104237 have lower limits of $\sim10^{-3.5}$~M$_\odot$ and $\sim10^{-3}$~M$_\odot$ respectively. This is also evident from Figure \ref{fig:mgas_vs_lum}, clearly showing that the different gas masses give similar \ce{C^18O} luminosities for the smallest disks. For the smallest radius bin size a lower limit on the size of the disk can be given as well. Disks which are smaller than R$_{90\%}$=20~au in \ce{^12CO} and have a mass of more than $10^{-4}$~M$_\odot$ do not occur in our models.

To constrain the disk masses even more, one can include other parameters as well. For example, for AK~Sco and HD~142666 the vertical extent of the disk from \citet{stapper2023} can be exploited. We implement this in a simple way, where for the very flat disks AK~Sco and HD~142666 we choose models with $h_c=0.05$~rad. While this does not improve the overall range in possible disk masses (as $h_c$ does not considerably change the overall luminosity, see Section \ref{subsec:13CO_C18O_luminosities}), for both flat disks we can rule out either lower mass models for AK~Sco or higher mass models for HD~142666. This difference is due to their relative position in Fig.~\ref{fig:observations_and_models}.

For two low mass disks, HD~9672 and HD~141569, we can determine their disk mass within an order of magnitude. Based on their position in Fig.~\ref{fig:observations_and_models} (the two disks with the lowest \ce{C^18O} luminosity or upper limits thereon), we can already see that the gas mass should be around $10^{-3}$-$10^{-4}$~M$_\odot$. No other models fall within the computed confidence intervals. Note that for HD~9672 we do not reproduce the size of the \ce{^13CO} disk well with the models. We will comment more on this in Section \ref{subsec:debris_disks}. This results in a well determined disk mass of $10^{-4}$~M$_\odot$ for HD~9672 and $10^{-3}$~M$_\odot$ for HD~141569.

\begin{figure*}[t]
    \centering
    \includegraphics[width=1.0\textwidth]{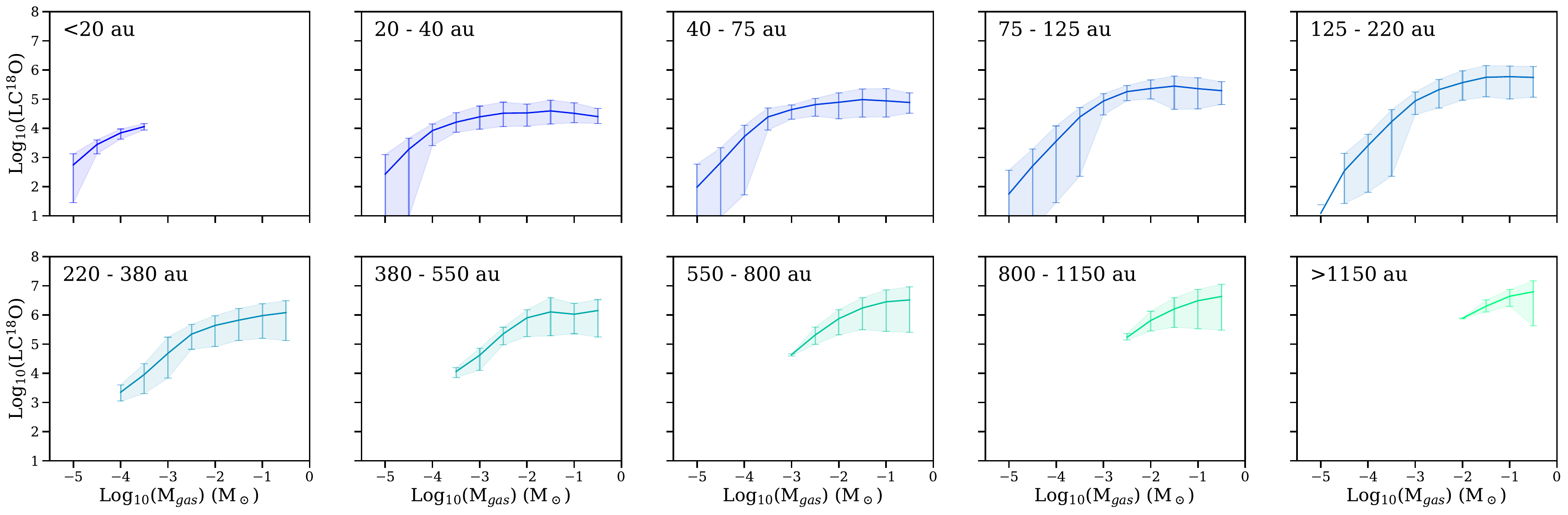}
    \caption{Range of \ce{C^18O} luminosities for different mass disks, selected for \ce{^12CO} $R_{90\%}$ radii from the DALI models presented in Fig.~\ref{fig:observations_and_models}. The shaded area indicates the minimum and maximum values, and the solid line is the mean value.}
    \label{fig:mgas_vs_lum}
\end{figure*}

Based on the stellar mass of the Herbig stars, and the disk mass and disk outer radius of \ce{^12CO} from the models, we can determine if a specific model would be gravitationally unstable around the Herbig star using the relation 

\begin{equation}
    \frac{M_d}{M_\star} > 0.06 \left(\frac{f}{1}\right) \left( \frac{T}{10\rm{~K}} \right)^{1/2} \left( \frac{r}{100 \rm{~au}} \right)^{1/2} \left( \frac{\rm{M}_\odot}{M_\star} \right)^{1/2}
    \label{eq:GI}
\end{equation}

\noindent
from \citet{Kratter2016}, where $T$ is the temperature of the disk, $r$ the outer radius of the disk (given by the measured \ce{^13CO} R$_{90\%}$ radius) and $M_\star$ the mass of the star, we set the pre-factor $f$ equal to 1. The temperature $T$ is determined with the luminosity of the Herbig star via the relation by \citet{Andrews2013}. For seven disks the radius of the disk combined with Eq.~(\ref{eq:GI}) is constraining enough to result in a lower mean disk gas mass by ruling out the potentially gravitationally unstable models. For the disks with radius upper limits especially, the compact high mass disks can be ruled out if assuming that these observed disks are not gravitationally unstable.

For eight of the Herbig disks, upper limits on the gas masses have been estimated from HD \citep{Kama2020}, see the downward facing triangles in Figure \ref{fig:gas_masses}. In general, the gas masses we find are consistent with the HD upper limits. For HD~163296 and HD~36112, the HD upper limits are roughly equal to our gas mass estimates, suggesting that the observations might have been close to detecting HD. For HD~100453, HD~169142 and HD~97048 the mean gas masses of the models are higher than the HD upper limits, implying that these disks are not as massive as their size and \ce{^13CO} and \ce{C^18O} luminosities suggest. A combination of modeling the CO emission and HD upper limits could add additional constraints to the obtained gas masses.

Figure \ref{fig:gas_masses} also shows 100$\times$ the dust mass from \citet{stapper2022}. For the five NOEMA targets we compute the dust masses in the same way as \citet{stapper2022}, see Appendix \ref{app:noema_data}. The middle panel in Fig.~\ref{fig:gas_masses} shows the resulting gas-to-dust ratio from the mean gas mass as the dark red diamonds and from the range in gas masses as the vertical gray line. The horizontal gray line indicates a gas-to-dust ratio of 100. For most disks we find that the total disk mass derived from the dust mass is consistent with a gas-to-dust ratio of 100. Some disks suggest a depletion of dust compared to the interstellar medium. Primarily for AB~Aur the $100\times$ dust mass falls well below the gas mass range, giving a gas-to-dust ratio that is two orders of magnitude higher than the canonical value. For some of the other higher mass disks, such as HD~169142, the dust mass also suggests a depletion of dust compared to gas of a factor of a few. This apparent depletion of dust might be related to the dust being optically thick, we comment on this in \S\ref{subsubsec:cdfs}. In general however, the dust mass does seem to indeed trace the total disk mass relatively well for Herbig disks, in contrast with the T~Tauri disks.

Lastly, we note that the size of the models are generally smaller than those of the observations for the same flux in a particular CO isotopologue. When selecting all models within a factor of two of the observations, we find that the \ce{^12CO}, \ce{^13CO} and \ce{C^18O} radii of the models are on average a factor 1.3, 1.4 and 1.9 times smaller than the observed radii. This is likely due to the models being smooth, with no substructures present. The observations do show some gas structures which increase the emitting area of the disk, and also present weak emission extended structures which increase their $90\%$ radius without much affecting the overall flux. Specifically gas cavities, the main gas structure seen in our data, are generally found to be smaller than continuum cavities \citep{vanderMarel2016, Leemker2022}, and thus are likely not affecting the gas as much as it would to the continuum. Comparing our found gas masses to existing disk specific modeling efforts taking into account the structure of the disk show only differences of a factor of a few. For example the MAPS Herbig disks HD~163296 and HD~31648 were found to have gas masses of 0.14~M$_\odot$ and 0.16~M$_\odot$ respectively \citep{Zhang2021}, within a factor of a few from our derived masses (other examples include \citealt{Tilling2012, Flaherty2015, Flaherty2020}). Scaling the observed radii by the aforementioned factors does indeed not affect the inferred range in disk masses for a given disk. However, when using Fig.~\ref{fig:mgas_vs_lum} one should keep this in mind when selecting which panel to use. Implementing gas and dust structures in the models exceeds the scope of this work, but future work could also consider the effect of these structures on the obtained dust and gas masses.

\subsubsection{Mass lower limits from \ce{C^18O}}
\label{subsubsec:c18o_lower_limits}
\begin{figure}[t]
    \centering
    \includegraphics[width=0.5\textwidth]{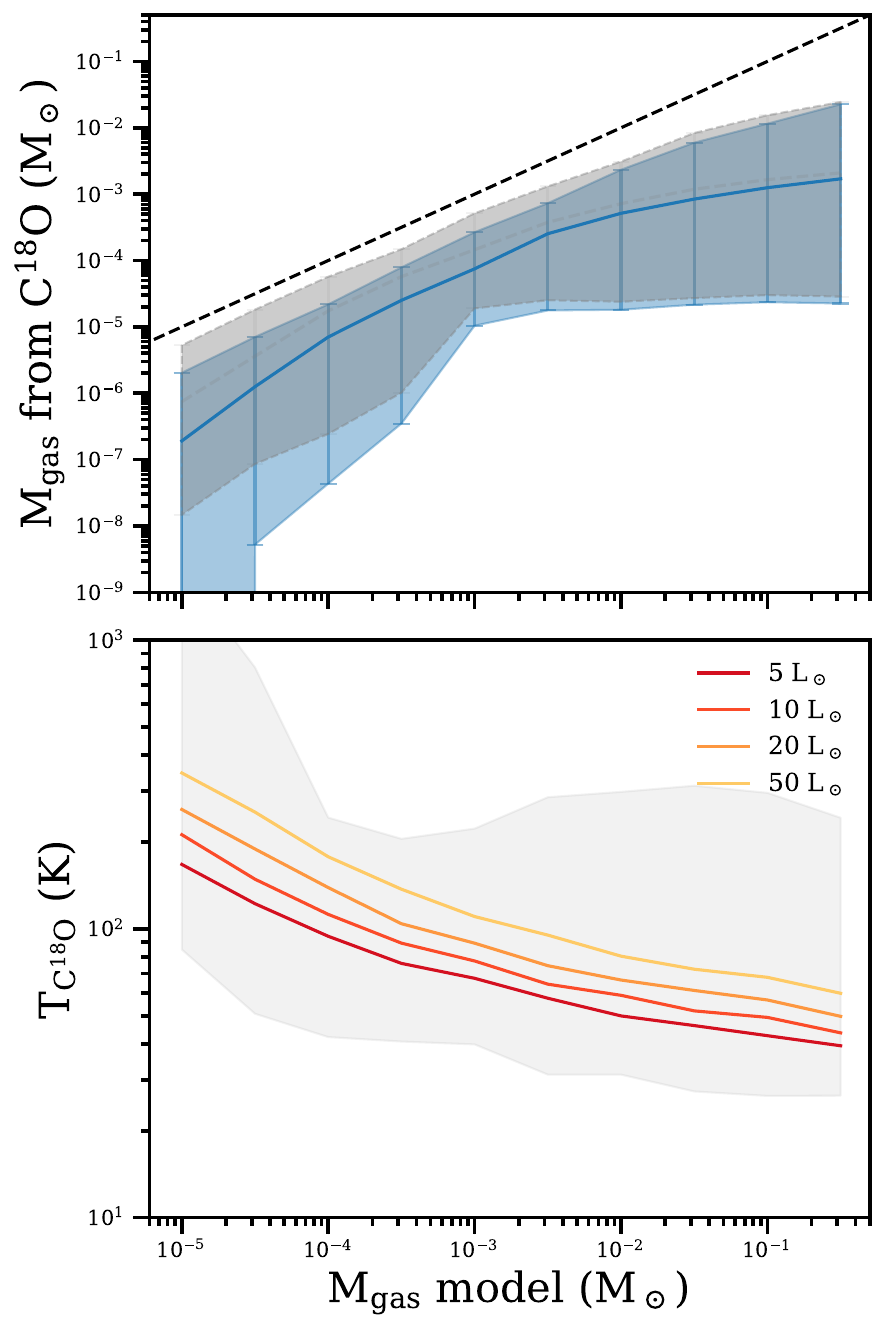}
    \caption{Gas mass of the models versus the retrieved gas mass from the \ce{C^18O} flux using Eq.~(\ref{eq:number_of_molecules}), and the mass-weighted average of the temperature in the \ce{C^18O} emitting region. For the blue shaded region a temperature of 40~K is used, the gray shaded region uses the temperature of each model shown in the bottom panel. The shaded areas are the minimum and maximum values, with the mean value indicated by the solid line. The dashed line indicates the one-to-one correspondence between the two masses. The typical range in temperatures is indicated by the gray shaded region in the bottom panel.}
    \label{fig:model_c18o_gas_mass}
\end{figure}

In many works \citep[see e.g.,][]{Hughes2008, Loomis2018, Miley2018, Booth2020}, the disk mass has been estimated by using a flux scaling relation, assuming the emission is optically thin. This formula has been used to obtain a lower limit on the gas mass from \ce{^12CO}, \ce{^13CO}, and \ce{C^18O} flux. As we find that most Herbig disks are optically thick in \ce{C^18O}, it is useful to obtain a measure of how much the total gas mass is underestimated when using \ce{C^18O} as a gas mass tracer. The total number of \ce{C^18O} molecules in the disk from the \ce{C^18O} flux can be calculated, assuming optically thin emission \citep[see e.g.,][]{Loomis2018}, with

\begin{equation}
    n_{\rm c^{18}o} = \frac{4\pi}{h c} \frac{F_\nu \Delta V d^2}{A_{ul} x_u},
    \label{eq:number_of_molecules}
\end{equation}

\noindent
where $h$, $c$, and $A_{ul}$ are the Planck constant, speed of light, and the Einstein A coefficient for spontaneous emission respectively. $x_u$ is the fractional population of the upper level, $d$ the distance to the source, and $F_\nu \Delta V$ is the velocity integrated flux over the disk. Using a ratio between CO and H$_2$ of $2.7\times10^{-4}$ \citep{Lacy1994}, and ratios of 77 and 560 of \ce{^12C}/\ce{^13C} and \ce{^16O}/\ce{^18O} in the local interstellar medium \citep{Wilson1994}, in combination with a factor of 2.4 for the mean molecular weight, the total disk mass can be calculated. An excitation temperature is necessary to compute the population levels using the Boltzmann equation as well. We assume $T_\text{ex}=40$~K which is higher than the \ce{C^18O} brightness temperatures of some of the Herbig disks (see e.g., \citealt{Zhang2021}), but is the same as other disks in our sample (e.g., HD~100546, HD~135344B, and HD~169142). We use the line properties (e.g., partition function values, $A_{ul}$, $E_u$, and $g_u$) from CDMS \citep{Endres2016}. Using the integrated \ce{C^18O} fluxes from Table \ref{tab:fluxes_and_radii}, we obtain lower limits on the total gas mass of the disk, see Figure \ref{fig:gas_masses}. The formula is also applied to our models, a comparison can be found in Fig.~\ref{fig:model_c18o_gas_mass} and the upper panel in Fig.~\ref{fig:gas_masses}.

Compared to the masses we find from the models, the masses obtained with Eq.~(\ref{eq:number_of_molecules}) are a factor of 10-100 times lower, see Fig.~\ref{fig:gas_masses}. In the top panel of Figure \ref{fig:model_c18o_gas_mass}, the blue shaded region indicates the retrieved masses using Eq.~(\ref{eq:number_of_molecules}) with a temperature of 40~K compared to the true mass of the model. This large difference between the two is primarily due to two reasons. First, for the most massive disks, the \ce{C^18O} emission is optically thick which reduces the obtained disk gas mass. The disks for which the retrieved masses are closest to the true disk mass are also the largest disks, indicating that these are (marginally) optically thin. For the highest mass disks, even the largest disks are not optically thin as the maximum retrieved mass bend towards relatively lower values. For the lowest mass disks on the other hand (e.g., HD~9672 and HD~141569) photodissociation becomes the dominant process reducing the CO abundance. Consequently, the mass of the disk is always underestimated when connecting \ce{C^18O} directly to the mass of the disk. As can be seen in both Figs.~\ref{fig:gas_masses} and \ref{fig:model_c18o_gas_mass}, the lowest mass disks have very low luminosities and retrieved masses due to photodissociation. The gray shaded area in the top panel of Fig.~\ref{fig:model_c18o_gas_mass} are the retrieved masses after using the gas temperature in the emitting region of \ce{C^18O} where 90\% of the emission is coming from as weighted by the mass in each cell, as indicated in the bottom panel. Both lower mass disks and smaller disks are warmer in the \ce{C^18O} emitting region, hence using a temperature of 40~K is not adequate. Especially for the lower mass disks this can underestimate the total disk mass with a factor of a few in addition to the lower CO abundance.

\begin{figure}[t]
    \centering
    \includegraphics[width=0.5\textwidth]{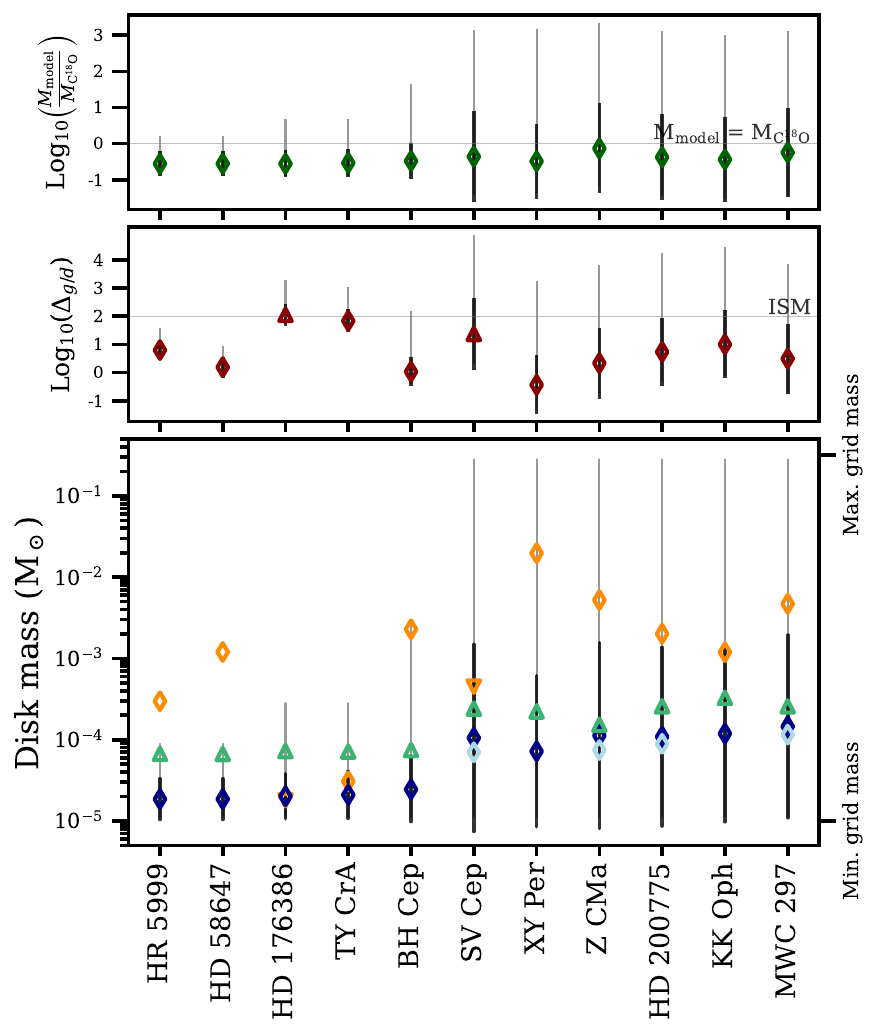}
    \caption{Same as Figure \ref{fig:gas_masses}, but now for the disks with only upper limits on the \ce{^13CO} and \ce{C^18O} fluxes.}
    \label{fig:gas_masses_non_detections}
\end{figure}

\subsubsection{Upper limits}
\label{subsubsec:mass_upper_limits_obs}
For the disks with upper limits on both isotopologues no selection can be made based on a region of luminosities in Figure \ref{fig:observations_and_models} nor the size of the disk can be used. Using the dust radii of \citet{stapper2022} in combination with the typical ratio of 2.7 between the gas and dust radii (see Section \ref{subsec:12CO_radius}) does not add any constraints either as for all disks only upper limits on the dust disk size are known. Hence, a selection of all models within the quadrant confined by the upper limits is made. Figure \ref{fig:gas_masses_non_detections} presents the resulting upper limits on the gas masses.

\begin{figure*}[t]
    \centering
    \includegraphics[width=\textwidth]{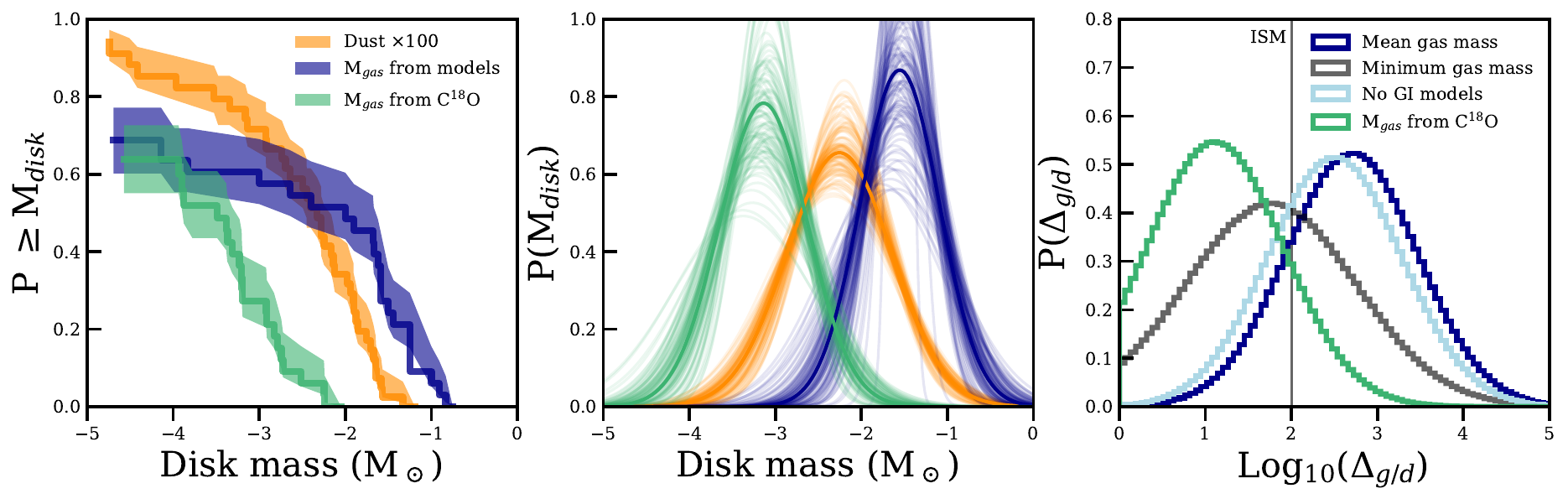}
    \caption{Cumulative distributions of the gas masses from Figs.~\ref{fig:gas_masses} and \ref{fig:gas_masses_non_detections} obtained from our models (dark blue) and obtained from the \ce{C^18O} flux (green), and dust masses $\times100$ (orange) from \citet{stapper2022}. The corresponding probability distributions are shown in the middle panel, obtained by fitting a lognormal distribution to the cumulative distributions. The faint lines indicate the possible range in fits. Sampling from the gas and dust distributions a distribution of possible gas-to-dust mass ratios can be made, as shown in the right most panel.}
    \label{fig:cdf}
\end{figure*}

\begin{figure}[t]
    \centering
    \includegraphics[width=0.4\textwidth]{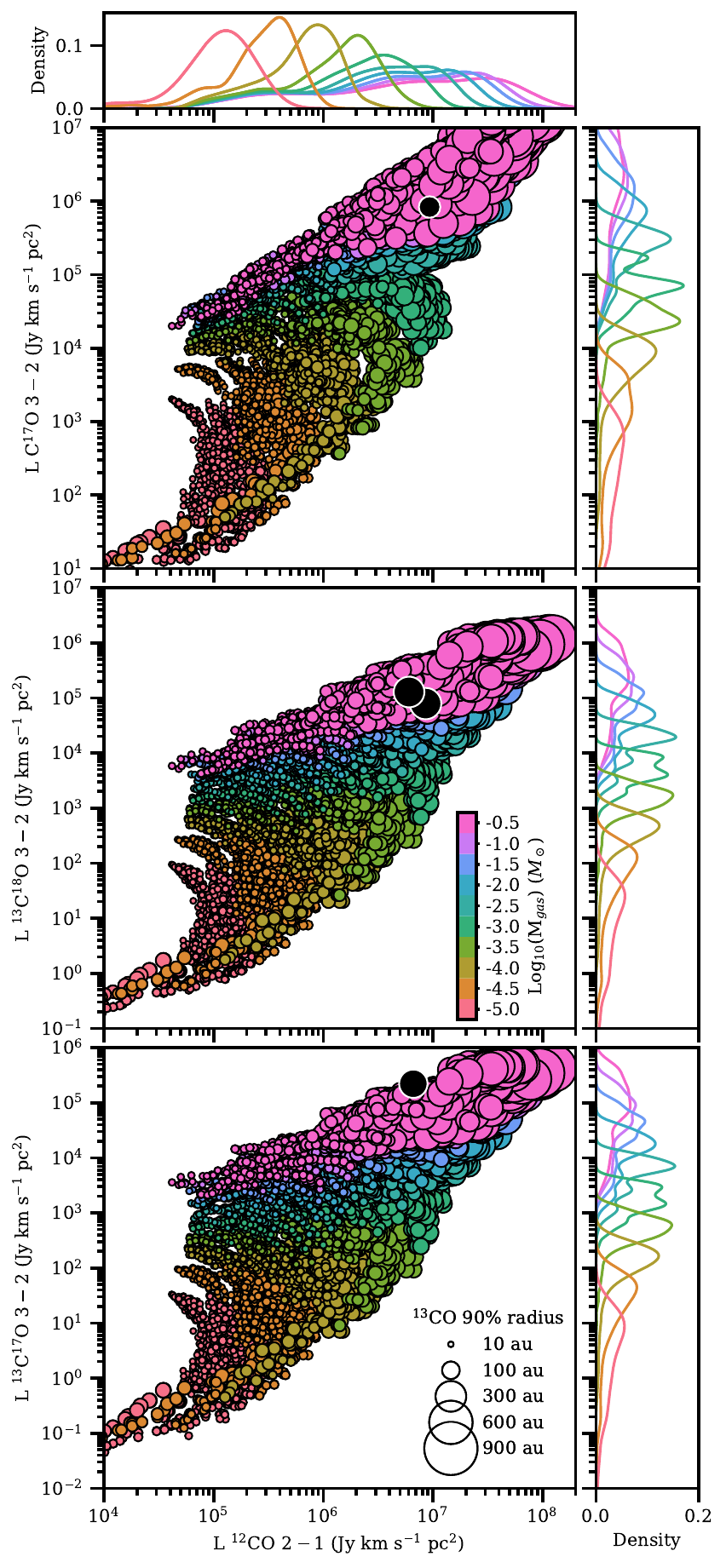}
    \caption{Similar to Fig.~\ref{fig:observations_and_models}, but now with \ce{C^17O}, \ce{^13C^18O} and \ce{^13C^17O} on the vertical axes and \ce{^12CO} on the horizontal axis. The black dots mark HD~100546 observations of \ce{C^17O} $J=3-2$ (priv.comm.), HD~163296 observations of \ce{^13C^17O} $J=3-2$ from \citet{Booth2019}, HD~31648 observations of \ce{^13C^18O} $J=3-2$ from \citet{Loomis2020}, and HD~142527 observations of \ce{^13C^18O} $J=3-2$ by \citet{Temmink2023}. Similar to Fig.~\ref{fig:observations_and_models} probability density curves are shown along the vertical and horizontal axes to indicate where most of the models of a particular mass reside.}
    \label{fig:rare_isos}
\end{figure}

Based on the parameters in Table \ref{tab:disk_params}, a few constraints on the possible disk mass can be made. This is especially true for the \ce{^13CO} and \ce{C^18O} upper limits on the left side of Fig.~\ref{fig:gas_masses_non_detections}, as these upper limits include models of all masses. While for the undetected or unresolved disks the inclination and radius do not give constraints, the luminosity gives enough constraints to lower the upper bound on some of the disks by multiple orders of magnitude, see Fig.~\ref{fig:gas_masses_non_detections}. We find that HD~58647 and HR~5999 can have at most a mass of $10^{-4}$~M$_\odot$ to explain the non-detection in both \ce{^13CO} and \ce{C^18O}. Interestingly, for both of these disks a higher dust mass is found, suggesting an increased abundance of dust compared to the ISM. For HD~176386 and TY~CrA an upper limit of $10^{-3.5}$~M$_\odot$ is found. Z~CMa has a higher upper limit of $10^{-2.5}$~M$_\odot$. The standard deviation, as given by the black lines in Fig.~\ref{fig:gas_masses_non_detections}, show tighter constraints. For the other upper limits no constraints on the gas mass could be made, even after selecting for the source-specific parameters. However, we do note that the $10^{-1}$~M$_\odot$ models are by far outnumbered by lower mass models as most, but not all, high mass models can be excluded. This is reflected in the mean gas mass and standard deviation, as these are much lower than the maximum mass possible, see the dark blue diamonds and black lines in Fig.~\ref{fig:gas_masses_non_detections}.

Based on the \ce{C^18O} upper limits, a maximum radius can be estimated for these disks. As Fig.~\ref{fig:mgas_vs_lum} shows, an increase in size increases the luminosity of the disk. Hence, upper limits on the \ce{^12CO} radius can be determined for the disks with no detections. We find upper limits of 550~au (HD~58647, HD~176386, HR~5999, TY~CrA and Z~CMa) and 800~au (BH~Cep, HD~200775, KK~Oph, MWC~297 and SV~Cep). Besides non-detections, we also have four disks which have upper limits on their radius, for which we obtain an additional minimum radius to explain the found luminosities, as a decrease in size decreases the luninosity of the disk (see Fig.~\ref{fig:mgas_vs_lum}). BO~Cep has at least a size of 125~au, and HD~139614 and V718~Sco have a size of at least 40~au. For HD~104237 we obtain the most stringent lower limit of 15~au.

Similarly, for HD~245185 and VV~Ser, which only have \ce{^12CO} observations, the mass and radius can also be constrained. \ce{^12CO} is optically thick for lower disk masses, resulting in the same luminosities for multiple orders of magnitude in mass even for the largest disks. These luminosities can be used to obtain a lower limit on the mass and size of HD~245185. To explain the observed \ce{^12CO} emission, the disk needs to be at least 125~au in size and have a mass lower limit of $2\times10^{-4}$~M$_\odot$. Due to the non-detection of \ce{^12CO} for VV~Ser, the radius can be relatively well constrained to be less than 220~au in size using the \ce{^12CO} equivalent to Fig.~\ref{fig:mgas_vs_lum}.

Lastly, \citet{Grant2023} have shown that the relationship between the accretion rate $\dot{M}$ and dust disk mass is largely flat at $\sim10^{-7}$~M$_\odot$~yr${-1}$ over three orders of magnitude in dust mass. Hence, some disks have very short inferred disk lifetimes. Our gas mass estimates do not resolve this problem, as the inferred disk masses from our models are either low (see Fig.~\ref{fig:gas_masses_non_detections}), or no observations are available.

\renewcommand{\arraystretch}{1.2}
\begin{table}[t!]
\caption{Log-normal distribution fit results for the dust and gas mass cumulative distributions shown in Fig.~\ref{fig:cdf}. The values are in log base 10.}
\centering
\begin{tabular}{l|cc|cc}
\hline\hline
                                      & \multicolumn{2}{c}{$M_\text{disk}$ ($M_\odot$)}  & \multicolumn{2}{c}{$\Delta_{g/d}$}   \\ \hline
                                      & $\mu$                   & $\sigma$               & $\mu$ & $\sigma$ \\ \hline
Dust $\times100$                      & -2.25$^{+0.05}_{-0.05}$ & 0.61$^{+0.06}_{-0.06}$ &       &          \\
$\overline{M}_{\rm gas, model}$       & -1.55$^{+0.06}_{-0.07}$ & 0.46$^{+0.12}_{-0.10}$ & 2.70  & 0.76     \\
$\overline{M}_{\rm gas, model}$ no GI & -1.75$^{+0.05}_{-0.06}$ & 0.48$^{+0.15}_{-0.18}$ & 2.50  & 0.78     \\
Min. $M_{\rm gas, model}$             & -2.49$^{+0.18}_{-0.20}$ & 0.78$^{+0.20}_{-0.17}$ & 1.76  & 0.98     \\
$M_{\rm gas, C^{18}O}$                & -3.14$^{+0.06}_{-0.07}$ & 0.51$^{+0.09}_{-0.08}$ & 1.11  & 0.79     \\ \hline
\end{tabular}\\
\label{tab:cdf_fit}
\end{table}

\subsubsection{Cumulative distributions}
\label{subsubsec:cdfs}
Figure \ref{fig:cdf} shows cumulative distributions of the dust masses from \citet{stapper2022}, together with the gas masses as found by the models and computed with Eq.~(\ref{eq:number_of_molecules}). Following \citet{stapper2022}, we obtained the cumulative distributions using the \texttt{Python} package \texttt{Lifelines} \citep{DavidsonPilon2021}. The shaded area indicates the $1\sigma$ confidence intervals. The dark blue distribution is made from the mean values in Figs.~\ref{fig:gas_masses} and \ref{fig:gas_masses_non_detections}. The observed disks for which the range in model disk masses extends down to the lowest mass in our model grid are considered as an upper limit. The dust distribution in orange is obtained by using the dust masses from \citet{stapper2022} in addition to the five NOEMA targets presented in this work (see Appendix \ref{app:noema_data}). The green cumulative distribution is obtained from computing the gas using Eq.~(\ref{eq:number_of_molecules}). The dust and gas distributions show a similar slope for the highest mass disks, indicating a relatively constant overestimate of the gas mass, or a constant underestimate of the dust mass. For lower mass disks, the distribution is set by the non-detections, resulting in a leveling-off at a $p\sim0.7$ of the distribution reflecting the non-detection rate of 31\% (11/35).

Using a bootstrapping method (see for details \citealt{stapper2022}), a lognormal distribution is fit to the cumulative distributions to obtain a probability distribution, see the middle panel in Fig.~\ref{fig:cdf}. The fitting results are presented in Table \ref{tab:cdf_fit}. The best fit distributions are shown as the solid line. The fainter lines are included to demonstrate the range in possible fits. The dust distribution is slightly shifted towards lower masses when adding the five NOEMA targets, but the confidence intervals still overlap (in \citet{stapper2022} $\mu$=-2.18±0.05 and $\sigma$=0.53±0.07). We find that the mean of the $100\times$ dust mass distribution lies $\sim0.7$~dex lower than the mean gas mass distribution, indicating that we do find more gas present than what would be assumed given $100\times$ dust mass in most Herbig disks. This was also the observed trend in Figs.~\ref{fig:gas_masses} and \ref{fig:gas_masses_non_detections}. Removing the gravitational unstable disks only moves the gas masses slightly down and makes the distribution slightly wider, see Table~\ref{tab:cdf_fit}. The distribution of the minimum gas masses from our models lies $\sim0.2$~dex below the mean of the dust mass distribution. Hence, the gas mass is likely higher than this value, consistent with a gas-to-dust ratio of close to or above 100.

To test this, we can sample the distributions obtained by fitting to the cumulative distributions and obtain a gas-to-dust ratio distribution by dividing the resulting gas mass values by the dust masses. After taking $10^7$ samples of each distribution, the gas-to-dust ratio distributions in the right panel of Fig.~\ref{fig:cdf} are obtained. The canonical value of 100 for the gas-to-dust ratio falls within one standard deviation from the mean values of the different gas distributions obtained from our models. The mean of the gas-to-dust ratio made with the minimum gas mass values from the models differs an order of magnitude with the distribution using the mean gas masses. The gas-to-dust ratio distribution obtained from the gas masses based on the \ce{C^18O} flux is even lower at almost two orders of magnitude lower than those found by our models. This emphasizes the necessity of models to determine the disk mass, otherwise fundamental properties such as the gas-to-dust ratio can be severely underestimated.

Lastly, the higher than 100 gas-to-dust ratio may be originating from either optically thick dust or an increased size of dust grains in these disks. As more mass has grown into larger sized grains, the total mass visible at millimeter wavelengths decreases. The findings by \citet{Liu2022} and \citet{Kaeufer2023} show that an order of magnitude in mass can be hidden in the most massive disks. Taking multiple continuum wavelength dust observations into account our gas mass estimates are indeed close to 100 times the dust mass \citep{Sierra2021}. Thus, this order of magnitude difference between $100\times$ the dust mass and the total disk mass is expected.

\section{Discussion}
\label{sec:discussion}

\begin{figure*}[t]
    \centering
    \includegraphics[width=\textwidth]{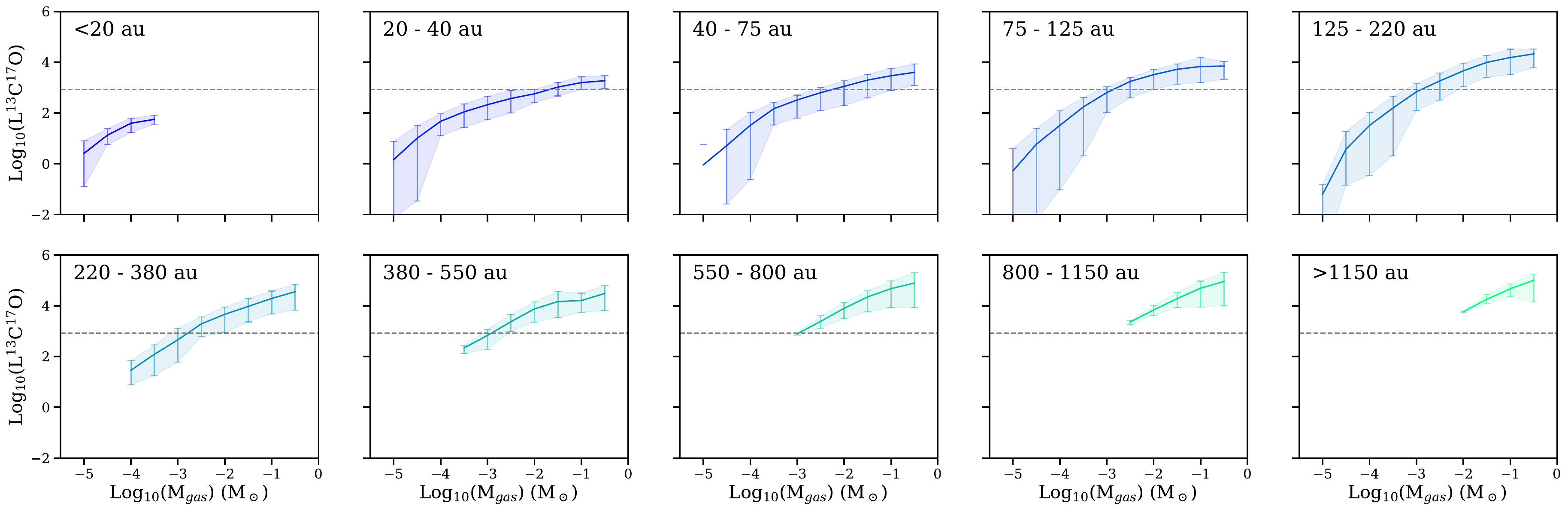}
    \caption{Same as Figure \ref{fig:mgas_vs_lum}, but for \ce{^13C^17O} $J=2-1$. The gray dashed line indicates a typical ALMA $3\sigma$ detection of 1~mJy~km~s$^{-1}$ at a distance of 150~pc, which is achievable within an hour of integration time. The $J=3-2$ transition can be found in Fig.~\ref{fig:mgas_vs_lum_C18O_J32}.}
    \label{fig:mgas_vs_lum_13C17O}
\end{figure*}

\subsection{Comparison between different CO mass tracers}
\label{subsec:different_tracers}
\subsubsection{Rare CO isotopologues}
\label{subsec:rare_isotopologues}
Because of the disks in Fig.~\ref{fig:observations_and_models} being optically thick in both \ce{^13CO} and \ce{C^18O}, a different way to measure the different masses of the disks is necessary. To be able to properly do this, one needs to have a handle on the size of the disk and. have an optically thin tracer. Here the rare(r) isotopologues \ce{C^17O}, \ce{^13C^17O} and \ce{^13C^18O} come into play.

Figure \ref{fig:rare_isos} shows the \ce{C^17O}, \ce{^13C^18O}, and \ce{^13C^17O} $J=3-2$ luminosities plotted against \ce{^12CO} $J=2-1$ of the DALI models, as the observations we compare these to have these transitions available. The optically thick tracer on the horizontal axis gives an indication of the size of the disk, while the optically thinner tracer on the vertical axis gives an indication of the mass of the disk. The same trends (e.g., models of the same mass curve downwards for an increase in radius) can be seen as found in Section \ref{subsec:obtaining_mass}. For \ce{^12CO} the smaller disks are now clearly separated into different areas of this parameter space. A clear stratification can be seen in the luminosities of the optically thinner isotopologues, where each mass has its own range in possible luminosities. However, even for these rare isotopologues, the most massive disks ($>10^{-1.5}$~M$_\odot$) have very similar luminosities. This may be due to the dust opacity playing an important role in reducing the luminosity of the rare isotopologues, which mostly emit from a layer close to the midplane.

We can use observations of rare isotopologues to see if these can assist us in obtaining a better measure of the disk mass. We use the following lines: \ce{C^17O} $J=3-2$ for HD~100546 (priv.comm.), \ce{^13C^18O} $J=2-1$ \citep{ZhangBosman2020} and \ce{^13C^17O} $J=3-2$ \citep{Booth2019} for HD~163296, \ce{^13C^18O} $J=3-2$ \citep{Loomis2018} for HD~31648, and \ce{^13C^18O} $J=3-2$ \citep{Temmink2023} for HD~142527. We select the closest 100 models in luminosity around the observations as shown in Fig.~\ref{fig:rare_isos}, which reproduce the observed luminosities of the rare isotopologues to within $15\%$, except for the \ce{^13C^17O} $J=3-2$ line of HD~163296 for which our models predict a factor of three lower luminosity. With these models we obtain ranges of possible disk masses very similar to the lower limits found with \ce{^13CO} and \ce{C^18O}. For the highest mass models, the luminosities of the very rare isotopologues are still very similar for different mass disks.

However, for compacter lower mass disks ($\lesssim10^{-2}$), for which no rare isotopologues have been observed yet, the luminosity would result in a well constrained gas mass. Figure~\ref{fig:mgas_vs_lum_13C17O} shows the luminosity of the \ce{^13C^17O} $J=2-1$ transition for different mass disks, selected again for different sizes in \ce{^12CO}. This Figure shows that the gas mass of a disk, if resolved in \ce{^12CO}, can be constrained to within an order of magnitude given its \ce{^13C^17O} flux. An integration time of one hour with ALMA typically gives a sensitivity of 1~mJy~km~s$^{-1}$ which at 150~pc gives a luminosity of $8.5\times10^{2}$~Jy~km~s$^{-1}$~pc$^2$. In Fig.~\ref{fig:mgas_vs_lum_13C17O} a $3\sigma$ detection is indicated with the dashed gray line. This shows that for the vast majority of disks, typically larger than $\sim300$~au (Fig.~\ref{fig:flux_and_radius_hists}) a detection is expected if the disk is more massive than $10^{-3}$~M$_\odot$ within an hour of observing time.

\subsubsection{Peeling the disk `onion'}
\label{subsec:peeling_the_onion}
As the previous section showed, an uncertainty of an order of magnitude is still present when determining the mass of the most massive disks. For some of those massive disks \ce{^13C^17O} is only marginally optically thin. Moreover, extrapolating the mass of the disk from the \ce{^13C^17O} emission which is mostly detected in the inner parts of the disk may not be accurate for the disk as a whole. Hence, this section will explore how the disk gas surface density and mass can be constructed by considering, from the inner disk to the outer disk, a series of CO isotopologues with increasing abundance, from \ce{^13C^17O} or \ce{^13C^18O} to \ce{^12CO}. Until now we neglected processes such as radial drift which can enhance the CO/H ratio inside the CO snowline \citep{Zhang2021}, and depleting it outside. This technique is able to take this depletion into account.

\begin{figure*}
    \centering
    \includegraphics[width=\textwidth]{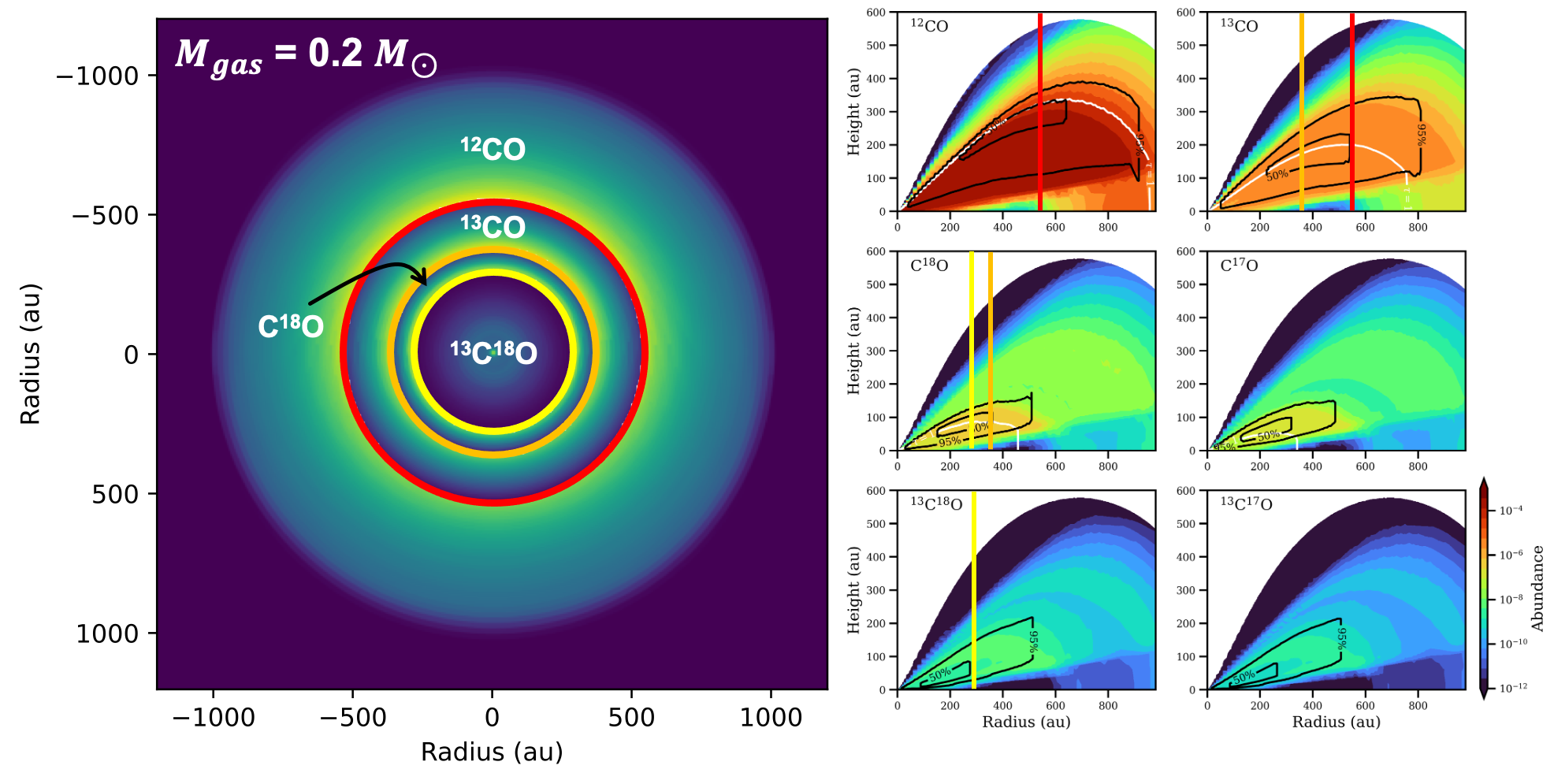}
    \caption{DALI model with a mass of $0.2$~M$_\odot$. The disk is divided into rings based on the $R_{90\%}$ of the isotopologue. The rarest isotopologue (\ce{^13C^18O}) is in the center, while \ce{^12CO} is in the outer ring. The six panels on the right show the abundance of the six ray-traced CO isotopologues. The vertical colored lines correspond to the colors of the circles on the left. The white lines indicate the $\tau=1$ surface, and the black lines indicate where 50\% and 95\% of the emission is coming from.}
    \label{fig:disk_onions}
\end{figure*}

The left panel of Figure \ref{fig:disk_onions} shows a disk model cut into sections of different CO isotopologues. The outer parts of the disk consist of the most abundant isotopologue, \ce{^12CO}. The closer to the star, the rarer the isotopologue to ensure that the tracer stays as optically thin as possible. The regions are determined by the $R_{90\%}$ radii of the isotopologues. The right panels show the abundance of the six CO isotopologues looked at in this work, together with their $\tau=1$ lines and areas within which 50\% and 95\% of the emission is coming from.

Using Eq.~(\ref{eq:number_of_molecules}) we compute the mass of each shell and add those together to obtain a measure of the mass of the disk. We obtain a total mass of 0.05~M$_\odot$ when combining the four regions shown in Fig.~\ref{fig:disk_onions}, which is a quarter of the mass of the model. This recovered mass comprises for 97\% by mass out of \ce{^13C^18O}, with the last 3\% made up of \ce{^13CO} due to the larger emitting area. \ce{^12CO} does not contribute any significant amount due to either high optical thickness or weak emission at large radii. Changing the rarest isotopologue into \ce{^13C^17O} does not change the estimated mass, for it is still dominated by the rarest isotopologue.

Doing the same analysis on the HD~163296 and HD~142527 disks, using an excitation temperature of 40~K (based on the bottom panel of Fig.~\ref{fig:model_c18o_gas_mass}) and the rare isotopologue observations by \citet{Booth2019} and \citet{Temmink2023}, a mass of respectively 0.05~M$_\odot$ and 0.18~M$_\odot$ were found. This is a factor of three lower than what has been found in MAPS for HD~163296 \citep{Zhang2021}, and based on the spiral present one would expect a factor of 1.5 higher disk mass for HD~142527 \citep{Yu2019}. However, these masses are consistent with those found using \ce{^13CO} and \ce{C^18O}, see Figure~\ref{fig:gas_masses}.

In conclusion, rare isotopologues do trace the overall disk mass better than \ce{C^18O}. However, more modeling of these rare isotopologues needs to be done to properly use them.

\begin{figure*}[b]
    \centering
    \includegraphics[width=\textwidth]{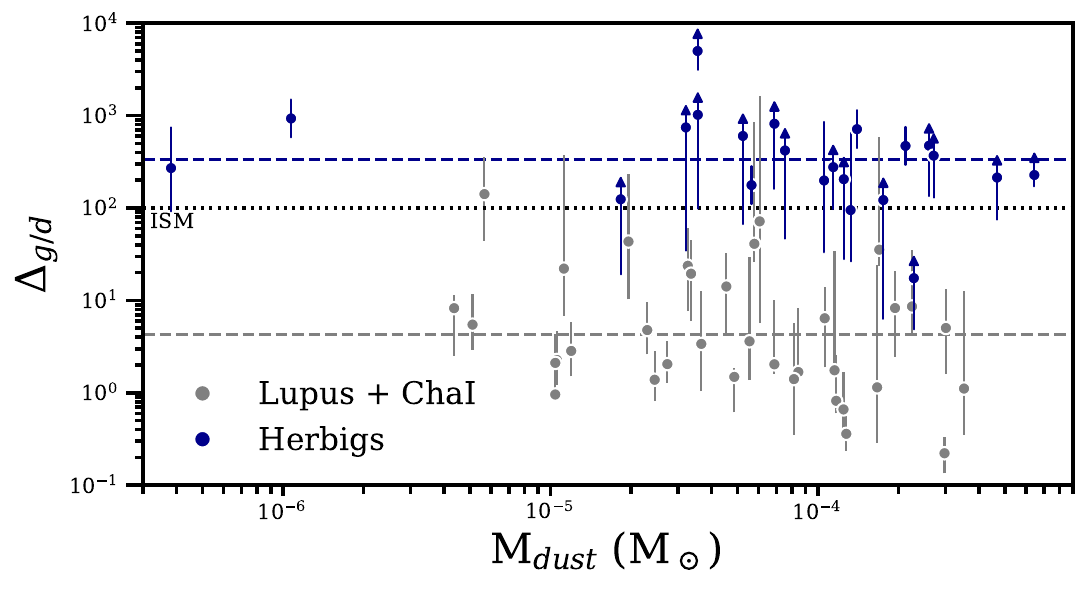}
     \caption{Gas to dust ratio of the Herbig disks compared to T~Tauri disks in Lupus and Chamaeleon~I \citep{Ansdell2016, Pascucci2016, Miotello2017, Long2017}. The corresponding horizontal dashed lines are the logarithmic mean values. The canonical value of 100 is indicated as the horizontal black dotted line.}
    \label{fig:gdr_compare}
\end{figure*}

\begin{figure*}[t]
    \centering
    \includegraphics[width=\textwidth]{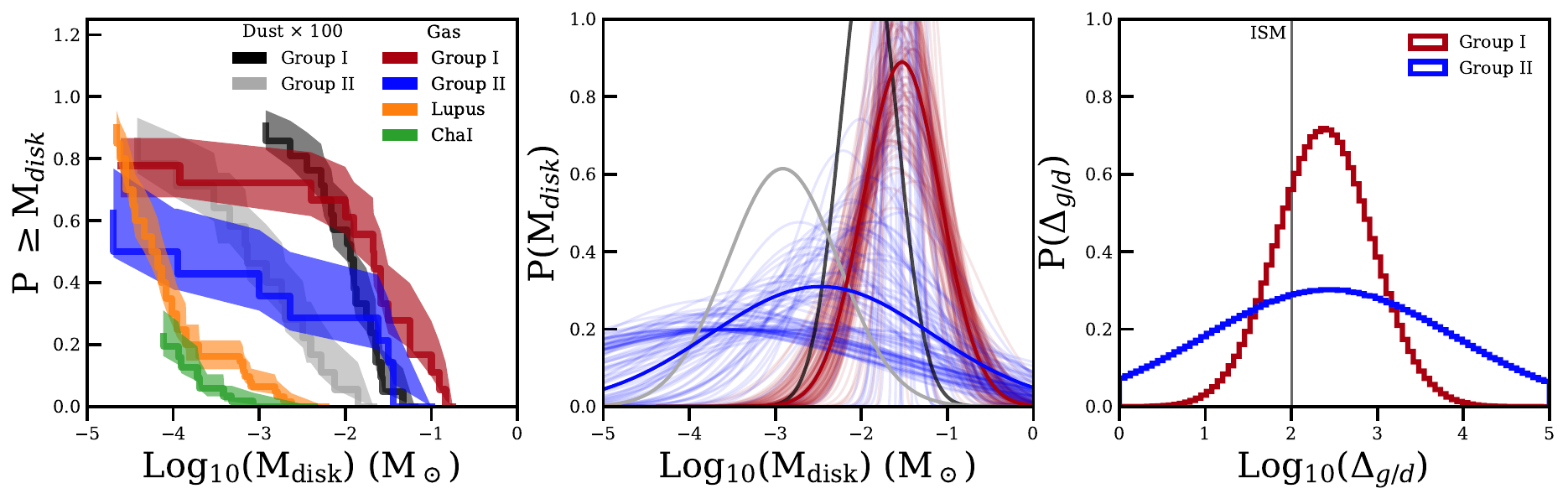}
     \caption{Cumulative distributions of the dust and gas masses of the Herbigs separated into group~I and group~II. The dust distributions are shown in the left and middle panel in dark gray for group~I and light gray for group~II. In addition, the gas distributions from Lupus (orange) and Chamaeleon~I (green) is shown \citep{Miotello2017, Long2017}. The fitted probability distributions are shown in the middle panel. The resulting gas-to-dust ratios are presented in the right panel.}
    \label{fig:cdf_groups}
\end{figure*}

\subsection{Comparison to other disk populations}
\label{subsec:comparisons}
\subsubsection{T~Tauri disks}
\label{subsec:TTauri}
Herbig disks are expected to be warmer than T~Tauri disks, therefore causing less CO depletion due to freeze-out. Using the dust mass estimates from \citet{Ansdell2016} for Lupus and \citet{Pascucci2016} for Chameaeleon~I together with the gas mass estimates by \citet{Miotello2017} and \citet{Long2017} (who use the models from \citealt{Miotello2016}), gas-to-dust ratios of T-Tauri disks are obtained. Figure \ref{fig:gdr_compare} shows these gas-to-dust ratios plotted against the inferred dust masses together with the gas-to-dust ratios obtained for the Herbig disks. This comparison is also done in \citet{Miotello2022}, but we increase the number of Herbigs by more than a factor of two.

As was mentioned before, the mean gas-to-dust ratio of the Herbig disks lies above the canonical ISM value of 100, but many of the disks are still consistent with a gas-to-dust ratio of 100. The T~Tauri disks on the other hand lie at least an order of magnitude below the gas-to-dust ratio of the ISM. Moreover, the data shown in Fig.~\ref{fig:gdr_compare} only includes detected disks. The many non-detections among the T~Tauri disks suggests a lack of CO gas in these disks. The fact that orders of magnitude differences in the gas-to-dust ratio are found, over multiple orders of magnitude in dust mass, confirms the expectation that the warmer Herbig disks lack the CO-conversion processes that occur in the cold T~Tauri disks.

\subsubsection{Group I versus group II}
\label{subsec:GI_vs_GII}

Herbig disks can be separated into two different groups based on their spectral energy distribution (SED): group~I have a rising SED in the far-infrared, while group~II disks do not \citep{Meeus2001}. \citet{stapper2022} showed a difference in disk dust mass between group I and group II disks, where the group II disks have a dust distribution very similar to T~Tauri disks. Moreover, group~I disks are generally large disks with large cavities (transition disks), while group~II disks are more compact \citep{stapper2022, Garufi2017}.

Similar to the distributions in Fig.~\ref{fig:cdf}, we can obtain cumulative distributions for the separate groups as well, see Fig.~\ref{fig:cdf_groups}. Both the gas distributions in red (group~I) and blue (group~II), and the dust distributions in dark gray (group~I) and light gray (group~II) are shown. The dust mass distributions are relatively well constrained, while the gas distributions much less so. The gas distributions are shown in the left most panel in Fig.~\ref{fig:cdf_groups} for Lupus \citep[orange;][]{Miotello2017} and Chamaeleon~I \citep[green;][]{Long2017}.

Comparing the group~I and group~II disks, the group~II disks are less massive than the group~I disks, which is consistent with their dust masses. The probability distributions show an offset in the mean of the distribution of an order of magnitude for both dust and gas (see Table \ref{tab:cdf_fit_groups}). Hence, regardless of the differences in dust and gas masses between the two groups, the relative gas-to-dust ratio remains the same. As the right most panel of Fig.~\ref{fig:cdf_groups} shows, the mean of the gas-to-dust ratio distributions are very similar (see Table~\ref{tab:cdf_fit_groups}). While \citet{stapper2022} found that the group~II disks have a very similar dust mass distribution as T~Tauri disks, compared to group~I disks they have an order of magnitude lower dust mass. We find here that the gas distribution of the group~II disks is shifted towards lower gas masses compared to the group~I disks by the same factor (see Table~\ref{tab:cdf_fit_groups}). This further supports that the gas-to-dust ratio is disk mass independent and rather a result from differences in temperature compared to T~Tauri disks.

Some group~I disks have been found to have little to no CO depletion (e.g., HD~100546 \citealt{Booth2023}, HD~169142 \citealt{Carney2018, Booth2023b}), while group~II disks do in the outer disk \citep{Zhang2021}. As discussed in \citet{stapper2022}, group~I disks show in general large cavities in millimeter emission, while group~II disks are more compact. These differences in CO depletion between the two groups could indicate a much higher impact due to radial drift where most CO stayed in the outer disk for group~I disks and the CO drifted inwards for group~II disks. This is further supported by observations of the metallicities of the hosting Herbig stars \citep{Kama2015, GuzmanDiaz2023}, for which low metallicities generally coincide with group~I disks. Hence, group~I disks might be the formation sites of giant exoplanets stopping radial drift, stopping the enrichment of the host star, trapping CO in the outer disk, and creating the quintessential large cavity structure often associated with these disks. The fact that we find higher gas masses of group~I disks compared to group~II disks only further supports this hypothesis.

\begin{figure}
    \centering
    \includegraphics[width=0.5\textwidth]{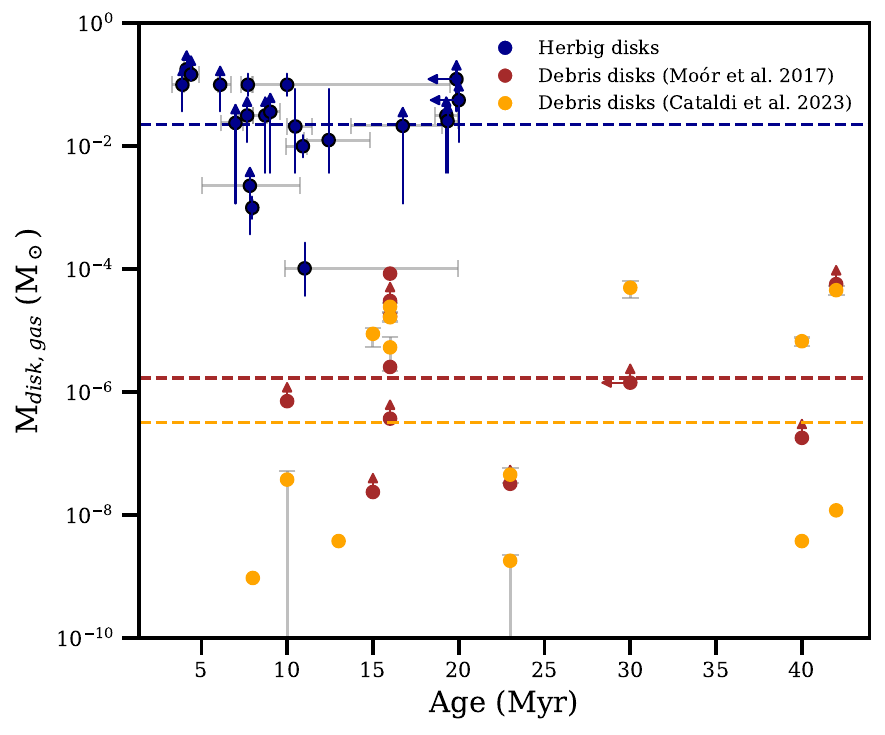}
    \caption{Comparison of the gas masses and ages of debris disks from \citet{Moor2017} and \citet{Cataldi2023} to our Herbig disk gas masses. The horizontal lines indicate the logarithmic means.}
    \label{fig:masses_debris_disks}
\end{figure}

\renewcommand{\arraystretch}{1.2}
\begin{table}[t!]
\caption{Log-normal distribution fit results for the dust and gas mass cumulative distributions shown in Fig.~\ref{fig:cdf_groups}. The values are in log base 10.}
\centering
\begin{tabular}{l|cc|cc}
\hline\hline
                              & \multicolumn{2}{c}{$M_\text{disk}$ ($M_\odot$)}  & \multicolumn{2}{c}{$\Delta_{g/d}$}   \\ \hline
                              & $\mu$                   & $\sigma$               & $\mu$ & $\sigma$ \\ \hline
\textbf{Group I}              &                         &                        &       &          \\
Dust $\times100$              & -1.91$^{+0.05}_{-0.05}$ & 0.33$^{+0.06}_{-0.06}$ &       &          \\
$\overline{M}_{gas, model}$   & -1.53$^{+0.08}_{-0.08}$ & 0.45$^{+0.10}_{-0.10}$ & 2.39  & 0.56     \\\hline
\textbf{Group II}             &                         &                        &       &          \\
Dust $\times100$              & -2.91$^{+0.14}_{-0.17}$ & 0.65$^{+0.16}_{-0.14}$ &       &          \\
$\overline{M}_{gas, model}$   & -2.47$^{+0.85}_{-1.11}$ & 1.29$^{+0.71}_{-1.01}$ & 2.44  & 1.44     \\\hline
\end{tabular}\\
\label{tab:cdf_fit_groups}
\end{table}

Regarding the sizes of the group~I and group~II disks, \citet{Brittain2023} report a comparison of the \ce{^13CO} and dust radii for 17 Herbig disks. They find that the group~II disks are the smallest disks in both gas and dust and that in general the ratio between the dust and gas radii are consistent with the Lupus disks. Using \ce{^12CO} radii we do not report a similar difference between the two groups in gas observations (see Fig.~\ref{fig:Rg_vs_Rd}) as most of the smallest disks lack any detection of \ce{^12CO}. Hence, this remains inconclusive. A uniform survey of Herbig disks would help in characterizing these differences.

\begin{figure*}[t]
    \centering
    \includegraphics[width=\textwidth]{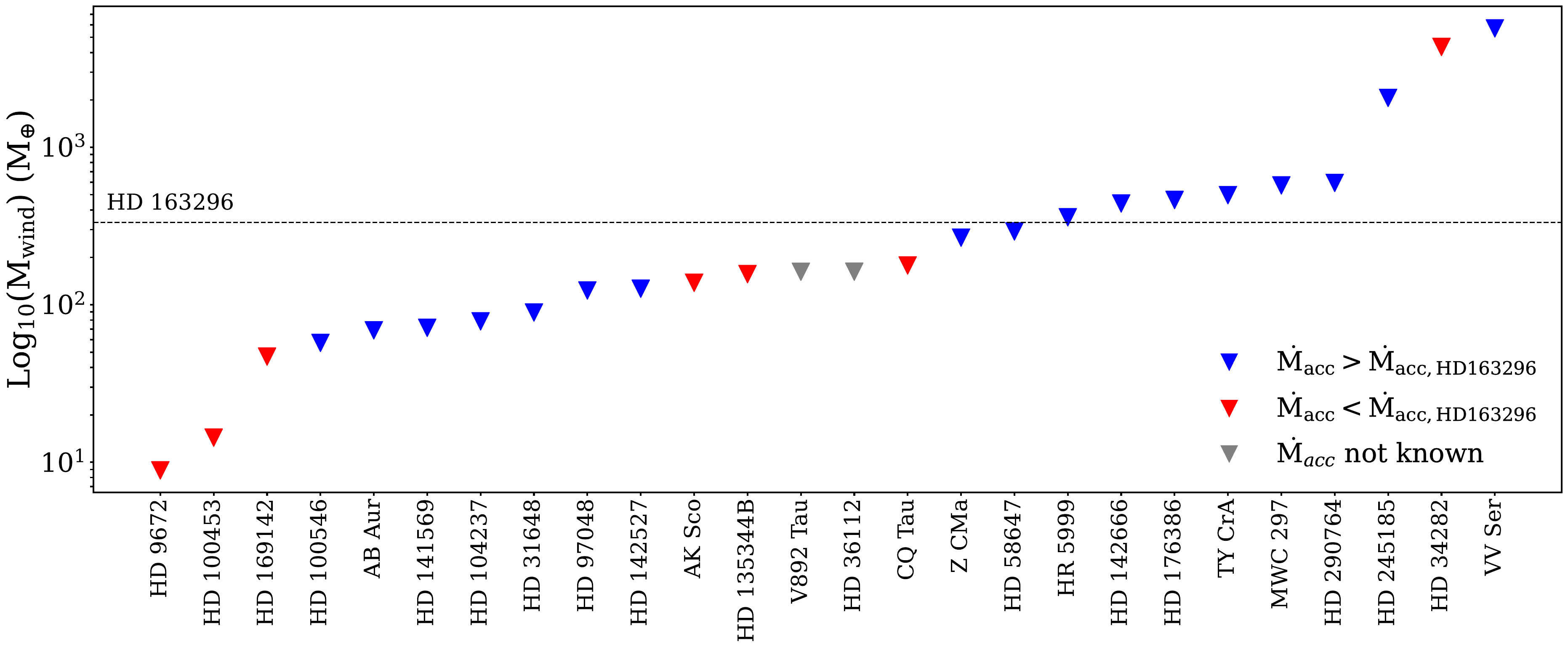}
    \caption{Upper limits on the disk wind mass. The mass of the disk wind of HD~163296 is shown as the dotted horizontal line \citep{Booth2021}. Disks which have an accretion rate higher or lower than that of HD~163296 are shown as blue or red respectively \citep{GuzmanDiaz2021}. HD~34282 has a lower limit on its accretion rate. Disks with no accretion rate measured are shown in gray.}
    \label{fig:wind_upper_limit}
\end{figure*}

\begin{figure}
    \centering
    \includegraphics[width=0.5\textwidth]{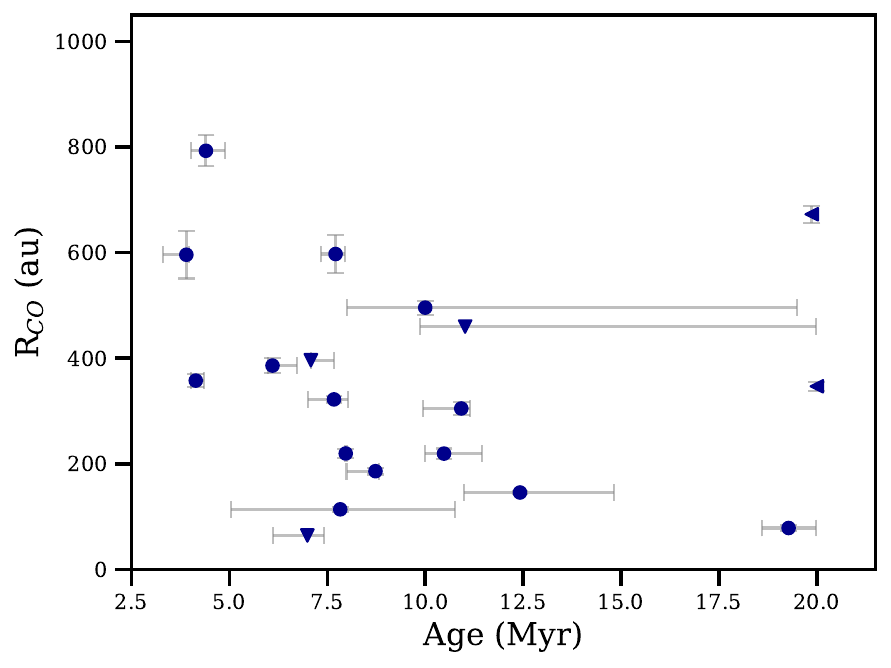}
    \caption{The gas radii ($R_{90\%}$) of the Herbig disks plotted against the age of the system (from \citealt{GuzmanDiaz2021}). The downward facing triangles indicate an upper limit on the gas radius, an upper limit on the age is indicated as a left facing triangle.}
    \label{fig:age_vs_radius}
\end{figure}

\subsubsection{Debris disks}
\label{subsec:debris_disks}
Debris disks are the final stage of planet-forming disks and are sustained by collisional processes producing secondary dust \citep{Hughes2018}. In contrast to later spectral type stars, debris disks around A-type stars (i.e., the evolutionary successors of Herbig disks) are more common to be detected in \ce{CO} gas compared to later spectral type stars \citep[e.g.,][]{Moor2017, Moor2020}. Whether this is primordial or secondary gas is still heavily debated \citep[see for an overview][]{Hughes2018}. What has been found is that the amount of \ce{CO} mass in the (debris) disks around A-type stars rapidly decreases between Herbig disks and debris disks around A-type stars \citep[e.g.,][]{Moor2020}. In this section we compare the \ce{CO} masses found by \citet{Moor2020} and \citet{Cataldi2023} to the disk masses we obtained for Herbig disks.

Using the previously adopted CO isotopologue and CO/\ce{H_2} ratios, we compute the total disk mass from the \ce{CO} mass estimates by \citet{Moor2017}, see Figure \ref{fig:masses_debris_disks}. \citet{Moor2020} showed that most debris disks with an A spectral type are two orders of magnitude less massive than Herbig disks in CO. Here we show that this difference may be even more dramatic, with a difference between the logarithmic means of the two population of four orders of magnitude. In Figure \ref{fig:masses_debris_disks} we additionally show the LTE models from \citet{Cataldi2023} who modeled the CO and CI emission of 14 debris disks to obtain a CO mass in the disk. After converting these into total disk mass, large differences between the Herbig disks and the debris disks can be seen, as is the case for the masses from \citet{Moor2017}. The fact that few disks are in between the Herbig disks and debris disks suggests that the disk needs to dissipate quickly over just a few million years. 

Lastly we note that if the disk consists of second generation gas, the assumption of CO/$H_2$=$10^{-4}$ might not be correct and should rather be closer to one, but the assumption does give an upper limit on the gas mass in the disk. In the case of CO/$H_2$=$1$ the debris disk masses shown in Fig.~\ref{fig:masses_debris_disks} are even lower.

\subsection{Viscous or wind driven evolution}
\label{subsec:viscous_vs_wind}
Finally, we discuss the viscous versus wind driven evolution of disks. Wind signatures have been found in embedded disks but for later stage disks, in particular Herbig disks, direct evidence for a disk wind has only been found in HD~163296 in \ce{^12CO} and \ce{^13CO} \citep{Klaassen2013, Booth2021}. As our work has compiled all Herbig disks with \ce{CO} observations available, we present in this section the results of a search for wind signatures in other Herbig disks.

Following previous works \citep{Klaassen2013, Booth2021}, the search for wind signatures is done with visibility spectra. To ensure that the visibility spectrum is dominated by the large-scale structure of the wind, only the short baselines are selected. The 20$\%$ shortest baselines are used for the analysis, with two exceptions: AK~Sco and HD~142666. Due to their relatively high resolution, a lower percentage is chosen such that all baselines smaller than respectively 80 and 100 meter are selected, corresponding to spatial scales of $\sim2.0''$. To obtain the visibility spectra, the time and baselines are averaged to obtain a better $S/N$. The velocity is sampled from around -50~km~s$^{-1}$ to 50~km~s$^{-1}$ relative to the system velocities, so that all the possible disk wind channels can be covered \citep{Pascucci2022}. 

No wind signatures have been found, either due to low-quality data or no presence of a disk wind. To assess this we obtain a measure of the noise from the visibility spectra of each disk scaled by the square-root of the velocity resolution relative to that of the HD~163296 observation. Using the estimate of the total mass of the disk wind from \citet{Booth2021} of $10^{-3}$~M$_\odot$ and the peak flux in their spectrum of $0.14$~Jy in \ce{^13CO}, we can scale the upper limits of $5\times$ the noise with the distance to the object relative to HD~163296 (see Table \ref{tab:disk_params}), assuming that CO/\ce{H_2} is not different in each disk wind. This results in an upper limit on the disk wind in Earth masses for each disk. Figure \ref{fig:wind_upper_limit} presents the results.

The upper limits on the majority of the disks fall below the mass of the disk wind of HD~163296. Of these disks, a third have accretion rates lower than HD~163296 (see the red scatter points, from \citealt{GuzmanDiaz2021}). This could result in the non-detections as a stronger disk wind induces a higher accretion rate. For these disks we can rule out the sensitivity of the data being the main reason for not detecting a disk wind similar to that of HD~163296. On the other hand, nine disks have upper limits above the mass of HD~163296, which does indicate a lack of sensitivity and a disk wind like that of HD~163296 could possibly still be present.

Comparing these results to the gas masses found in Section \ref{subsec:obtaining_mass}, it is clear that only a few disks are likely more massive than HD~163296, most of which have a better upper limit on the disk wind mass than lower mass disks. For example HD~290764 and HD~245185 may be good candidates for deeper follow-up surveys of disk winds in Herbig disks due to their high gas mass and accretion rates, but relatively high upper limits. Another option would be HD~34282, which has a high disk mass but a lower accretion rate compared to HD~163296. Notably, \citet{Pegues2023} identify a tentative extended structure in \ce{^12CO} in HD~34282, while not seen in our data this further substantiates a follow-up. For some other disks such as HD~142666 longer and lower resolution observations to obtain better short baseline coverage and higher sensitivities would also be useful in order to confirm a lack of disk wind signatures.

Using the gas radii in combination with the ages from \citet{GuzmanDiaz2021}, we determine if the radii of Herbig disks evolve over their lifetime, and if so, whether a specific evolutionary scenario is favorable. Viscously evolving disks are expected to increase in size over time, while wind driven evolution results in a reduction of the size of the disk \citep[e.g.,][]{Manara2022}. \citet{Trapman2022} modeled the evolution of the CO radius in the wind driven case, showing that the size of the disk is indeed expected to decrease in this specific tracer. Though effects of external photoevaporation are important to keep in mind \citep{Trapman2023}. Figure \ref{fig:age_vs_radius} shows the gas radius plotted against the age of the system. The Herbig disks younger than 12~Myr do seem to correspond well with the wind evolution shown by \citet{Trapman2022}, the largest and highest mass disks decrease in size by a factor of $\sim4$ between 2 and 10 Myr. For the three disks on the right hand side of Fig.~\ref{fig:age_vs_radius}, two (HD~34282 and HD~169142) only have upper limits on their age. However, other works do put these at younger ages of $\sim10$~Myr \citep[e.g.,][]{Vioque2018}. Samples of older ages are missing, which is needed to constrain the specific evolutionary scenario. More sensitive observations of a larger sample are needed to make progress.

\section{Conclusion}
\label{sec:conclusion}
In this work we analyze the \ce{^12CO} ,\ce{^13CO} and \ce{C^18O} $J=2-1$ or $J=3-2$ emission in 35 Herbig disks, 30 with ALMA archival data and 5 new data sets observed with NOEMA. We compare the integrated line luminosities and R$_{90\%}$ radii of \ce{^12CO} and \ce{^13CO} with a large grid of \texttt{DALI} models \citep{Bruderer2012, Bruderer2013} to obtain a measure of the gas mass. We can conclude the following:

\begin{enumerate}
     \item We detect \ce{^12CO} emission in 20 out of the 27 disks which have the line covered, for \ce{^13CO} in 22 disks out of 33 disks, and for \ce{C^18O} in 21 disks out of 33 disks. In total, 15 disks are resolved in \ce{^12CO}, 16 in \ce{^13CO}, and 15 in \ce{C^18O}. For all resolved disks, the \ce{^12CO} emission extends to larger radii compared to \ce{^13CO}, which in turn is larger than the \ce{C^18O} disk.
    \item The main model parameters affecting the luminosities of the \ce{^13CO} and \ce{C^18O} lines in the models are the mass and size of the disk. In addition, the power-law index of the surface density affects the line luminosities due to changing the distribution of the mass in the disk. For the most massive disks we find that the effect of the stellar luminosity, the vertical distribution of the disk mass, and the inclination of the disk have negligible impact on the line luminosities. For low mass disks however, increasing the vertical distribution and stellar luminosity decreases the luminosity of the isotopologue lines by one to two orders of magnitude.
    \item The two deciding processes influencing the line luminosity of the disk is are the line optical depth as seen from the observer and the self-shielding capacity of the CO isotopologues. Hence, one can make disks with similar luminosities with very different masses if the disk is optically thick in both \ce{^13CO} and \ce{C^18O}. When enlarging the disk, first \ce{C^18O} becomes optically thin and starts to photodissociate, reducing the \ce{C^18O} line luminosity. At even larger sizes, \ce{^13CO} becomes optically thin as well and subsequently its luminosity reduces.
    \item We find that most of the detected Herbig disks are optically thick in both \ce{^13CO} and \ce{C^18O}. For almost all disks we can only find a lower disk mass which can reproduce the observed line luminosities. The R$_{90\%}$ size of the disk is an essential observable to constrain the gas mass, as only the most massive disks can be large in \ce{^12CO}.
    \item Comparing the gas masses to those obtained from the number of \ce{C^18O} molecules with a simple CO/H$_2$ conversion shows that the disk mass is generally underestimated by at least an order of magnitude. This shows that to obtain the mass of a disk, modeling the disk is vital.
    \item Combining the gas masses of the disks with the dust masses from \citet{stapper2022}, we find that Herbig disks are consistent with the canonical gas-to-dust ratio of 100. In general the ratio is even higher, which may be caused by a combination of dust optical depth and grain growth.
    \item Comparing the gas radii with the dust radii from \citet{stapper2022} we find a ratio of 2.7, higher compared to the disks in Lupus (a factor of 2.0, \citealt{Ansdell2018}). Still, the majority of the disks fall well below the factor of four, indicating that this difference may only be due to line optical depth effects rather than radial drift.
    \item To distinguish different disk masses for optically thick disks, a combination of \ce{^12CO} tracing the size of the disk and \ce{^13C^17O} tracing the mass of the disk would make this possible for a large range in masses of disks. However, for the most massive disks dust opacity may inhibit tracing the disk mass with such a rare isotopologue.
    \item Comparing the Herbig gas-to-dust ratios with those in T~Tauri disks \citep{Ansdell2016, Pascucci2016, Miotello2017, Long2017}, we find that Herbig disks have a gas-to-dust ratio of almost two orders of magnitude higher over a range of multiple orders of magnitude in dust mass. Hence, this disparity could be caused by a fundamental difference in chemistry due to Herbig disks being much warmer, as proposed in previous works.
    \item The gas and dust masses of the two different \citet{Meeus2001} groups are found both differ by an order of magnitude. This results in the same gas-to-dust ratios for both groups, even though group~II disks have similar dust masses as T~Tauri disks \citep{stapper2022}. This further supports the idea that the gas-to-dust ratio is disk mass independent and that the lack of CO emission in T~Tauri disks is not due to their lower mass disks, but rather a temperature difference compared to Herbig disks.
    \item The total masses of debris disks are found to be at least four orders of magnitude lower than those of Herbig disks, indicating a rapid dissipation of the material in the disk within a few Myr. Full chemical modeling of debris disks is necessary to explore this further.
    \item A search for disk wind signatures such as those found for HD~163296 \citep{Booth2021} in the \ce{^12CO} data has resulted in no additional detections. Most disks have sufficiently sensitive data in which a disk wind analogous to that in HD~163296 would have been detected. This lack of disk wind may be in some of the disks related to a difference in accretion rate.
    \item Comparing the gas radii to the age of the system we find that the data seem to support a disk wind driven evolution, but data of older age systems are lacking.
\end{enumerate}

\noindent
In conclusion, the warmer Herbig disks have significantly larger gas-to-dust mass ratios compared to the colder T~Tauri stars, close to, or exceeding, the canonical value of 100.

\begin{acknowledgements}
The research of LMS is supported by the Netherlands Research School for Astronomy (NOVA). This paper makes use of the following ALMA data: 2012.1.00158.S, 2012.1.00303.S, 2012.1.00698.S, 2012.1.00870.S, 2013.1.00498.S, 2015.1.00192.S, 2015.1.00222.S, 2015.1.00986.S, 2015.1.01058.S, 2015.1.01353.S, 2015.1.01600.S, 2016.1.00110.S, 2016.1.00204.S, 2016.1.00344.S, 2016.1.00484.L, 2016.1.00724.S, 2017.1.00466.S, 2017.1.00940.S, 2017.1.01404.S, 2017.1.01419.S, 2017.1.01607.S, 2018.1.00814.S, 2018.1.01055.L, 2018.1.01222.S, 2019.1.00218.S, 2019.1.00579.S. ALMA is a partnership of ESO (representing its member states), NSF (USA) and NINS (Japan), together with NRC (Canada), MOST and ASIAA (Taiwan), and KASI (Republic of Korea), in cooperation with the Republic of Chile. The Joint ALMA Observatory is operated by ESO, AUI/NRAO and NAOJ. This work is based on observations carried out under project number S21AS with the IRAM NOEMA Interferometer. IRAM is supported by INSU/CNRS (France), MPG (Germany) and IGN (Spain). This work makes use of the following software: The Common Astronomy Software Applications (CASA) package \citep{McMullin2007}, Dust And LInes \citep[DALI,][]{Bruderer2012, Bruderer2013}, Python version 3.9, astropy \citep{astropy2013, astropy2018}, cmasher \citep{cmasher}, lifelines \citep{DavidsonPilon2021}, matplotlib \citep{Hunter2007}, numpy \citep{Harris2020}, pandas \citep{reback2020pandas}, scipy \citep{2020SciPy-NMeth} and seaborn \citep{Waskom2021}. Lastly,  we thank the referee for their insightful comments which have improved this paper.
\end{acknowledgements}

\bibliographystyle{aa}
\bibliography{references.bib}

\begin{thebibliography}{128}
\expandafter\ifx\csname natexlab\endcsname\relax\def\natexlab#1{#1}\fi

\bibitem[{{Ag{\'u}ndez} {et~al.}(2018){Ag{\'u}ndez}, {Roueff}, {Le Petit}, \& {Le Bourlot}}]{Agundez2018}
{Ag{\'u}ndez}, M., {Roueff}, E., {Le Petit}, F., \& {Le Bourlot}, J. 2018, \href{http://dx.doi.org/10.1051/0004-6361/201732518}{\color{blue}\aap}, \href{https://ui.adsabs.harvard.edu/abs/2018A&A...616A..19A}{616, A19}

\bibitem[{{Anderson} {et~al.}(2022){Anderson}, {Williams}, {van der Marel}, {Law}, {Ricci}, {Tobin}, \& {Tong}}]{Anderson2022}
{Anderson}, A.~R., {Williams}, J.~P., {van der Marel}, N., {et~al.} 2022, \href{http://dx.doi.org/10.3847/1538-4357/ac8ff0}{\color{blue}\apj}, \href{https://ui.adsabs.harvard.edu/abs/2022ApJ...938...55A}{938, 55}

\bibitem[{{Andrews} {et~al.}(2013){Andrews}, {Rosenfeld}, {Kraus}, \& {Wilner}}]{Andrews2013}
{Andrews}, S.~M., {Rosenfeld}, K.~A., {Kraus}, A.~L., \& {Wilner}, D.~J. 2013, \href{http://dx.doi.org/10.1088/0004-637X/771/2/129}{\color{blue}\apj}, \href{https://ui.adsabs.harvard.edu/abs/2013ApJ...771..129A}{771, 129}

\bibitem[{{Andrews} {et~al.}(2011){Andrews}, {Wilner}, {Espaillat}, {Hughes}, {Dullemond}, {McClure}, {Qi}, \& {Brown}}]{Andrews2011}
{Andrews}, S.~M., {Wilner}, D.~J., {Espaillat}, C., {et~al.} 2011, \href{http://dx.doi.org/10.1088/0004-637X/732/1/42}{\color{blue}\apj}, \href{https://ui.adsabs.harvard.edu/abs/2011ApJ...732...42A}{732, 42}

\bibitem[{{Ansdell} {et~al.}(2017){Ansdell}, {Williams}, {Manara}, {Miotello}, {Facchini}, {van der Marel}, {Testi}, \& {van Dishoeck}}]{Ansdell2017}
{Ansdell}, M., {Williams}, J.~P., {Manara}, C.~F., {et~al.} 2017, \href{http://dx.doi.org/10.3847/1538-3881/aa69c0}{\color{blue}\aj}, \href{https://ui.adsabs.harvard.edu/abs/2017AJ....153..240A}{153, 240}

\bibitem[{{Ansdell} {et~al.}(2018){Ansdell}, {Williams}, {Trapman}, {van Terwisga}, {Facchini}, {Manara}, {van der Marel}, {Miotello}, {Tazzari}, {Hogerheijde}, {Guidi}, {Testi}, \& {van Dishoeck}}]{Ansdell2018}
{Ansdell}, M., {Williams}, J.~P., {Trapman}, L., {et~al.} 2018, \href{http://dx.doi.org/10.3847/1538-4357/aab890}{\color{blue}\apj}, \href{https://ui.adsabs.harvard.edu/abs/2018ApJ...859...21A}{859, 21}

\bibitem[{{Ansdell} {et~al.}(2016){Ansdell}, {Williams}, {van der Marel}, {Carpenter}, {Guidi}, {Hogerheijde}, {Mathews}, {Manara}, {Miotello}, {Natta}, {Oliveira}, {Tazzari}, {Testi}, {van Dishoeck}, \& {van Terwisga}}]{Ansdell2016}
{Ansdell}, M., {Williams}, J.~P., {van der Marel}, N., {et~al.} 2016, \href{http://dx.doi.org/10.3847/0004-637X/828/1/46}{\color{blue}\apj}, \href{https://ui.adsabs.harvard.edu/abs/2016ApJ...828...46A}{828, 46}

\bibitem[{{Astropy Collaboration} {et~al.}(2018){Astropy Collaboration}, {Price-Whelan}, {Sip{\H{o}}cz}, {G{\"u}nther}, {Lim}, {Crawford}, {Conseil}, {Shupe}, {Craig}, {Dencheva}, {Ginsburg}, {Vand erPlas}, {Bradley}, {P{\'e}rez-Su{\'a}rez}, {de Val-Borro}, {Aldcroft}, {Cruz}, {Robitaille}, {Tollerud}, {Ardelean}, {Babej}, {Bach}, {Bachetti}, {Bakanov}, {Bamford}, {Barentsen}, {Barmby}, {Baumbach}, {Berry}, {Biscani}, {Boquien}, {Bostroem}, {Bouma}, {Brammer}, {Bray}, {Breytenbach}, {Buddelmeijer}, {Burke}, {Calderone}, {Cano Rodr{\'\i}guez}, {Cara}, {Cardoso}, {Cheedella}, {Copin}, {Corrales}, {Crichton}, {D'Avella}, {Deil}, {Depagne}, {Dietrich}, {Donath}, {Droettboom}, {Earl}, {Erben}, {Fabbro}, {Ferreira}, {Finethy}, {Fox}, {Garrison}, {Gibbons}, {Goldstein}, {Gommers}, {Greco}, {Greenfield}, {Groener}, {Grollier}, {Hagen}, {Hirst}, {Homeier}, {Horton}, {Hosseinzadeh}, {Hu}, {Hunkeler}, {Ivezi{\'c}}, {Jain}, {Jenness}, {Kanarek}, {Kendrew}, {Kern}, {Kerzendorf}, {Khvalko}, {King}, {Kirkby}, {Kulkarni},
  {Kumar}, {Lee}, {Lenz}, {Littlefair}, {Ma}, {Macleod}, {Mastropietro}, {McCully}, {Montagnac}, {Morris}, {Mueller}, {Mumford}, {Muna}, {Murphy}, {Nelson}, {Nguyen}, {Ninan}, {N{\"o}the}, {Ogaz}, {Oh}, {Parejko}, {Parley}, {Pascual}, {Patil}, {Patil}, {Plunkett}, {Prochaska}, {Rastogi}, {Reddy Janga}, {Sabater}, {Sakurikar}, {Seifert}, {Sherbert}, {Sherwood-Taylor}, {Shih}, {Sick}, {Silbiger}, {Singanamalla}, {Singer}, {Sladen}, {Sooley}, {Sornarajah}, {Streicher}, {Teuben}, {Thomas}, {Tremblay}, {Turner}, {Terr{\'o}n}, {van Kerkwijk}, {de la Vega}, {Watkins}, {Weaver}, {Whitmore}, {Woillez}, {Zabalza}, \& {Astropy Contributors}}]{astropy2018}
{Astropy Collaboration}, {Price-Whelan}, A.~M., {Sip{\H{o}}cz}, B.~M., {et~al.} 2018, \href{http://dx.doi.org/10.3847/1538-3881/aabc4f}{\color{blue}\aj}, \href{https://ui.adsabs.harvard.edu/abs/2018AJ....156..123A}{156, 123}

\bibitem[{{Astropy Collaboration} {et~al.}(2013){Astropy Collaboration}, {Robitaille}, {Tollerud}, {Greenfield}, {Droettboom}, {Bray}, {Aldcroft}, {Davis}, {Ginsburg}, {Price-Whelan}, {Kerzendorf}, {Conley}, {Crighton}, {Barbary}, {Muna}, {Ferguson}, {Grollier}, {Parikh}, {Nair}, {Unther}, {Deil}, {Woillez}, {Conseil}, {Kramer}, {Turner}, {Singer}, {Fox}, {Weaver}, {Zabalza}, {Edwards}, {Azalee Bostroem}, {Burke}, {Casey}, {Crawford}, {Dencheva}, {Ely}, {Jenness}, {Labrie}, {Lim}, {Pierfederici}, {Pontzen}, {Ptak}, {Refsdal}, {Servillat}, \& {Streicher}}]{astropy2013}
{Astropy Collaboration}, {Robitaille}, T.~P., {Tollerud}, E.~J., {et~al.} 2013, \href{http://dx.doi.org/10.1051/0004-6361/201322068}{\color{blue}\aap}, \href{http://adsabs.harvard.edu/abs/2013A%26A...558A..33A}{558, A33}

\bibitem[{{Barenfeld} {et~al.}(2016){Barenfeld}, {Carpenter}, {Ricci}, \& {Isella}}]{Barenfeld2016}
{Barenfeld}, S.~A., {Carpenter}, J.~M., {Ricci}, L., \& {Isella}, A. 2016, \href{http://dx.doi.org/10.3847/0004-637X/827/2/142}{\color{blue}\apj}, \href{https://ui.adsabs.harvard.edu/abs/2016ApJ...827..142B}{827, 142}

\bibitem[{{Beckwith} {et~al.}(1990){Beckwith}, {Sargent}, {Chini}, \& {Guesten}}]{Beckwith1990}
{Beckwith}, S. V.~W., {Sargent}, A.~I., {Chini}, R.~S., \& {Guesten}, R. 1990, \href{http://dx.doi.org/10.1086/115385}{\color{blue}\aj}, \href{https://ui.adsabs.harvard.edu/abs/1990AJ.....99..924B}{99, 924}

\bibitem[{{Bergin} {et~al.}(2013){Bergin}, {Cleeves}, {Gorti}, {Zhang}, {Blake}, {Green}, {Andrews}, {Evans}, {Henning}, {{\"O}berg}, {Pontoppidan}, {Qi}, {Salyk}, \& {van Dishoeck}}]{Bergin2013}
{Bergin}, E.~A., {Cleeves}, L.~I., {Gorti}, U., {et~al.} 2013, \href{http://dx.doi.org/10.1038/nature11805}{\color{blue}\nat}, \href{https://ui.adsabs.harvard.edu/abs/2013Natur.493..644B}{493, 644}

\bibitem[{{Bohlin} {et~al.}(1978){Bohlin}, {Savage}, \& {Drake}}]{Bohlin1978}
{Bohlin}, R.~C., {Savage}, B.~D., \& {Drake}, J.~F. 1978, \href{http://dx.doi.org/10.1086/156357}{\color{blue}\apj}, \href{https://ui.adsabs.harvard.edu/abs/1978ApJ...224..132B}{224, 132}

\bibitem[{{Booth} \& {Ilee}(2020)}]{Booth2020}
{Booth}, A.~S. \& {Ilee}, J.~D. 2020, \href{http://dx.doi.org/10.1093/mnrasl/slaa014}{\color{blue}\mnras}, \href{https://ui.adsabs.harvard.edu/abs/2020MNRAS.493L.108B}{493, L108}

\bibitem[{{Booth} {et~al.}(2023{\natexlab{a}}){Booth}, {Ilee}, {Walsh}, {Kama}, {Keyte}, {van Dishoeck}, \& {Nomura}}]{Booth2023}
{Booth}, A.~S., {Ilee}, J.~D., {Walsh}, C., {et~al.} 2023{\natexlab{a}}, \href{http://dx.doi.org/10.1051/0004-6361/202244472}{\color{blue}\aap}, \href{https://ui.adsabs.harvard.edu/abs/2023A&A...669A..53B}{669, A53}

\bibitem[{{Booth} {et~al.}(2023{\natexlab{b}}){Booth}, {Law}, {Temmink}, {Leemker}, \& {Macias}}]{Booth2023b}
{Booth}, A.~S., {Law}, C.~J., {Temmink}, M., {Leemker}, M., \& {Macias}, E. 2023{\natexlab{b}}, \href{https://ui.adsabs.harvard.edu/abs/2023arXiv230807910B}{\href{http://dx.doi.org/10.48550/arXiv.2308.07910}{\color{blue}arXiv e-prints}, arXiv:2308.07910}

\bibitem[{{Booth} {et~al.}(2021){Booth}, {Tabone}, {Ilee}, {Walsh}, {Aikawa}, {Andrews}, {Bae}, {Bergin}, {Bergner}, {Bosman}, {Calahan}, {Cataldi}, {Cleeves}, {Czekala}, {Guzm{\'a}n}, {Huang}, {Law}, {Le Gal}, {Long}, {Loomis}, {M{\'e}nard}, {Nomura}, {{\"O}berg}, {Qi}, {Schwarz}, {Teague}, {Tsukagoshi}, {Wilner}, {Yamato}, \& {Zhang}}]{Booth2021}
{Booth}, A.~S., {Tabone}, B., {Ilee}, J.~D., {et~al.} 2021, \href{http://dx.doi.org/10.3847/1538-4365/ac1ad4}{\color{blue}\apjs}, \href{https://ui.adsabs.harvard.edu/abs/2021ApJS..257...16B}{257, 16}

\bibitem[{{Booth} {et~al.}(2019){Booth}, {Walsh}, {Ilee}, {Notsu}, {Qi}, {Nomura}, \& {Akiyama}}]{Booth2019}
{Booth}, A.~S., {Walsh}, C., {Ilee}, J.~D., {et~al.} 2019, \href{http://dx.doi.org/10.3847/2041-8213/ab3645}{\color{blue}\apjl}, \href{https://ui.adsabs.harvard.edu/abs/2019ApJ...882L..31B}{882, L31}

\bibitem[{{Bosman} {et~al.}(2018){Bosman}, {Walsh}, \& {van Dishoeck}}]{Bosman2018}
{Bosman}, A.~D., {Walsh}, C., \& {van Dishoeck}, E.~F. 2018, \href{http://dx.doi.org/10.1051/0004-6361/201833497}{\color{blue}\aap}, \href{https://ui.adsabs.harvard.edu/abs/2018A&A...618A.182B}{618, A182}

\bibitem[{{Brittain} {et~al.}(2023){Brittain}, {Kamp}, {Meeus}, {Oudmaijer}, \& {Waters}}]{Brittain2023}
{Brittain}, S.~D., {Kamp}, I., {Meeus}, G., {Oudmaijer}, R.~D., \& {Waters}, L.~B.~F.~M. 2023, \href{http://dx.doi.org/10.1007/s11214-023-00949-z}{\color{blue}\ssr}, \href{https://ui.adsabs.harvard.edu/abs/2023SSRv..219....7B}{219, 7}

\bibitem[{{Bruderer}(2013)}]{Bruderer2013}
{Bruderer}, S. 2013, \href{http://dx.doi.org/10.1051/0004-6361/201321171}{\color{blue}\aap}, \href{https://ui.adsabs.harvard.edu/abs/2013A&A...559A..46B}{559, A46}

\bibitem[{{Bruderer} {et~al.}(2012){Bruderer}, {van Dishoeck}, {Doty}, \& {Herczeg}}]{Bruderer2012}
{Bruderer}, S., {van Dishoeck}, E.~F., {Doty}, S.~D., \& {Herczeg}, G.~J. 2012, \href{http://dx.doi.org/10.1051/0004-6361/201118218}{\color{blue}\aap}, \href{https://ui.adsabs.harvard.edu/abs/2012A&A...541A..91B}{541, A91}

\bibitem[{{Carney} {et~al.}(2018){Carney}, {Fedele}, {Hogerheijde}, {Favre}, {Walsh}, {Bruderer}, {Miotello}, {Murillo}, {Klaassen}, {Henning}, \& {van Dishoeck}}]{Carney2018}
{Carney}, M.~T., {Fedele}, D., {Hogerheijde}, M.~R., {et~al.} 2018, \href{http://dx.doi.org/10.1051/0004-6361/201732384}{\color{blue}\aap}, \href{https://ui.adsabs.harvard.edu/abs/2018A&A...614A.106C}{614, A106}

\bibitem[{{Cataldi} {et~al.}(2023){Cataldi}, {Aikawa}, {Iwasaki}, {Marino}, {Brandeker}, {Hales}, {Henning}, {Higuchi}, {Hughes}, {Janson}, {Kral}, {Matr{\`a}}, {Mo{\'o}r}, {Olofsson}, {Redfield}, \& {Roberge}}]{Cataldi2023}
{Cataldi}, G., {Aikawa}, Y., {Iwasaki}, K., {et~al.} 2023, \href{https://ui.adsabs.harvard.edu/abs/2023arXiv230512093C}{\href{http://dx.doi.org/10.48550/arXiv.2305.12093}{\color{blue}arXiv e-prints}, arXiv:2305.12093}

\bibitem[{{Cazzoletti} {et~al.}(2019){Cazzoletti}, {Manara}, {Baobab Liu}, {van Dishoeck}, {Facchini}, {Alcal{\`a}}, {Ansdell}, {Testi}, {Williams}, {Carrasco-Gonz{\'a}lez}, {Dong}, {Forbrich}, {Fukagawa}, {Galv{\'a}n-Madrid}, {Hirano}, {Hogerheijde}, {Hasegawa}, {Muto}, {Pinilla}, {Takami}, {Tamura}, {Tazzari}, \& {Wisniewski}}]{Cazzoletti2019}
{Cazzoletti}, P., {Manara}, C.~F., {Baobab Liu}, H., {et~al.} 2019, \href{http://dx.doi.org/10.1051/0004-6361/201935273}{\color{blue}\aap}, \href{https://ui.adsabs.harvard.edu/abs/2019A&A...626A..11C}{626, A11}

\bibitem[{{Cazzoletti} {et~al.}(2018){Cazzoletti}, {van Dishoeck}, {Pinilla}, {Tazzari}, {Facchini}, {van der Marel}, {Benisty}, {Garufi}, \& {P{\'e}rez}}]{Cazzoletti2018}
{Cazzoletti}, P., {van Dishoeck}, E.~F., {Pinilla}, P., {et~al.} 2018, \href{http://dx.doi.org/10.1051/0004-6361/201834006}{\color{blue}\aap}, \href{https://ui.adsabs.harvard.edu/abs/2018A&A...619A.161C}{619, A161}

\bibitem[{{Czekala} {et~al.}(2015){Czekala}, {Andrews}, {Jensen}, {Stassun}, {Torres}, \& {Wilner}}]{Czekala2015}
{Czekala}, I., {Andrews}, S.~M., {Jensen}, E.~L.~N., {et~al.} 2015, \href{http://dx.doi.org/10.1088/0004-637X/806/2/154}{\color{blue}\apj}, \href{https://ui.adsabs.harvard.edu/abs/2015ApJ...806..154C}{806, 154}

\bibitem[{Davidson-Pilon {et~al.}(2021)Davidson-Pilon, Kalderstam, Jacobson, Reed, Kuhn, Zivich, Williamson, AbdealiJK, Datta, Fiore-Gartland, Parij, WIlson, Gabriel, Moneda, Moncada-Torres, Stark, Gadgil, Jona, Singaravelan, Besson, Peña, Anton, Klintberg, GrowthJeff, Noorbakhsh, Begun, Kumar, Hussey, Seabold, \& Golland}]{DavidsonPilon2021}
Davidson-Pilon, C., Kalderstam, J., Jacobson, N., {et~al.} 2021, CamDavidsonPilon/lifelines: 0.25.10

\bibitem[{{Draine} \& {Li}(2007)}]{Draine2007}
{Draine}, B.~T. \& {Li}, A. 2007, \href{http://dx.doi.org/10.1086/511055}{\color{blue}\apj}, \href{https://ui.adsabs.harvard.edu/abs/2007ApJ...657..810D}{657, 810}

\bibitem[{{Dullemond} {et~al.}(2001){Dullemond}, {Dominik}, \& {Natta}}]{Dullemond2001}
{Dullemond}, C.~P., {Dominik}, C., \& {Natta}, A. 2001, \href{http://dx.doi.org/10.1086/323057}{\color{blue}\apj}, \href{https://ui.adsabs.harvard.edu/abs/2001ApJ...560..957D}{560, 957}

\bibitem[{{Eisner} {et~al.}(2018){Eisner}, {Arce}, {Ballering}, {Bally}, {Andrews}, {Boyden}, {Di Francesco}, {Fang}, {Johnstone}, {Kim}, {Mann}, {Matthews}, {Pascucci}, {Ricci}, {Sheehan}, \& {Williams}}]{Eisner2018}
{Eisner}, J.~A., {Arce}, H.~G., {Ballering}, N.~P., {et~al.} 2018, \href{http://dx.doi.org/10.3847/1538-4357/aac3e2}{\color{blue}\apj}, \href{https://ui.adsabs.harvard.edu/abs/2018ApJ...860...77E}{860, 77}

\bibitem[{{Endres} {et~al.}(2016){Endres}, {Schlemmer}, {Schilke}, {Stutzki}, \& {M{\"u}ller}}]{Endres2016}
{Endres}, C.~P., {Schlemmer}, S., {Schilke}, P., {Stutzki}, J., \& {M{\"u}ller}, H. S.~P. 2016, \href{http://dx.doi.org/10.1016/j.jms.2016.03.005}{\color{blue}Journal of Molecular Spectroscopy}, \href{https://ui.adsabs.harvard.edu/abs/2016JMoSp.327...95E}{327, 95}

\bibitem[{{Fedele} {et~al.}(2017){Fedele}, {Carney}, {Hogerheijde}, {Walsh}, {Miotello}, {Klaassen}, {Bruderer}, {Henning}, \& {van Dishoeck}}]{Fedele2017}
{Fedele}, D., {Carney}, M., {Hogerheijde}, M.~R., {et~al.} 2017, \href{http://dx.doi.org/10.1051/0004-6361/201629860}{\color{blue}\aap}, \href{https://ui.adsabs.harvard.edu/abs/2017A&A...600A..72F}{600, A72}

\bibitem[{{Fedele} {et~al.}(2021){Fedele}, {Toci}, {Maud}, \& {Lodato}}]{Fedele2021}
{Fedele}, D., {Toci}, C., {Maud}, L., \& {Lodato}, G. 2021, \href{http://dx.doi.org/10.1051/0004-6361/202141278}{\color{blue}\aap}, \href{https://ui.adsabs.harvard.edu/abs/2021A&A...651A..90F}{651, A90}

\bibitem[{{Flaherty} {et~al.}(2020){Flaherty}, {Hughes}, {Simon}, {Qi}, {Bai}, {Bulatek}, {Andrews}, {Wilner}, \& {K{\'o}sp{\'a}l}}]{Flaherty2020}
{Flaherty}, K., {Hughes}, A.~M., {Simon}, J.~B., {et~al.} 2020, \href{http://dx.doi.org/10.3847/1538-4357/ab8cc5}{\color{blue}\apj}, \href{https://ui.adsabs.harvard.edu/abs/2020ApJ...895..109F}{895, 109}

\bibitem[{{Flaherty} {et~al.}(2015){Flaherty}, {Hughes}, {Rosenfeld}, {Andrews}, {Chiang}, {Simon}, {Kerzner}, \& {Wilner}}]{Flaherty2015}
{Flaherty}, K.~M., {Hughes}, A.~M., {Rosenfeld}, K.~A., {et~al.} 2015, \href{http://dx.doi.org/10.1088/0004-637X/813/2/99}{\color{blue}\apj}, \href{https://ui.adsabs.harvard.edu/abs/2015ApJ...813...99F}{813, 99}

\bibitem[{{Garufi} {et~al.}(2017){Garufi}, {Meeus}, {Benisty}, {Quanz}, {Banzatti}, {Kama}, {Canovas}, {Eiroa}, {Schmid}, {Stolker}, {Pohl}, {Rigliaco}, {M{\'e}nard}, {Meyer}, {van Boekel}, \& {Dominik}}]{Garufi2017}
{Garufi}, A., {Meeus}, G., {Benisty}, M., {et~al.} 2017, \href{http://dx.doi.org/10.1051/0004-6361/201630320}{\color{blue}\aap}, \href{https://ui.adsabs.harvard.edu/abs/2017A&A...603A..21G}{603, A21}

\bibitem[{{Grant} {et~al.}(2023){Grant}, {Stapper}, {Hogerheijde}, {van Dishoeck}, {Brittain}, \& {Vioque}}]{Grant2023}
{Grant}, S.~L., {Stapper}, L.~M., {Hogerheijde}, M.~R., {et~al.} 2023, \href{http://dx.doi.org/10.3847/1538-3881/acf128}{\color{blue}\aj}, \href{https://ui.adsabs.harvard.edu/abs/2023AJ....166..147G}{166, 147}

\bibitem[{{Guzm{\'a}n-D{\'\i}az} {et~al.}(2021){Guzm{\'a}n-D{\'\i}az}, {Mendigut{\'\i}a}, {Montesinos}, {Oudmaijer}, {Vioque}, {Rodrigo}, {Solano}, {Meeus}, \& {Marcos-Arenal}}]{GuzmanDiaz2021}
{Guzm{\'a}n-D{\'\i}az}, J., {Mendigut{\'\i}a}, I., {Montesinos}, B., {et~al.} 2021, \href{http://dx.doi.org/10.1051/0004-6361/202039519}{\color{blue}\aap}, \href{https://ui.adsabs.harvard.edu/abs/2021A&A...650A.182G}{650, A182}

\bibitem[{{Guzm{\'a}n-D{\'\i}az} {et~al.}(2023){Guzm{\'a}n-D{\'\i}az}, {Montesinos}, {Mendigut{\'\i}a}, {Kama}, {Meeus}, {Vioque}, {Oudmaijer}, \& {Villaver}}]{GuzmanDiaz2023}
{Guzm{\'a}n-D{\'\i}az}, J., {Montesinos}, B., {Mendigut{\'\i}a}, I., {et~al.} 2023, \href{http://dx.doi.org/10.1051/0004-6361/202245427}{\color{blue}\aap}, \href{https://ui.adsabs.harvard.edu/abs/2023A&A...671A.140G}{671, A140}

\bibitem[{Harris {et~al.}(2020)Harris, Millman, van~der Walt, Gommers, Virtanen, Cournapeau, Wieser, Taylor, Berg, Smith, Kern, Picus, Hoyer, van Kerkwijk, Brett, Haldane, del R{\'{i}}o, Wiebe, Peterson, G{\'{e}}rard-Marchant, Sheppard, Reddy, Weckesser, Abbasi, Gohlke, \& Oliphant}]{Harris2020}
Harris, C.~R., Millman, K.~J., van~der Walt, S.~J., {et~al.} 2020, \href{http://dx.doi.org/10.1038/s41586-020-2649-2}{\color{blue}Nature}, 585, 585

\bibitem[{{Hartmann} {et~al.}(1998){Hartmann}, {Calvet}, {Gullbring}, \& {D'Alessio}}]{Hartmann1998}
{Hartmann}, L., {Calvet}, N., {Gullbring}, E., \& {D'Alessio}, P. 1998, \href{http://dx.doi.org/10.1086/305277}{\color{blue}\apj}, \href{https://ui.adsabs.harvard.edu/abs/1998ApJ...495..385H}{495, 385}

\bibitem[{{Hendler} {et~al.}(2020){Hendler}, {Pascucci}, {Pinilla}, {Tazzari}, {Carpenter}, {Malhotra}, \& {Testi}}]{Hendler2020}
{Hendler}, N., {Pascucci}, I., {Pinilla}, P., {et~al.} 2020, \href{http://dx.doi.org/10.3847/1538-4357/ab70ba}{\color{blue}\apj}, \href{https://ui.adsabs.harvard.edu/abs/2020ApJ...895..126H}{895, 126}

\bibitem[{{Huang} {et~al.}(2018){Huang}, {Andrews}, {Dullemond}, {Isella}, {P{\'e}rez}, {Guzm{\'a}n}, {{\"O}berg}, {Zhu}, {Zhang}, {Bai}, {Benisty}, {Birnstiel}, {Carpenter}, {Hughes}, {Ricci}, {Weaver}, \& {Wilner}}]{Huang2018}
{Huang}, J., {Andrews}, S.~M., {Dullemond}, C.~P., {et~al.} 2018, \href{http://dx.doi.org/10.3847/2041-8213/aaf740}{\color{blue}\apjl}, \href{https://ui.adsabs.harvard.edu/abs/2018ApJ...869L..42H}{869, L42}

\bibitem[{{Hughes} {et~al.}(2018){Hughes}, {Duch{\^e}ne}, \& {Matthews}}]{Hughes2018}
{Hughes}, A.~M., {Duch{\^e}ne}, G., \& {Matthews}, B.~C. 2018, \href{http://dx.doi.org/10.1146/annurev-astro-081817-052035}{\color{blue}\araa}, \href{https://ui.adsabs.harvard.edu/abs/2018ARA&A..56..541H}{56, 541}

\bibitem[{{Hughes} {et~al.}(2017){Hughes}, {Lieman-Sifry}, {Flaherty}, {Daley}, {Roberge}, {K{\'o}sp{\'a}l}, {Mo{\'o}r}, {Kamp}, {Wilner}, {Andrews}, {Kastner}, \& {{\'A}brah{\'a}m}}]{Hughes2017}
{Hughes}, A.~M., {Lieman-Sifry}, J., {Flaherty}, K.~M., {et~al.} 2017, \href{http://dx.doi.org/10.3847/1538-4357/aa6b04}{\color{blue}\apj}, \href{https://ui.adsabs.harvard.edu/abs/2017ApJ...839...86H}{839, 86}

\bibitem[{{Hughes} {et~al.}(2008){Hughes}, {Wilner}, {Kamp}, \& {Hogerheijde}}]{Hughes2008}
{Hughes}, A.~M., {Wilner}, D.~J., {Kamp}, I., \& {Hogerheijde}, M.~R. 2008, \href{http://dx.doi.org/10.1086/588520}{\color{blue}\apj}, \href{https://ui.adsabs.harvard.edu/abs/2008ApJ...681..626H}{681, 626}

\bibitem[{Hunter(2007)}]{Hunter2007}
Hunter, J.~D. 2007, \href{http://dx.doi.org/10.1109/MCSE.2007.55}{\color{blue}Computing in Science \& Engineering}, 9, 9

\bibitem[{{Isella} {et~al.}(2010){Isella}, {Natta}, {Wilner}, {Carpenter}, \& {Testi}}]{Isella2010}
{Isella}, A., {Natta}, A., {Wilner}, D., {Carpenter}, J.~M., \& {Testi}, L. 2010, \href{http://dx.doi.org/10.1088/0004-637X/725/2/1735}{\color{blue}\apj}, \href{https://ui.adsabs.harvard.edu/abs/2010ApJ...725.1735I}{725, 1735}

\bibitem[{{Kaeufer} {et~al.}(2023){Kaeufer}, {Woitke}, {Min}, {Kamp}, \& {Pinte}}]{Kaeufer2023}
{Kaeufer}, T., {Woitke}, P., {Min}, M., {Kamp}, I., \& {Pinte}, C. 2023, \href{https://ui.adsabs.harvard.edu/abs/2023arXiv230204629K}{\href{http://dx.doi.org/10.48550/arXiv.2302.04629}{\color{blue}arXiv e-prints}, arXiv:2302.04629}

\bibitem[{{Kama} {et~al.}(2016){Kama}, {Bruderer}, {van Dishoeck}, {Hogerheijde}, {Folsom}, {Miotello}, {Fedele}, {Belloche}, {G{\"u}sten}, \& {Wyrowski}}]{Kama2016}
{Kama}, M., {Bruderer}, S., {van Dishoeck}, E.~F., {et~al.} 2016, \href{http://dx.doi.org/10.1051/0004-6361/201526991}{\color{blue}\aap}, \href{https://ui.adsabs.harvard.edu/abs/2016A&A...592A..83K}{592, A83}

\bibitem[{{Kama} {et~al.}(2015){Kama}, {Folsom}, \& {Pinilla}}]{Kama2015}
{Kama}, M., {Folsom}, C.~P., \& {Pinilla}, P. 2015, \href{http://dx.doi.org/10.1051/0004-6361/201527094}{\color{blue}\aap}, \href{https://ui.adsabs.harvard.edu/abs/2015A&A...582L..10K}{582, L10}

\bibitem[{{Kama} {et~al.}(2020){Kama}, {Trapman}, {Fedele}, {Bruderer}, {Hogerheijde}, {Miotello}, {van Dishoeck}, {Clarke}, \& {Bergin}}]{Kama2020}
{Kama}, M., {Trapman}, L., {Fedele}, D., {et~al.} 2020, \href{http://dx.doi.org/10.1051/0004-6361/201937124}{\color{blue}\aap}, \href{https://ui.adsabs.harvard.edu/abs/2020A&A...634A..88K}{634, A88}

\bibitem[{{Kataoka} {et~al.}(2016){Kataoka}, {Tsukagoshi}, {Momose}, {Nagai}, {Muto}, {Dullemond}, {Pohl}, {Fukagawa}, {Shibai}, {Hanawa}, \& {Murakawa}}]{Kataoka2016}
{Kataoka}, A., {Tsukagoshi}, T., {Momose}, M., {et~al.} 2016, \href{http://dx.doi.org/10.3847/2041-8205/831/2/L12}{\color{blue}\apjl}, \href{https://ui.adsabs.harvard.edu/abs/2016ApJ...831L..12K}{831, L12}

\bibitem[{{Klaassen} {et~al.}(2013){Klaassen}, {Juhasz}, {Mathews}, {Mottram}, {De Gregorio-Monsalvo}, {van Dishoeck}, {Takahashi}, {Akiyama}, {Chapillon}, {Espada}, {Hales}, {Hogerheijde}, {Rawlings}, {Schmalzl}, \& {Testi}}]{Klaassen2013}
{Klaassen}, P.~D., {Juhasz}, A., {Mathews}, G.~S., {et~al.} 2013, \href{http://dx.doi.org/10.1051/0004-6361/201321129}{\color{blue}\aap}, \href{https://ui.adsabs.harvard.edu/abs/2013A&A...555A..73K}{555, A73}

\bibitem[{{Kratter} \& {Lodato}(2016)}]{Kratter2016}
{Kratter}, K. \& {Lodato}, G. 2016, \href{http://dx.doi.org/10.1146/annurev-astro-081915-023307}{\color{blue}\araa}, \href{https://ui.adsabs.harvard.edu/abs/2016ARA&A..54..271K}{54, 271}

\bibitem[{{Kraus} {et~al.}(2017){Kraus}, {Kreplin}, {Fukugawa}, {Muto}, {Sitko}, {Young}, {Bate}, {Grady}, {Harries}, {Monnier}, {Willson}, \& {Wisniewski}}]{Kraus2017}
{Kraus}, S., {Kreplin}, A., {Fukugawa}, M., {et~al.} 2017, \href{http://dx.doi.org/10.3847/2041-8213/aa8edc}{\color{blue}\apjl}, \href{https://ui.adsabs.harvard.edu/abs/2017ApJ...848L..11K}{848, L11}

\bibitem[{{Krijt} {et~al.}(2018){Krijt}, {Schwarz}, {Bergin}, \& {Ciesla}}]{Krijt2018}
{Krijt}, S., {Schwarz}, K.~R., {Bergin}, E.~A., \& {Ciesla}, F.~J. 2018, \href{http://dx.doi.org/10.3847/1538-4357/aad69b}{\color{blue}\apj}, \href{https://ui.adsabs.harvard.edu/abs/2018ApJ...864...78K}{864, 78}

\bibitem[{{Lacy} {et~al.}(1994){Lacy}, {Knacke}, {Geballe}, \& {Tokunaga}}]{Lacy1994}
{Lacy}, J.~H., {Knacke}, R., {Geballe}, T.~R., \& {Tokunaga}, A.~T. 1994, \href{http://dx.doi.org/10.1086/187395}{\color{blue}\apjl}, \href{https://ui.adsabs.harvard.edu/abs/1994ApJ...428L..69L}{428, L69}

\bibitem[{{Law} {et~al.}(2022){Law}, {Crystian}, {Teague}, {{\"O}berg}, {Rich}, {Andrews}, {Bae}, {Flaherty}, {Guzm{\'a}n}, {Huang}, {Ilee}, {Kastner}, {Loomis}, {Long}, {P{\'e}rez}, {P{\'e}rez}, {Qi}, {Rosotti}, {Ru{\'\i}z-Rodr{\'\i}guez}, {Tsukagoshi}, \& {Wilner}}]{Law2022}
{Law}, C.~J., {Crystian}, S., {Teague}, R., {et~al.} 2022, \href{http://dx.doi.org/10.3847/1538-4357/ac6c02}{\color{blue}\apj}, \href{https://ui.adsabs.harvard.edu/abs/2022ApJ...932..114L}{932, 114}

\bibitem[{{Law} {et~al.}(2021){Law}, {Teague}, {Loomis}, {Bae}, {{\"O}berg}, {Czekala}, {Andrews}, {Aikawa}, {Alarc{\'o}n}, {Bergin}, {Bergner}, {Booth}, {Bosman}, {Calahan}, {Cataldi}, {Cleeves}, {Furuya}, {Guzm{\'a}n}, {Huang}, {Ilee}, {Le Gal}, {Liu}, {Long}, {M{\'e}nard}, {Nomura}, {P{\'e}rez}, {Qi}, {Schwarz}, {Soto}, {Tsukagoshi}, {Yamato}, {van't Hoff}, {Walsh}, {Wilner}, \& {Zhang}}]{Law2021}
{Law}, C.~J., {Teague}, R., {Loomis}, R.~A., {et~al.} 2021, \href{http://dx.doi.org/10.3847/1538-4365/ac1439}{\color{blue}\apjs}, \href{https://ui.adsabs.harvard.edu/abs/2021ApJS..257....4L}{257, 4}

\bibitem[{{Law} {et~al.}(2023){Law}, {Teague}, {{\"O}berg}, {Rich}, {Andrews}, {Bae}, {Benisty}, {Facchini}, {Flaherty}, {Isella}, {Jin}, {Hashimoto}, {Huang}, {Loomis}, {Long}, {Mu{\~n}oz-Romero}, {Paneque-Carre{\~n}o}, {P{\'e}rez}, {Qi}, {Schwarz}, {Stadler}, {Tsukagoshi}, {Wilner}, \& {van der Plas}}]{Law2023}
{Law}, C.~J., {Teague}, R., {{\"O}berg}, K.~I., {et~al.} 2023, \href{http://dx.doi.org/10.3847/1538-4357/acb3c4}{\color{blue}\apj}, \href{https://ui.adsabs.harvard.edu/abs/2023ApJ...948...60L}{948, 60}

\bibitem[{{Leemker} {et~al.}(2022){Leemker}, {Booth}, {van Dishoeck}, {P{\'e}rez-S{\'a}nchez}, {Szul{\'a}gyi}, {Bosman}, {Bruderer}, {Facchini}, {Hogerheijde}, {Paneque-Carre{\~n}o}, \& {Sturm}}]{Leemker2022}
{Leemker}, M., {Booth}, A.~S., {van Dishoeck}, E.~F., {et~al.} 2022, \href{http://dx.doi.org/10.1051/0004-6361/202243229}{\color{blue}\aap}, \href{https://ui.adsabs.harvard.edu/abs/2022A&A...663A..23L}{663, A23}

\bibitem[{{Liu} {et~al.}(2019){Liu}, {Dipierro}, {Ragusa}, {Lodato}, {Herczeg}, {Long}, {Harsono}, {Boehler}, {Menard}, {Johnstone}, {Pascucci}, {Pinilla}, {Salyk}, {van der Plas}, {Cabrit}, {Fischer}, {Hendler}, {Manara}, {Nisini}, {Rigliaco}, {Avenhaus}, {Banzatti}, \& {Gully-Santiago}}]{Liu2019}
{Liu}, Y., {Dipierro}, G., {Ragusa}, E., {et~al.} 2019, \href{http://dx.doi.org/10.1051/0004-6361/201834157}{\color{blue}\aap}, \href{https://ui.adsabs.harvard.edu/abs/2019A&A...622A..75L}{622, A75}

\bibitem[{{Liu} {et~al.}(2022){Liu}, {Linz}, {Fang}, {Henning}, {Wolf}, {Flock}, {Rosotti}, {Wang}, \& {Li}}]{Liu2022}
{Liu}, Y., {Linz}, H., {Fang}, M., {et~al.} 2022, \href{http://dx.doi.org/10.1051/0004-6361/202244505}{\color{blue}\aap}, \href{https://ui.adsabs.harvard.edu/abs/2022A&A...668A.175L}{668, A175}

\bibitem[{{Long} {et~al.}(2021){Long}, {Andrews}, {Vega}, {Wilner}, {Chandler}, {Ragusa}, {Teague}, {P{\'e}rez}, {Calvet}, {Carpenter}, {Henning}, {Kwon}, {Linz}, \& {Ricci}}]{Long2021}
{Long}, F., {Andrews}, S.~M., {Vega}, J., {et~al.} 2021, \href{http://dx.doi.org/10.3847/1538-4357/abff53}{\color{blue}\apj}, \href{https://ui.adsabs.harvard.edu/abs/2021ApJ...915..131L}{915, 131}

\bibitem[{{Long} {et~al.}(2017){Long}, {Herczeg}, {Pascucci}, {Drabek-Maunder}, {Mohanty}, {Testi}, {Apai}, {Hendler}, {Henning}, {Manara}, \& {Mulders}}]{Long2017}
{Long}, F., {Herczeg}, G.~J., {Pascucci}, I., {et~al.} 2017, \href{http://dx.doi.org/10.3847/1538-4357/aa78fc}{\color{blue}\apj}, \href{https://ui.adsabs.harvard.edu/abs/2017ApJ...844...99L}{844, 99}

\bibitem[{{Loomis} {et~al.}(2018){Loomis}, {Cleeves}, {{\"O}berg}, {Aikawa}, {Bergner}, {Furuya}, {Guzman}, \& {Walsh}}]{Loomis2018}
{Loomis}, R.~A., {Cleeves}, L.~I., {{\"O}berg}, K.~I., {et~al.} 2018, \href{http://dx.doi.org/10.3847/1538-4357/aac169}{\color{blue}\apj}, \href{https://ui.adsabs.harvard.edu/abs/2018ApJ...859..131L}{859, 131}

\bibitem[{{Loomis} {et~al.}(2020){Loomis}, {{\"O}berg}, {Andrews}, {Bergin}, {Bergner}, {Blake}, {Cleeves}, {Czekala}, {Huang}, {Le Gal}, {M{\'e}nard}, {Pegues}, {Qi}, {Walsh}, {Williams}, \& {Wilner}}]{Loomis2020}
{Loomis}, R.~A., {{\"O}berg}, K.~I., {Andrews}, S.~M., {et~al.} 2020, \href{http://dx.doi.org/10.3847/1538-4357/ab7cc8}{\color{blue}\apj}, \href{https://ui.adsabs.harvard.edu/abs/2020ApJ...893..101L}{893, 101}

\bibitem[{{Lynden-Bell} \& {Pringle}(1974)}]{LyndenBell1974}
{Lynden-Bell}, D. \& {Pringle}, J.~E. 1974, \href{http://dx.doi.org/10.1093/mnras/168.3.603}{\color{blue}\mnras}, \href{https://ui.adsabs.harvard.edu/abs/1974MNRAS.168..603L}{168, 603}

\bibitem[{{Manara} {et~al.}(2022){Manara}, {Ansdell}, {Rosotti}, {Hughes}, {Armitage}, {Lodato}, \& {Williams}}]{Manara2022}
{Manara}, C.~F., {Ansdell}, M., {Rosotti}, G.~P., {et~al.} 2022, \href{https://ui.adsabs.harvard.edu/abs/2022arXiv220309930M}{\href{http://dx.doi.org/10.48550/arXiv.2203.09930}{\color{blue}arXiv e-prints}, arXiv:2203.09930}

\bibitem[{{McClure} {et~al.}(2016){McClure}, {Bergin}, {Cleeves}, {van Dishoeck}, {Blake}, {Evans}, {Green}, {Henning}, {{\"O}berg}, {Pontoppidan}, \& {Salyk}}]{McClure2016}
{McClure}, M.~K., {Bergin}, E.~A., {Cleeves}, L.~I., {et~al.} 2016, \href{http://dx.doi.org/10.3847/0004-637X/831/2/167}{\color{blue}\apj}, \href{https://ui.adsabs.harvard.edu/abs/2016ApJ...831..167M}{831, 167}

\bibitem[{{McMullin} {et~al.}(2007){McMullin}, {Waters}, {Schiebel}, {Young}, \& {Golap}}]{McMullin2007}
{McMullin}, J.~P., {Waters}, B., {Schiebel}, D., {Young}, W., \& {Golap}, K. 2007, in Astronomical Society of the Pacific Conference Series, Vol. 376, Astronomical Data Analysis Software and Systems XVI, ed. R.~A. {Shaw}, F.~{Hill}, \& D.~J. {Bell}, \href{https://ui.adsabs.harvard.edu/abs/2007ASPC..376..127M}{127}

\bibitem[{{Meeus} {et~al.}(2001){Meeus}, {Waters}, {Bouwman}, {van den Ancker}, {Waelkens}, \& {Malfait}}]{Meeus2001}
{Meeus}, G., {Waters}, L.~B.~F.~M., {Bouwman}, J., {et~al.} 2001, \href{http://dx.doi.org/10.1051/0004-6361:20000144}{\color{blue}\aap}, \href{https://ui.adsabs.harvard.edu/abs/2001A&A...365..476M}{365, 476}

\bibitem[{{Miley} {et~al.}(2018){Miley}, {Pani{\'c}}, {Wyatt}, \& {Kennedy}}]{Miley2018}
{Miley}, J.~M., {Pani{\'c}}, O., {Wyatt}, M., \& {Kennedy}, G.~M. 2018, \href{http://dx.doi.org/10.1051/0004-6361/201833381}{\color{blue}\aap}, \href{https://ui.adsabs.harvard.edu/abs/2018A&A...615L..10M}{615, L10}

\bibitem[{{Miotello} {et~al.}(2014){Miotello}, {Bruderer}, \& {van Dishoeck}}]{Miotello2014}
{Miotello}, A., {Bruderer}, S., \& {van Dishoeck}, E.~F. 2014, \href{http://dx.doi.org/10.1051/0004-6361/201424712}{\color{blue}\aap}, \href{https://ui.adsabs.harvard.edu/abs/2014A&A...572A..96M}{572, A96}

\bibitem[{{Miotello} {et~al.}(2022){Miotello}, {Kamp}, {Birnstiel}, {Cleeves}, \& {Kataoka}}]{Miotello2022}
{Miotello}, A., {Kamp}, I., {Birnstiel}, T., {Cleeves}, L.~I., \& {Kataoka}, A. 2022, \href{https://ui.adsabs.harvard.edu/abs/2022arXiv220309818M}{\href{http://dx.doi.org/10.48550/arXiv.2203.09818}{\color{blue}arXiv e-prints}, arXiv:2203.09818}

\bibitem[{{Miotello} {et~al.}(2021){Miotello}, {Rosotti}, {Ansdell}, {Facchini}, {Manara}, {Williams}, \& {Bruderer}}]{Miotello2021}
{Miotello}, A., {Rosotti}, G., {Ansdell}, M., {et~al.} 2021, \href{http://dx.doi.org/10.1051/0004-6361/202140550}{\color{blue}\aap}, \href{https://ui.adsabs.harvard.edu/abs/2021A&A...651A..48M}{651, A48}

\bibitem[{{Miotello} {et~al.}(2016){Miotello}, {van Dishoeck}, {Kama}, \& {Bruderer}}]{Miotello2016}
{Miotello}, A., {van Dishoeck}, E.~F., {Kama}, M., \& {Bruderer}, S. 2016, \href{http://dx.doi.org/10.1051/0004-6361/201628159}{\color{blue}\aap}, \href{https://ui.adsabs.harvard.edu/abs/2016A&A...594A..85M}{594, A85}

\bibitem[{{Miotello} {et~al.}(2017){Miotello}, {van Dishoeck}, {Williams}, {Ansdell}, {Guidi}, {Hogerheijde}, {Manara}, {Tazzari}, {Testi}, {van der Marel}, \& {van Terwisga}}]{Miotello2017}
{Miotello}, A., {van Dishoeck}, E.~F., {Williams}, J.~P., {et~al.} 2017, \href{http://dx.doi.org/10.1051/0004-6361/201629556}{\color{blue}\aap}, \href{https://ui.adsabs.harvard.edu/abs/2017A&A...599A.113M}{599, A113}

\bibitem[{{Mo{\'o}r} {et~al.}(2017){Mo{\'o}r}, {Cur{\'e}}, {K{\'o}sp{\'a}l}, {{\'A}brah{\'a}m}, {Csengeri}, {Eiroa}, {Gunawan}, {Henning}, {Hughes}, {Juh{\'a}sz}, {Pawellek}, \& {Wyatt}}]{Moor2017}
{Mo{\'o}r}, A., {Cur{\'e}}, M., {K{\'o}sp{\'a}l}, {\'A}., {et~al.} 2017, \href{http://dx.doi.org/10.3847/1538-4357/aa8e4e}{\color{blue}\apj}, \href{https://ui.adsabs.harvard.edu/abs/2017ApJ...849..123M}{849, 123}

\bibitem[{{Mo{\'o}r} {et~al.}(2020){Mo{\'o}r}, {K{\'o}sp{\'a}l}, {{\'A}brah{\'a}m}, \& {Pawellek}}]{Moor2020}
{Mo{\'o}r}, A., {K{\'o}sp{\'a}l}, {\'A}., {{\'A}brah{\'a}m}, P., \& {Pawellek}, N. 2020, in Origins: From the Protosun to the First Steps of Life, ed. B.~G. {Elmegreen}, L.~V. {T{\'o}th}, \& M.~{G{\"u}del}, Vol. 345, \href{https://ui.adsabs.harvard.edu/abs/2020IAUS..345..349M}{349--350}

\bibitem[{{Muro-Arena} {et~al.}(2020){Muro-Arena}, {Benisty}, {Ginski}, {Dominik}, {Facchini}, {Villenave}, {van Boekel}, {Chauvin}, {Garufi}, {Henning}, {Janson}, {Keppler}, {Matter}, {M{\'e}nard}, {Stolker}, {Zurlo}, {Blanchard}, {Maurel}, {Moeller-Nilsson}, {Petit}, {Roux}, {Sevin}, \& {Wildi}}]{MuroArena2020}
{Muro-Arena}, G.~A., {Benisty}, M., {Ginski}, C., {et~al.} 2020, \href{http://dx.doi.org/10.1051/0004-6361/201936509}{\color{blue}\aap}, \href{https://ui.adsabs.harvard.edu/abs/2020A&A...635A.121M}{635, A121}

\bibitem[{{Okamoto} {et~al.}(2009){Okamoto}, {Kataza}, {Honda}, {Fujiwara}, {Momose}, {Ohashi}, {Fujiyoshi}, {Sakon}, {Sako}, {Yamashita}, {Miyata}, \& {Onaka}}]{Okamoto2009}
{Okamoto}, Y.~K., {Kataza}, H., {Honda}, M., {et~al.} 2009, \href{http://dx.doi.org/10.1088/0004-637X/706/1/665}{\color{blue}\apj}, \href{https://ui.adsabs.harvard.edu/abs/2009ApJ...706..665O}{706, 665}

\bibitem[{pandas~development team(2020)}]{reback2020pandas}
pandas~development team, T. 2020, pandas-dev/pandas: Pandas

\bibitem[{{Pascucci} {et~al.}(2022){Pascucci}, {Cabrit}, {Edwards}, {Gorti}, {Gressel}, \& {Suzuki}}]{Pascucci2022}
{Pascucci}, I., {Cabrit}, S., {Edwards}, S., {et~al.} 2022, \href{https://ui.adsabs.harvard.edu/abs/2022arXiv220310068P}{\href{http://dx.doi.org/10.48550/arXiv.2203.10068}{\color{blue}arXiv e-prints}, arXiv:2203.10068}

\bibitem[{{Pascucci} {et~al.}(2016){Pascucci}, {Testi}, {Herczeg}, {Long}, {Manara}, {Hendler}, {Mulders}, {Krijt}, {Ciesla}, {Henning}, {Mohanty}, {Drabek-Maunder}, {Apai}, {Sz{\H{u}}cs}, {Sacco}, \& {Olofsson}}]{Pascucci2016}
{Pascucci}, I., {Testi}, L., {Herczeg}, G.~J., {et~al.} 2016, \href{http://dx.doi.org/10.3847/0004-637X/831/2/125}{\color{blue}\apj}, \href{https://ui.adsabs.harvard.edu/abs/2016ApJ...831..125P}{831, 125}

\bibitem[{{Pegues} {et~al.}(2023){Pegues}, {{\"O}berg}, {Qi}, {Andrews}, {Huang}, {Law}, {Le Gal}, {Matr{\`a}}, \& {Wilner}}]{Pegues2023}
{Pegues}, J., {{\"O}berg}, K.~I., {Qi}, C., {et~al.} 2023, \href{http://dx.doi.org/10.3847/1538-4357/acbf31}{\color{blue}\apj}, \href{https://ui.adsabs.harvard.edu/abs/2023ApJ...948...57P}{948, 57}

\bibitem[{{Pineda} {et~al.}(2019){Pineda}, {Szul{\'a}gyi}, {Quanz}, {van Dishoeck}, {Garufi}, {Meru}, {Mulders}, {Testi}, {Meyer}, \& {Reggiani}}]{Pineda2019}
{Pineda}, J.~E., {Szul{\'a}gyi}, J., {Quanz}, S.~P., {et~al.} 2019, \href{http://dx.doi.org/10.3847/1538-4357/aaf389}{\color{blue}\apj}, \href{https://ui.adsabs.harvard.edu/abs/2019ApJ...871...48P}{871, 48}

\bibitem[{{Richards} {et~al.}(2022){Richards}, {Moravec}, {Etoka}, {Fomalont}, {P{\'e}rez-S{\'a}nchez}, {Toribio}, \& {Laing}}]{Richards2022}
{Richards}, A.~M.~S., {Moravec}, E., {Etoka}, S., {et~al.} 2022, \href{https://ui.adsabs.harvard.edu/abs/2022arXiv220705591R}{\href{http://dx.doi.org/10.48550/arXiv.2207.05591}{\color{blue}arXiv e-prints}, arXiv:2207.05591}

\bibitem[{{Rosotti} {et~al.}(2020){Rosotti}, {Benisty}, {Juh{\'a}sz}, {Teague}, {Clarke}, {Dominik}, {Dullemond}, {Klaassen}, {Matr{\`a}}, \& {Stolker}}]{Rosotti2020}
{Rosotti}, G.~P., {Benisty}, M., {Juh{\'a}sz}, A., {et~al.} 2020, \href{http://dx.doi.org/10.1093/mnras/stz3090}{\color{blue}\mnras}, \href{https://ui.adsabs.harvard.edu/abs/2020MNRAS.491.1335R}{491, 1335}

\bibitem[{{Salinas} {et~al.}(2017){Salinas}, {Hogerheijde}, {Mathews}, {{\"O}berg}, {Qi}, {Williams}, \& {Wilner}}]{Salinas2017}
{Salinas}, V.~N., {Hogerheijde}, M.~R., {Mathews}, G.~S., {et~al.} 2017, \href{http://dx.doi.org/10.1051/0004-6361/201731223}{\color{blue}\aap}, \href{https://ui.adsabs.harvard.edu/abs/2017A&A...606A.125S}{606, A125}

\bibitem[{{Sch{\"o}ier} {et~al.}(2005){Sch{\"o}ier}, {van der Tak}, {van Dishoeck}, \& {Black}}]{Schoier2005}
{Sch{\"o}ier}, F.~L., {van der Tak}, F.~F.~S., {van Dishoeck}, E.~F., \& {Black}, J.~H. 2005, \href{http://dx.doi.org/10.1051/0004-6361:20041729}{\color{blue}\aap}, \href{https://ui.adsabs.harvard.edu/abs/2005A&A...432..369S}{432, 369}

\bibitem[{{Schwarz} {et~al.}(2016){Schwarz}, {Bergin}, {Cleeves}, {Blake}, {Zhang}, {{\"O}berg}, {van Dishoeck}, \& {Qi}}]{Schwarz2016}
{Schwarz}, K.~R., {Bergin}, E.~A., {Cleeves}, L.~I., {et~al.} 2016, \href{http://dx.doi.org/10.3847/0004-637X/823/2/91}{\color{blue}\apj}, \href{https://ui.adsabs.harvard.edu/abs/2016ApJ...823...91S}{823, 91}

\bibitem[{{Sierra} {et~al.}(2021){Sierra}, {P{\'e}rez}, {Zhang}, {Law}, {Guzm{\'a}n}, {Qi}, {Bosman}, {{\"O}berg}, {Andrews}, {Long}, {Teague}, {Booth}, {Walsh}, {Wilner}, {M{\'e}nard}, {Cataldi}, {Czekala}, {Bae}, {Huang}, {Bergner}, {Ilee}, {Benisty}, {Le Gal}, {Loomis}, {Tsukagoshi}, {Liu}, {Yamato}, \& {Aikawa}}]{Sierra2021}
{Sierra}, A., {P{\'e}rez}, L.~M., {Zhang}, K., {et~al.} 2021, \href{http://dx.doi.org/10.3847/1538-4365/ac1431}{\color{blue}\apjs}, \href{https://ui.adsabs.harvard.edu/abs/2021ApJS..257...14S}{257, 14}

\bibitem[{{Stapper} {et~al.}(2022){Stapper}, {Hogerheijde}, {van Dishoeck}, \& {Mentel}}]{stapper2022}
{Stapper}, L.~M., {Hogerheijde}, M.~R., {van Dishoeck}, E.~F., \& {Mentel}, R. 2022, \href{http://dx.doi.org/10.1051/0004-6361/202142164}{\color{blue}\aap}, \href{https://ui.adsabs.harvard.edu/abs/2022A&A...658A.112S}{658, A112}

\bibitem[{{Stapper} {et~al.}(2023){Stapper}, {Hogerheijde}, {van Dishoeck}, \& {Paneque-Carre{\~n}o}}]{stapper2023}
{Stapper}, L.~M., {Hogerheijde}, M.~R., {van Dishoeck}, E.~F., \& {Paneque-Carre{\~n}o}, T. 2023, \href{http://dx.doi.org/10.1051/0004-6361/202245137}{\color{blue}\aap}, \href{https://ui.adsabs.harvard.edu/abs/2023A&A...669A.158S}{669, A158}

\bibitem[{{Sturm} {et~al.}(2023){Sturm}, {Booth}, {McClure}, {Leemker}, \& {van Dishoeck}}]{Sturm2023}
{Sturm}, J.~A., {Booth}, A.~S., {McClure}, M.~K., {Leemker}, M., \& {van Dishoeck}, E.~F. 2023, \href{http://dx.doi.org/10.1051/0004-6361/202244227}{\color{blue}\aap}, \href{https://ui.adsabs.harvard.edu/abs/2023A&A...670A..12S}{670, A12}

\bibitem[{{Tang} {et~al.}(2017){Tang}, {Guilloteau}, {Dutrey}, {Muto}, {Shen}, {Gu}, {Inutsuka}, {Momose}, {Pietu}, {Fukagawa}, {Chapillon}, {Ho}, {di Folco}, {Corder}, {Ohashi}, \& {Hashimoto}}]{Tang2017}
{Tang}, Y.-W., {Guilloteau}, S., {Dutrey}, A., {et~al.} 2017, \href{http://dx.doi.org/10.3847/1538-4357/aa6af7}{\color{blue}\apj}, \href{https://ui.adsabs.harvard.edu/abs/2017ApJ...840...32T}{840, 32}

\bibitem[{{Teague} {et~al.}(2021){Teague}, {Bae}, {Aikawa}, {Andrews}, {Bergin}, {Bergner}, {Boehler}, {Booth}, {Bosman}, {Cataldi}, {Czekala}, {Guzm{\'a}n}, {Huang}, {Ilee}, {Law}, {Le Gal}, {Long}, {Loomis}, {M{\'e}nard}, {{\"O}berg}, {P{\'e}rez}, {Schwarz}, {Sierra}, {Walsh}, {Wilner}, {Yamato}, \& {Zhang}}]{Teague2021}
{Teague}, R., {Bae}, J., {Aikawa}, Y., {et~al.} 2021, \href{http://dx.doi.org/10.3847/1538-4365/ac1438}{\color{blue}\apjs}, \href{https://ui.adsabs.harvard.edu/abs/2021ApJS..257...18T}{257, 18}

\bibitem[{{Temmink} {et~al.}(2023){Temmink}, {Booth}, {van der Marel}, \& {van Dishoeck}}]{Temmink2023}
{Temmink}, M., {Booth}, A.~S., {van der Marel}, N., \& {van Dishoeck}, E.~F. 2023, \href{https://ui.adsabs.harvard.edu/abs/2023arXiv230406382T}{\href{http://dx.doi.org/10.48550/arXiv.2304.06382}{\color{blue}arXiv e-prints}, arXiv:2304.06382}

\bibitem[{{Testi} {et~al.}(2022){Testi}, {Natta}, {Manara}, {de Gregorio Monsalvo}, {Lodato}, {Lopez}, {Muzic}, {Pascucci}, {Sanchis}, {Miranda}, {Scholz}, {De Simone}, \& {Williams}}]{Testi2022}
{Testi}, L., {Natta}, A., {Manara}, C.~F., {et~al.} 2022, \href{http://dx.doi.org/10.1051/0004-6361/202141380}{\color{blue}\aap}, \href{https://ui.adsabs.harvard.edu/abs/2022A&A...663A..98T}{663, A98}

\bibitem[{{Tilling} {et~al.}(2012){Tilling}, {Woitke}, {Meeus}, {Mora}, {Montesinos}, {Riviere-Marichalar}, {Eiroa}, {Thi}, {Isella}, {Roberge}, {Martin-Zaidi}, {Kamp}, {Pinte}, {Sandell}, {Vacca}, {M{\'e}nard}, {Mendigut{\'\i}a}, {Duch{\^e}ne}, {Dent}, {Aresu}, {Meijerink}, \& {Spaans}}]{Tilling2012}
{Tilling}, I., {Woitke}, P., {Meeus}, G., {et~al.} 2012, \href{http://dx.doi.org/10.1051/0004-6361/201116919}{\color{blue}\aap}, \href{https://ui.adsabs.harvard.edu/abs/2012A&A...538A..20T}{538, A20}

\bibitem[{{Trapman} {et~al.}(2020){Trapman}, {Ansdell}, {Hogerheijde}, {Facchini}, {Manara}, {Miotello}, {Williams}, \& {Bruderer}}]{Trapman2020}
{Trapman}, L., {Ansdell}, M., {Hogerheijde}, M.~R., {et~al.} 2020, \href{http://dx.doi.org/10.1051/0004-6361/201834537}{\color{blue}\aap}, \href{https://ui.adsabs.harvard.edu/abs/2020A&A...638A..38T}{638, A38}

\bibitem[{{Trapman} {et~al.}(2019){Trapman}, {Facchini}, {Hogerheijde}, {van Dishoeck}, \& {Bruderer}}]{Trapman2019}
{Trapman}, L., {Facchini}, S., {Hogerheijde}, M.~R., {van Dishoeck}, E.~F., \& {Bruderer}, S. 2019, \href{http://dx.doi.org/10.1051/0004-6361/201834723}{\color{blue}\aap}, \href{https://ui.adsabs.harvard.edu/abs/2019A&A...629A..79T}{629, A79}

\bibitem[{{Trapman} {et~al.}(2017){Trapman}, {Miotello}, {Kama}, {van Dishoeck}, \& {Bruderer}}]{Trapman2017}
{Trapman}, L., {Miotello}, A., {Kama}, M., {van Dishoeck}, E.~F., \& {Bruderer}, S. 2017, \href{http://dx.doi.org/10.1051/0004-6361/201630308}{\color{blue}\aap}, \href{https://ui.adsabs.harvard.edu/abs/2017A&A...605A..69T}{605, A69}

\bibitem[{{Trapman} {et~al.}(2023){Trapman}, {Rosotti}, {Zhang}, \& {Tabone}}]{Trapman2023}
{Trapman}, L., {Rosotti}, G., {Zhang}, K., \& {Tabone}, B. 2023, \href{http://dx.doi.org/10.3847/1538-4357/ace7d1}{\color{blue}\apj}, \href{https://ui.adsabs.harvard.edu/abs/2023ApJ...954...41T}{954, 41}

\bibitem[{{Trapman} {et~al.}(2022){Trapman}, {Tabone}, {Rosotti}, \& {Zhang}}]{Trapman2022}
{Trapman}, L., {Tabone}, B., {Rosotti}, G., \& {Zhang}, K. 2022, \href{http://dx.doi.org/10.3847/1538-4357/ac3ed5}{\color{blue}\apj}, \href{https://ui.adsabs.harvard.edu/abs/2022ApJ...926...61T}{926, 61}

\bibitem[{{van der Marel} {et~al.}(2016){van der Marel}, {Cazzoletti}, {Pinilla}, \& {Garufi}}]{vanderMarel2016}
{van der Marel}, N., {Cazzoletti}, P., {Pinilla}, P., \& {Garufi}, A. 2016, \href{http://dx.doi.org/10.3847/0004-637X/832/2/178}{\color{blue}\apj}, \href{https://ui.adsabs.harvard.edu/abs/2016ApJ...832..178V}{832, 178}

\bibitem[{{van der Plas} {et~al.}(2017){van der Plas}, {M{\'e}nard}, {Canovas}, {Avenhaus}, {Casassus}, {Pinte}, {Caceres}, \& {Cieza}}]{vanderPlas2017a}
{van der Plas}, G., {M{\'e}nard}, F., {Canovas}, H., {et~al.} 2017, \href{http://dx.doi.org/10.1051/0004-6361/201731392}{\color{blue}\aap}, \href{https://ui.adsabs.harvard.edu/abs/2017A&A...607A..55V}{607, A55}

\bibitem[{{van der Velden}(2020)}]{cmasher}
{van der Velden}, E. 2020, \href{http://dx.doi.org/10.21105/joss.02004}{\color{blue}The Journal of Open Source Software}, \href{https://ui.adsabs.harvard.edu/abs/2020JOSS....5.2004V}{5, 2004}

\bibitem[{{van Terwisga} {et~al.}(2022){van Terwisga}, {Hacar}, {van Dishoeck}, {Oonk}, \& {Portegies Zwart}}]{vanTerwisga2022}
{van Terwisga}, S.~E., {Hacar}, A., {van Dishoeck}, E.~F., {Oonk}, R., \& {Portegies Zwart}, S. 2022, \href{http://dx.doi.org/10.1051/0004-6361/202141913}{\color{blue}\aap}, \href{https://ui.adsabs.harvard.edu/abs/2022A&A...661A..53V}{661, A53}

\bibitem[{{Vioque} {et~al.}(2018){Vioque}, {Oudmaijer}, {Baines}, {Mendigut{\'\i}a}, \& {P{\'e}rez-Mart{\'\i}nez}}]{Vioque2018}
{Vioque}, M., {Oudmaijer}, R.~D., {Baines}, D., {Mendigut{\'\i}a}, I., \& {P{\'e}rez-Mart{\'\i}nez}, R. 2018, \href{http://dx.doi.org/10.1051/0004-6361/201832870}{\color{blue}\aap}, \href{https://ui.adsabs.harvard.edu/abs/2018A&A...620A.128V}{620, A128}

\bibitem[{Virtanen {et~al.}(2020)Virtanen, Gommers, Oliphant, Haberland, Reddy, Cournapeau, Burovski, Peterson, Weckesser, Bright, {van der Walt}, Brett, Wilson, Millman, Mayorov, Nelson, Jones, Kern, Larson, Carey, Polat, Feng, Moore, {VanderPlas}, Laxalde, Perktold, Cimrman, Henriksen, Quintero, Harris, Archibald, Ribeiro, Pedregosa, {van Mulbregt}, \& {SciPy 1.0 Contributors}}]{2020SciPy-NMeth}
Virtanen, P., Gommers, R., Oliphant, T.~E., {et~al.} 2020, \href{http://dx.doi.org/10.1038/s41592-019-0686-2}{\color{blue}Nature Methods}, \href{https://rdcu.be/b08Wh}{17, 261}

\bibitem[{{Visser} {et~al.}(2009){Visser}, {van Dishoeck}, \& {Black}}]{Visser2009}
{Visser}, R., {van Dishoeck}, E.~F., \& {Black}, J.~H. 2009, \href{http://dx.doi.org/10.1051/0004-6361/200912129}{\color{blue}\aap}, \href{https://ui.adsabs.harvard.edu/abs/2009A&A...503..323V}{503, 323}

\bibitem[{{Walsh} {et~al.}(2016){Walsh}, {Juh{\'a}sz}, {Meeus}, {Dent}, {Maud}, {Aikawa}, {Millar}, \& {Nomura}}]{Walsh2016}
{Walsh}, C., {Juh{\'a}sz}, A., {Meeus}, G., {et~al.} 2016, \href{http://dx.doi.org/10.3847/0004-637X/831/2/200}{\color{blue}\apj}, \href{https://ui.adsabs.harvard.edu/abs/2016ApJ...831..200W}{831, 200}

\bibitem[{{Walsh} {et~al.}(2014){Walsh}, {Juh{\'a}sz}, {Pinilla}, {Harsono}, {Mathews}, {Dent}, {Hogerheijde}, {Birnstiel}, {Meeus}, {Nomura}, {Aikawa}, {Millar}, \& {Sandell}}]{Walsh2014}
{Walsh}, C., {Juh{\'a}sz}, A., {Pinilla}, P., {et~al.} 2014, \href{http://dx.doi.org/10.1088/2041-8205/791/1/L6}{\color{blue}\apjl}, \href{https://ui.adsabs.harvard.edu/abs/2014ApJ...791L...6W}{791, L6}

\bibitem[{Waskom(2021)}]{Waskom2021}
Waskom, M.~L. 2021, \href{http://dx.doi.org/10.21105/joss.03021}{\color{blue}Journal of Open Source Software}, 6, 6

\bibitem[{{White} {et~al.}(2016){White}, {Boley}, {Hughes}, {Flaherty}, {Ford}, {Wilner}, {Corder}, \& {Payne}}]{White2016}
{White}, J.~A., {Boley}, A.~C., {Hughes}, A.~M., {et~al.} 2016, \href{http://dx.doi.org/10.3847/0004-637X/829/1/6}{\color{blue}\apj}, \href{https://ui.adsabs.harvard.edu/abs/2016ApJ...829....6W}{829, 6}

\bibitem[{{Wilson} \& {Rood}(1994)}]{Wilson1994}
{Wilson}, T.~L. \& {Rood}, R. 1994, \href{http://dx.doi.org/10.1146/annurev.aa.32.090194.001203}{\color{blue}\araa}, \href{https://ui.adsabs.harvard.edu/abs/1994ARA&A..32..191W}{32, 191}

\bibitem[{{W{\"o}lfer} {et~al.}(2021){W{\"o}lfer}, {Facchini}, {Kurtovic}, {Teague}, {van Dishoeck}, {Benisty}, {Ercolano}, {Lodato}, {Miotello}, {Rosotti}, {Testi}, \& {Ubeira Gabellini}}]{Wolfer2021}
{W{\"o}lfer}, L., {Facchini}, S., {Kurtovic}, N.~T., {et~al.} 2021, \href{http://dx.doi.org/10.1051/0004-6361/202039469}{\color{blue}\aap}, \href{https://ui.adsabs.harvard.edu/abs/2021A&A...648A..19W}{648, A19}

\bibitem[{{Xin} {et~al.}(2023){Xin}, {Espaillat}, {Rilinger}, {Ribas}, \& {Mac{\'\i}as}}]{Xin2023}
{Xin}, Z., {Espaillat}, C.~C., {Rilinger}, A.~M., {Ribas}, {\'A}., \& {Mac{\'\i}as}, E. 2023, \href{http://dx.doi.org/10.3847/1538-4357/aca52b}{\color{blue}\apj}, \href{https://ui.adsabs.harvard.edu/abs/2023ApJ...942....4X}{942, 4}

\bibitem[{{Yang} {et~al.}(2010){Yang}, {Stancil}, {Balakrishnan}, \& {Forrey}}]{Yang2010}
{Yang}, B., {Stancil}, P.~C., {Balakrishnan}, N., \& {Forrey}, R.~C. 2010, \href{http://dx.doi.org/10.1088/0004-637X/718/2/1062}{\color{blue}\apj}, \href{https://ui.adsabs.harvard.edu/abs/2010ApJ...718.1062Y}{718, 1062}

\bibitem[{{Yu} {et~al.}(2021){Yu}, {Teague}, {Bae}, \& {{\"O}berg}}]{Yu2021}
{Yu}, H., {Teague}, R., {Bae}, J., \& {{\"O}berg}, K. 2021, \href{http://dx.doi.org/10.3847/2041-8213/ac283e}{\color{blue}\apjl}, \href{https://ui.adsabs.harvard.edu/abs/2021ApJ...920L..33Y}{920, L33}

\bibitem[{{Yu} {et~al.}(2019){Yu}, {Ho}, \& {Zhu}}]{Yu2019}
{Yu}, S.-Y., {Ho}, L.~C., \& {Zhu}, Z. 2019, \href{http://dx.doi.org/10.3847/1538-4357/ab1d65}{\color{blue}\apj}, \href{https://ui.adsabs.harvard.edu/abs/2019ApJ...877..100Y}{877, 100}

\bibitem[{{Zhang} {et~al.}(2021){Zhang}, {Booth}, {Law}, {Bosman}, {Schwarz}, {Bergin}, {{\"O}berg}, {Andrews}, {Guzm{\'a}n}, {Walsh}, {Qi}, {van't Hoff}, {Long}, {Wilner}, {Huang}, {Czekala}, {Ilee}, {Cataldi}, {Bergner}, {Aikawa}, {Teague}, {Bae}, {Loomis}, {Calahan}, {Alarc{\'o}n}, {M{\'e}nard}, {Le Gal}, {Sierra}, {Yamato}, {Nomura}, {Tsukagoshi}, {P{\'e}rez}, {Trapman}, {Liu}, \& {Furuya}}]{Zhang2021}
{Zhang}, K., {Booth}, A.~S., {Law}, C.~J., {et~al.} 2021, \href{http://dx.doi.org/10.3847/1538-4365/ac1580}{\color{blue}\apjs}, \href{https://ui.adsabs.harvard.edu/abs/2021ApJS..257....5Z}{257, 5}

\bibitem[{{Zhang} {et~al.}(2020{\natexlab{a}}){Zhang}, {Bosman}, \& {Bergin}}]{ZhangBosman2020}
{Zhang}, K., {Bosman}, A.~D., \& {Bergin}, E.~A. 2020{\natexlab{a}}, \href{http://dx.doi.org/10.3847/2041-8213/ab77ca}{\color{blue}\apjl}, \href{https://ui.adsabs.harvard.edu/abs/2020ApJ...891L..16Z}{891, L16}

\bibitem[{{Zhang} {et~al.}(2020{\natexlab{b}}){Zhang}, {Schwarz}, \& {Bergin}}]{Zhang2020}
{Zhang}, K., {Schwarz}, K.~R., \& {Bergin}, E.~A. 2020{\natexlab{b}}, \href{http://dx.doi.org/10.3847/2041-8213/ab7823}{\color{blue}\apjl}, \href{https://ui.adsabs.harvard.edu/abs/2020ApJ...891L..17Z}{891, L17}

\end{thebibliography}

\appendix
\section{Data sets used}
\label{app:data_sets}
In Table \ref{tab:project_codes} the project codes of the used data sets are listed together with their spatial and velocity resolution and the line-free rms noise.

\setlength{\tabcolsep}{1pt}
\begin{table*}[]
\caption{Data sets and corresponding parameters for each Herbig disk. The rms noise is for a line-free channel at the given velocity resolution. When the CO isotopologue observations are coming from different projects, multiple project codes are listed.}
\tiny\centering
\begin{tabular}{l|ccc|ccc|ccc|l}
\hline\hline
& \multicolumn{3}{c|}{\ce{^12CO}} & \multicolumn{3}{c|}{\ce{^13CO}} & \multicolumn{3}{c}{\ce{C^18O}} \\ \hline
\makecell{Name \\ \hspace{1mm} \\ \hspace{1mm}\\ \hspace{1mm}} & \makecell{Spat.res. \\ ($''$) \\ \hspace{1mm}} & \makecell{Vel.res. \\ (km~s$^{{-1}}$) \\ \hspace{1mm}} & \makecell{rms \\ (mJy\\beam$^{-1}$)} & \makecell{Spat.res. \\ ($''$) \\ \hspace{1mm}} & \makecell{Vel.res. \\ (km~s$^{{-1}}$) \\ \hspace{1mm}} & \makecell{rms \\ (mJy\\beam$^{-1}$)} & \makecell{Spat.res. \\ ($''$) \\ \hspace{1mm}} & \makecell{Vel.res. \\ (km~s$^{{-1}}$) \\ \hspace{1mm}} & \makecell{rms \\ (mJy\\beam$^{-1}$)} & \makecell{Project codes \\ \hspace{1mm} \\ \hspace{1mm}} \\ \hline
AB Aur & 0.39 $\times$ 0.28 (10$\degree$) & 0.20 & 16.83  &  &  &  &  &  &  & 2012.1.00303.S$^{*}$\hspace{-4.4pt}\\
       &                    &      &        & 1.14 $\times$ 0.70 (14$\degree$) & 0.20 & 5.67 & 1.14 $\times$ 0.71 (13$\degree$) & 0.20 & 4.57 & 2019.1.00579.S\\
AK Sco & 0.15 $\times$ 0.11 (-63$\degree$) & 1.00 & 1.35  & 0.16 $\times$ 0.12 (-63$\degree$) & 1.00 & 1.53  & 0.16 $\times$ 0.13 (-75$\degree$) & 1.00 & 0.95  & 2016.1.00204.S\\
CQ Tau & 0.11 $\times$ 0.08 (18$\degree$) & 0.32 & 1.53  & 0.12 $\times$ 0.09 (26$\degree$) & 0.66 & 1.32  & 0.12 $\times$ 0.09 (25$\degree$) & 0.67 & 0.91  & 2017.1.01404.S\\
HD 100453 & 0.29 $\times$ 0.21 (40$\degree$) & 0.20 & 3.26  & 0.31 $\times$ 0.23 (39$\degree$) & 0.20 & 2.74  & 0.30 $\times$ 0.22 (38$\degree$) & 0.20 & 2.05  & 2015.1.00192.S\\
HD 100546 & 0.27 $\times$ 0.17 (-85$\degree$) & 0.20 & 3.65  & 0.27 $\times$ 0.19 (-10$\degree$) & 0.20 & 3.44  & 0.27 $\times$ 0.19 (-10$\degree$) & 0.20 & 2.61  & 2016.1.00344.S\\
HD 104237 & 0.31 $\times$ 0.22 (-12$\degree$) & 0.70 & 3.28  & 0.32 $\times$ 0.22 (-14$\degree$) & 0.70 & 2.43  & 0.32 $\times$ 0.22 (-13$\degree$) & 0.70 & 1.92  & 2017.1.01419.S\\
HD 135344B &  &  &   & 0.34 $\times$ 0.30 (75$\degree$) & 0.20 & 7.64  & 0.35 $\times$ 0.30 (84$\degree$) & 0.20 & 10.43  & 2012.1.00158.S$^{*}$\hspace{-4.4pt}\\
  & 0.36 $\times$ 0.29 (-68$\degree$) & 0.20 & 11.79 &  &  &  &  &  &  & 2012.1.00870.S\\
HD 139614 &  & & & 0.79 $\times$ 0.60 (-49$\degree$) & 0.40 & 12.13  & 0.81 $\times$ 0.60 (-47$\degree$) & 0.40 & 8.03  & 2015.1.01600.S\\
HD 141569 & 0.41 $\times$ 0.33 (-61$\degree$) & 0.42 & 5.99  &  &  &   &  &  &   & 2012.1.00698.S$^{*}$\hspace{-4.4pt}\\
          &                    &      &       & 0.76 $\times$ 0.66 (82$\degree$) & 0.40 & 15.19 & 0.72 $\times$ 0.67 (-66$\degree$) & 0.40 & 10.91 & 2015.1.01600.S\\
HD 142527 & 0.93 $\times$ 0.81 (-85$\degree$) & 0.20 & 6.35  & 0.97 $\times$ 0.84 (-87$\degree$) & 0.20 & 6.68  & 0.98 $\times$ 0.85 (-88$\degree$) & 0.20 & 4.92  & 2015.1.01353.S\\
HD 142666 &  &  &   & 1.05 $\times$ 0.82 (87$\degree$) & 0.20 & 16.18  & 1.04 $\times$ 0.85 (87$\degree$) & 0.20 & 10.46  & 2015.1.01600.S\\
  & 0.21 $\times$ 0.20 (-70$\degree$) & 0.32 & 2.27 &  &  &  &  &  &  & 2016.1.00484.L\\
HD 163296 & 0.65 $\times$ 0.56 (71$\degree$) & 0.20 & 3.63  & 0.68 $\times$ 0.59 (68$\degree$) & 0.20 & 2.98  & 0.69 $\times$ 0.59 (72$\degree$) & 0.20 & 2.11  & 2018.1.01055.L\\
HD 169142 & 0.37 $\times$ 0.31 (-77$\degree$) & 0.20 & 3.61  & 0.39 $\times$ 0.33 (88$\degree$) & 0.20 & 3.67  & 0.39 $\times$ 0.33 (-85$\degree$) & 0.20 & 2.64  & 2016.1.00344.S\\
HD 176386 & 0.40 $\times$ 0.30 (77$\degree$) & 0.30 & 17.88  & 0.42 $\times$ 0.31 (79$\degree$) & 0.30 & 17.33  & 0.42 $\times$ 0.31 (78$\degree$) & 0.30 & 14.01  & 2015.1.01058.S\\
HD 245185 & 0.42 $\times$ 0.39 (-85$\degree$) & 0.70 & 8.05  &  & & &  & &  & 2017.1.00466.S\\
HD 290764 & 0.08 $\times$ 0.07 (-80$\degree$) & 1.00 & 2.04  &  &  &   &  &  &  & 2015.1.00986.S$^{*}$\hspace{-4.4pt}\\
          &                    &      &       & 0.21 $\times$ 0.14 (-62$\degree$) & 1.00 & 1.60 & 0.23 $\times$ 0.14 (-63$\degree$) & 1.00 & 2.06 & 2017.1.01607.S$^{*}$\hspace{-4.4pt}\\
HD 31648 & 1.16 $\times$ 0.88 (15$\degree$) & 0.20 & 7.35  & 1.22 $\times$ 0.92 (13$\degree$) & 0.20 & 7.64  & 1.22 $\times$ 0.92 (14$\degree$) & 0.20 & 5.89  & 2016.1.00724.S\\
HD 34282 & 0.27 $\times$ 0.24 (68$\degree$) & 0.20 & 4.50  & 0.28 $\times$ 0.27 (63$\degree$) & 0.20 & 3.32  & 0.28 $\times$ 0.27 (59$\degree$) & 0.20 & 2.46  & 2015.1.00192.S\\
HD 36112 & 0.19 $\times$ 0.14 (-17$\degree$) & 1.40 & 1.56  & 0.20 $\times$ 0.15 (-16$\degree$) & 1.40 & 1.90  & 0.20 $\times$ 0.15 (-16$\degree$) & 1.40 & 1.28  & 2017.1.00940.S\\
HD 58647 & 0.54 $\times$ 0.45 (73$\degree$) & 0.20 & 6.03  & 0.64 $\times$ 0.54 (74$\degree$) & 0.20 & 4.69  & 0.65 $\times$ 0.55 (77$\degree$) & 0.20 & 3.87  & 2018.1.00814.S\\
HD 9672 & 1.70 $\times$ 1.15 (-75$\degree$) & 0.64 & 3.61  & 1.75 $\times$ 1.25 (-75$\degree$) & 0.66 & 3.92  & 1.78 $\times$ 1.20 (-74$\degree$) & 0.67 & 2.93  & 2018.1.01222.S\\
HD 97048 & 1.22 $\times$ 0.76 (-31$\degree$) & 0.30 & 7.81  & 1.30 $\times$ 0.79 (-32$\degree$) & 0.30 & 7.17  & 1.31 $\times$ 0.80 (-31$\degree$) & 0.30 & 5.89  & 2015.1.00192.S\\
HR 5999 & 0.28 $\times$ 0.27 (87$\degree$) & 0.40 & 12.62  & 0.29 $\times$ 0.29 (-82$\degree$) & 0.40 & 12.86  & 0.29 $\times$ 0.29 (-74$\degree$) & 0.40 & 9.57  & 2015.1.00222.S\\
KK Oph &  & & & 0.77 $\times$ 0.68 (73$\degree$) & 0.40 & 28.90  & 0.68 $\times$ 0.63 (-61$\degree$) & 0.40 & 11.98  & 2015.1.01600.S\\
MWC 297 & 0.41 $\times$ 0.36 (71$\degree$) & 0.20 & 7.29  & 0.43 $\times$ 0.38 (74$\degree$) & 0.20 & 5.83  & 0.43 $\times$ 0.38 (75$\degree$) & 0.20 & 4.65  & 2018.1.00814.S\\
TY CrA & 0.40 $\times$ 0.30 (78$\degree$) & 0.30 & 17.60  & 0.42 $\times$ 0.31 (78$\degree$) & 0.30 & 17.60  & 0.42 $\times$ 0.31 (77$\degree$) & 0.30 & 13.67  & 2015.1.01058.S\\
V718 Sco &  & & & 1.01 $\times$ 0.82 (-67$\degree$) & 0.40 & 10.52  & 0.99 $\times$ 0.85 (-63$\degree$) & 0.40 & 8.00  & 2015.1.01600.S\\
V892 Tau & 0.26 $\times$ 0.19 (4$\degree$) & 0.32 & 21.12  & 0.27 $\times$ 0.19 (4$\degree$) & 0.66 & 5.57  & 0.28 $\times$ 0.20 (5$\degree$) & 0.67 & 3.93  & 2013.1.00498.S\\
VV Ser & 1.73 $\times$ 1.21 (-63$\degree$) & 0.40 & 34.83  &  & & &  & &  & 2019.1.00218.S\\
Z CMa & 0.22 $\times$ 0.20 (-78$\degree$) & 0.20 & 5.55  & 0.23 $\times$ 0.21 (-79$\degree$) & 0.20 & 6.15  & 0.23 $\times$ 0.21 (-76$\degree$) & 0.20 & 4.68  & 2016.1.00110.S\\
BH Cep &  & & & 1.10 $\times$ 0.82 (31$\degree$) & 0.40 & 9.42  & 1.11 $\times$ 0.82 (32$\degree$) & 0.40 & 9.61  & S21AS\\
BO Cep &  & & & 1.03 $\times$ 0.85 (64$\degree$) & 0.40 & 11.31  & 1.04 $\times$ 0.85 (61$\degree$) & 0.40 & 10.77  & S21AS\\
HD 200775 &  & & & 1.09 $\times$ 0.85 (51$\degree$) & 0.40 & 11.47  & 1.09 $\times$ 0.86 (51$\degree$) & 0.40 & 11.24  & S21AS\\
SV Cep &  & & & 1.01 $\times$ 0.83 (71$\degree$) & 0.40 & 10.83  & 1.01 $\times$ 0.84 (70$\degree$) & 0.40 & 10.69  & S21AS\\
XY Per &  & & & 1.09 $\times$ 0.66 (8$\degree$) & 0.40 & 10.33  & 1.10 $\times$ 0.66 (8$\degree$) & 0.40 & 10.44  & S21AS\\
\hline
\end{tabular}
\label{tab:project_codes}
\textbf{Notes.} The Project codes with an asterisk are for the $J=3-2$ transition, all others are for the $J=2-1$ transition.
\end{table*}

\section{NOEMA continuum data}
\label{app:noema_data}

\begin{figure*}[h]
    \centering
    \includegraphics[width=\textwidth]{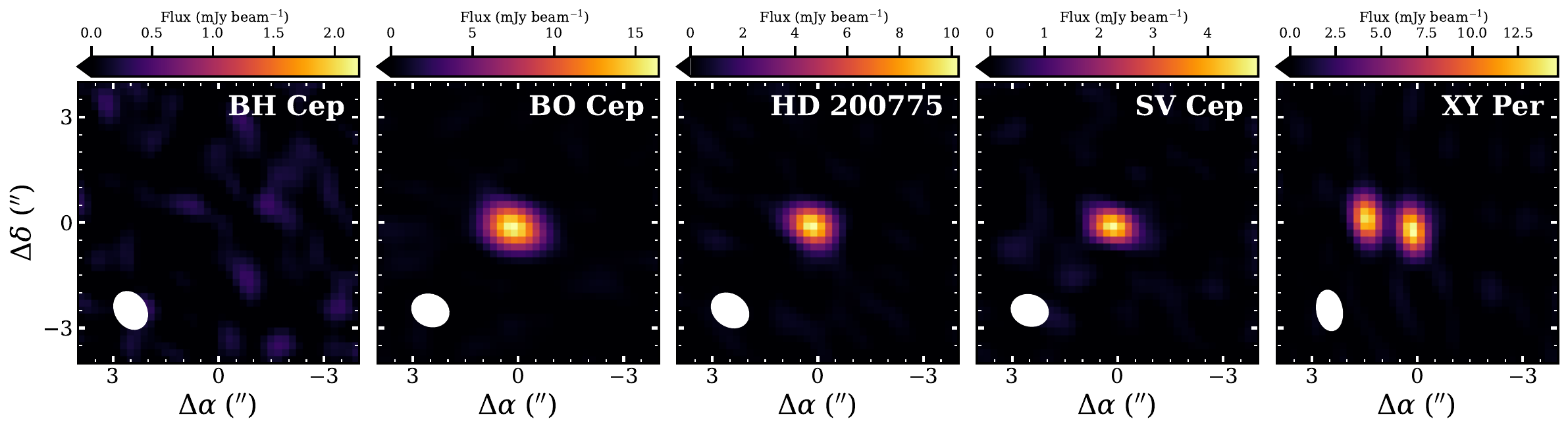}
    \caption{Continuum images of the five northern Herbig disks observed with NOEMA.}
    \label{fig:NOEMA_continuum}
\end{figure*}

Figure \ref{fig:NOEMA_continuum} presents the continuum images of the five northern Herbig disks observed with NOEMA during the summer semester of 2021 (PI: Cridland, Booth; Project code: S21AS). XY~Per was observed on the 17th of November 2021, for 3.4 hours. Both the bandpass and gain calibrator used was 3c84, while the flux calibrator was MWC349. BH~Cep, BO~Cep, HD~200775, and SV~Cep were observed on the 18th of October, for 1.6 hours. The bandpass calibrator was 3C454.3, the gain calibrator was 2010+723, and the flux calibrator was MWC349. The observations were done in the C configuration, with 9 antennas.

The data imaging was done using the \texttt{Common Astronomy Software Applications} (CASA) version 5.8.0 \citep{McMullin2007}. For the data one round of phase-only self-calibration was done. After applying the resulting calibration table to all spectral windows, the data were cleaned using the \texttt{hogbom} algorithm and imaged using the multi-frequency synthesis spectral definition mode and a Briggs robust weighting of 0.5. The resulting beam and rms of each continuum observation can be found in Table~\ref{tab:NOEMA_dust_masses}. Four out of five Herbig disks have been detected. XY~Per is a binary with the A component being a Herbig star.

The integrated fluxes and their corresponding dust masses obtained by following the same procedure as \citet{stapper2022} can be found in Table \ref{tab:NOEMA_dust_masses}. Because the disks are unresolved, the median upper limits on their size are $245$~au and $373$~au for the 68\% and 90\% radii respectively. Comparing the dust masses to the distribution found by \citet{stapper2022}, we find that only BO~Cep is more massive than the mean dust mass, which is likely related to it being the only disk in our sample observed with NOEMA for which the \ce{^13CO} and \ce{C^18O} isotopologues are detected. XY~Per is slightly less massive than the mean dust mass from \citet{stapper2022}, but still well above the dust mass for which we still detect the CO isotopologues, which we do not detect in this disk possibly caused by the binary nature of this object. SV~Cep and HD~200775 both have relatively low disk masses. Interestingly, in the mid-infrared HD~200775 was found to have diffuse emission going out to $\sim700$~au ($2''$, corrected for the most recent distance estimate) in both north and south direction, and a large tail extending to the north-east ($\sim10''$, \citealt{Okamoto2009}). We do not resolve the continuum emission with a beam of $1.1''\times0.8''$, and thus also do not see any of these structures. Lastly, for BH~Cep we determined an upper limit on the dust mass of 1.5~M$_\oplus$ which is higher than some of the detection made by ALMA \citep{stapper2022}.

\begin{table}[]
\caption{Northern Herbigs continuum data parameters, flux measurements and mass estimates observed with NOEMA.}
\centering
\begin{tabular}{lccll}
\hline\hline
\makecell{Name \\ \hspace{1mm}}  & \makecell{Spat.res. \\ ($''$)} & \makecell{rms \\ mJy beam$^{-1}$}  & \makecell{Flux \\ (mJy)} & \makecell{Dust mass \\ (M$_\oplus$)} \\\hline
BH~Cep    & 1.05 $\times$ 0.80 (31$\degree$) & 0.34 & <1.0 & <1.5 \\
BO~Cep    & 1.00 $\times$ 0.82 (67$\degree$) & 0.33 & 25.7 & 59±6 \\
SV~Cep    & 0.98 $\times$ 0.80 (73$\degree$) & 0.37 &  5.4 & 6.7±0.7 \\
HD~200775 & 1.05 $\times$ 0.81 (54$\degree$) & 0.49 & 10.7 & 4.0±0.4 \\
XY~Per    & 1.07 $\times$ 0.64 (10$\degree$) & 0.73 & 15.5 & 16±1.6 \\\hline
\end{tabular}
\label{tab:NOEMA_dust_masses}
\end{table}

\section{Figures for the $J=3-2$ transition}
\label{app:other_transitions}
\begin{figure*}
    \centering
    \includegraphics[width=\textwidth]{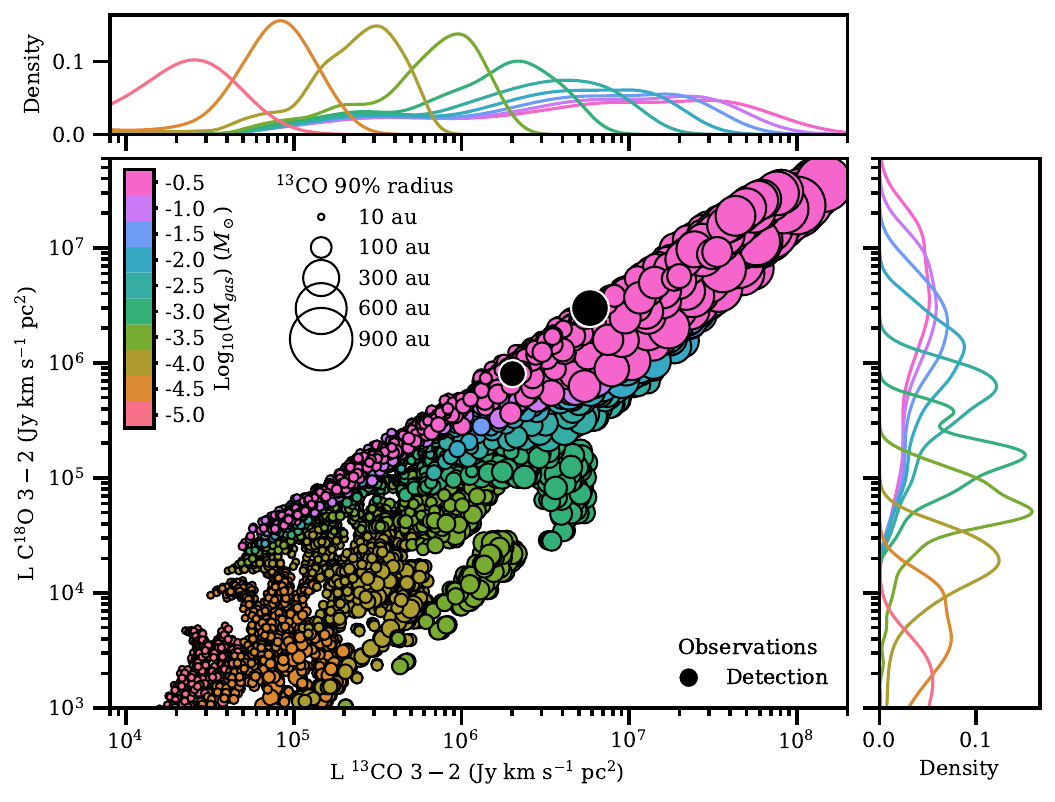}
    \caption{Same as Figure \ref{fig:observations_and_models}, but for the $J=3-2$ transition.}
    \label{fig:observations_and_models_J32}
\end{figure*}

\begin{figure*}
    \centering
    \includegraphics[width=\textwidth]{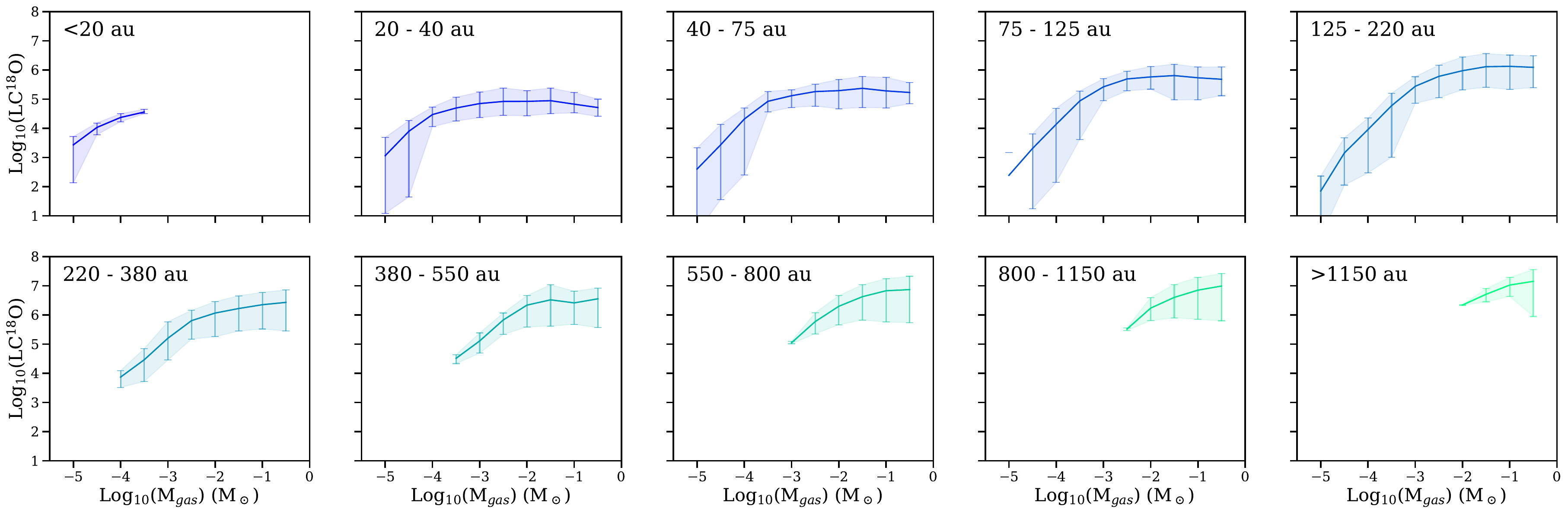}
    \caption{Same as Figure \ref{fig:mgas_vs_lum}, but for the $J=3-2$ transition.}
    \label{fig:mgas_vs_lum_C18O_J32}
\end{figure*}

\begin{figure*}
    \centering
    \includegraphics[width=\textwidth]{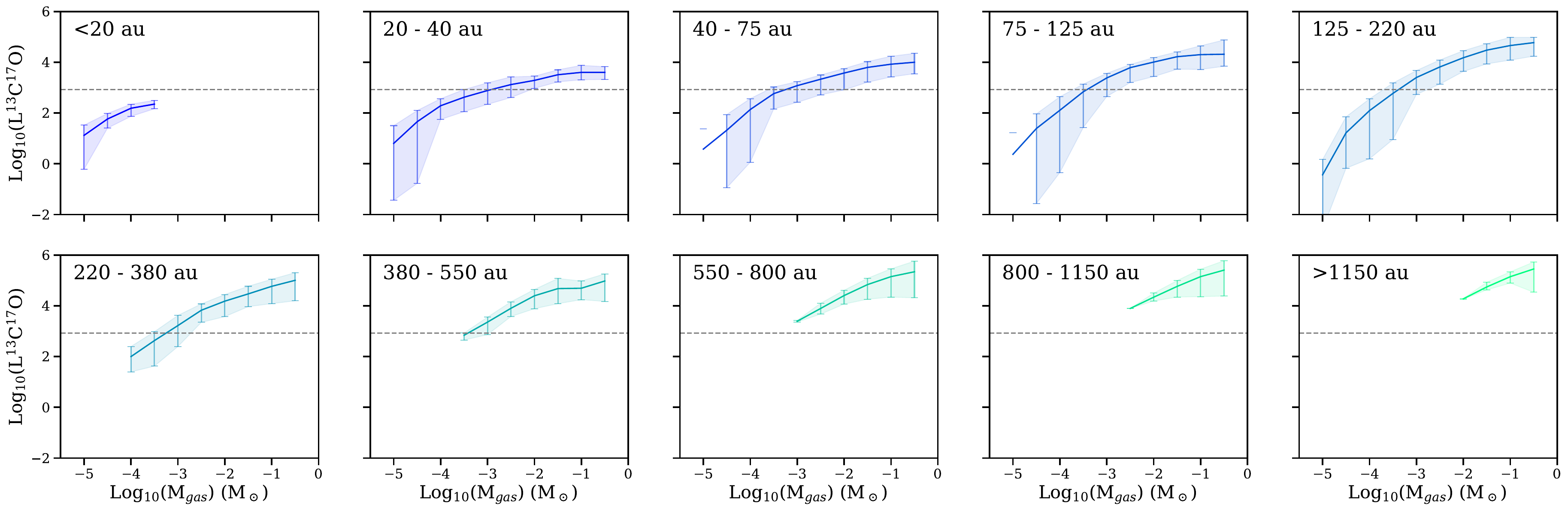}
    \caption{Same as Figure \ref{fig:mgas_vs_lum_13C17O}, but for the $J=3-2$ transition.}
    \label{fig:mgas_vs_lum_13C17O_J32}
\end{figure*}

Figure \ref{fig:observations_and_models_J32} presents the data and models for the $J=3-2$ transition, similar to Fig.~\ref{fig:observations_and_models}. The two disks plotted are HD~135344B and HD~290764. Figures \ref{fig:mgas_vs_lum_C18O_J32} and \ref{fig:mgas_vs_lum_13C17O_J32} are the same as Figs.~\ref{fig:mgas_vs_lum} and \ref{fig:mgas_vs_lum_13C17O} respectively but for the $J=3-2$ transition.

\section{Parameter overview plots}
\label{app:parameter_overview}
\begin{figure*}
    \centering
    \includegraphics[width=0.9\textwidth]{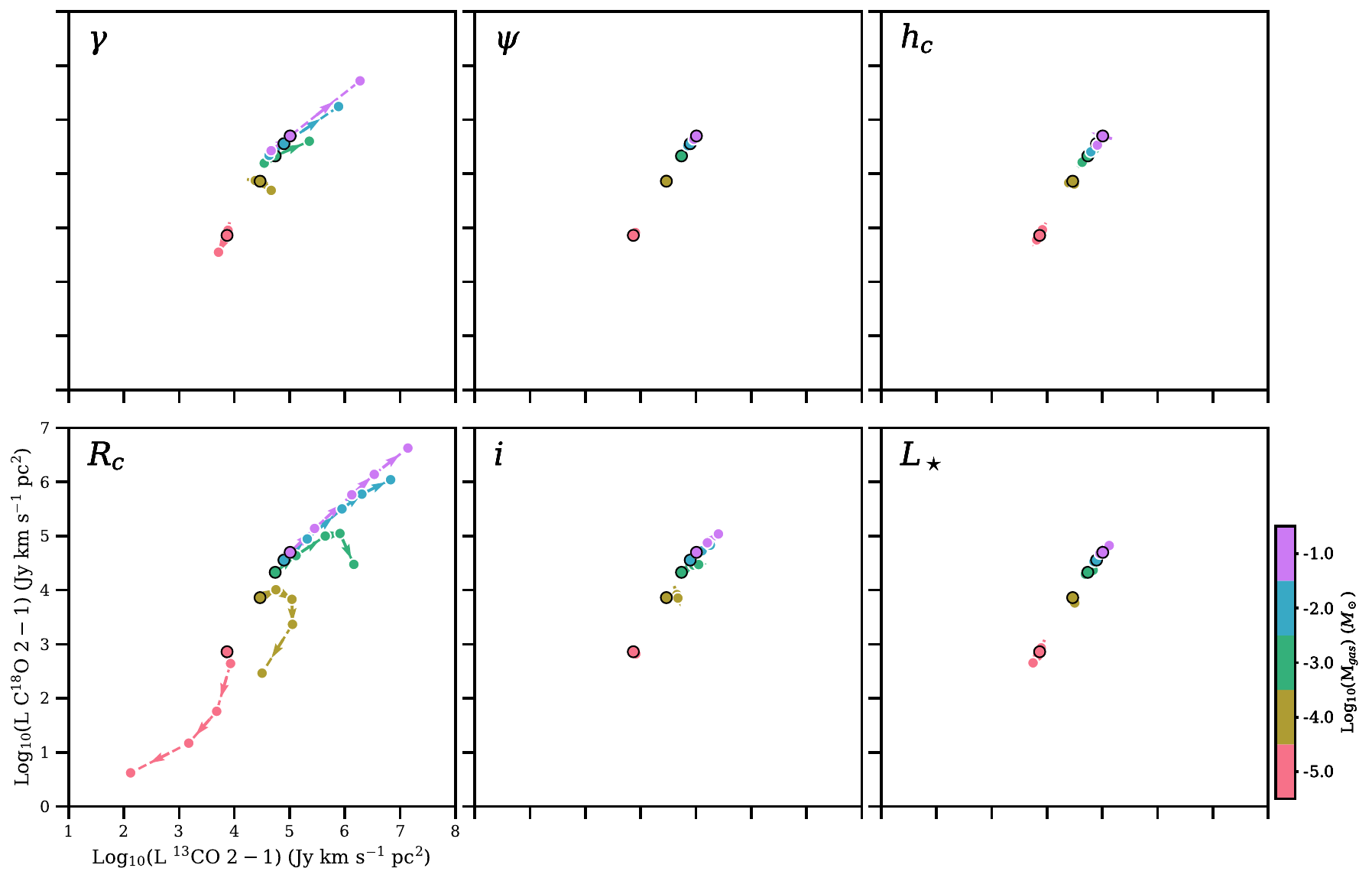}
    \caption{Same as Fig.~\ref{fig:parameter_overview}, but with $R_c=5$~au}
    \label{fig:parameter_overview_Rc_5}
\end{figure*}

\begin{figure*}
    \centering
    \includegraphics[width=0.9\textwidth]{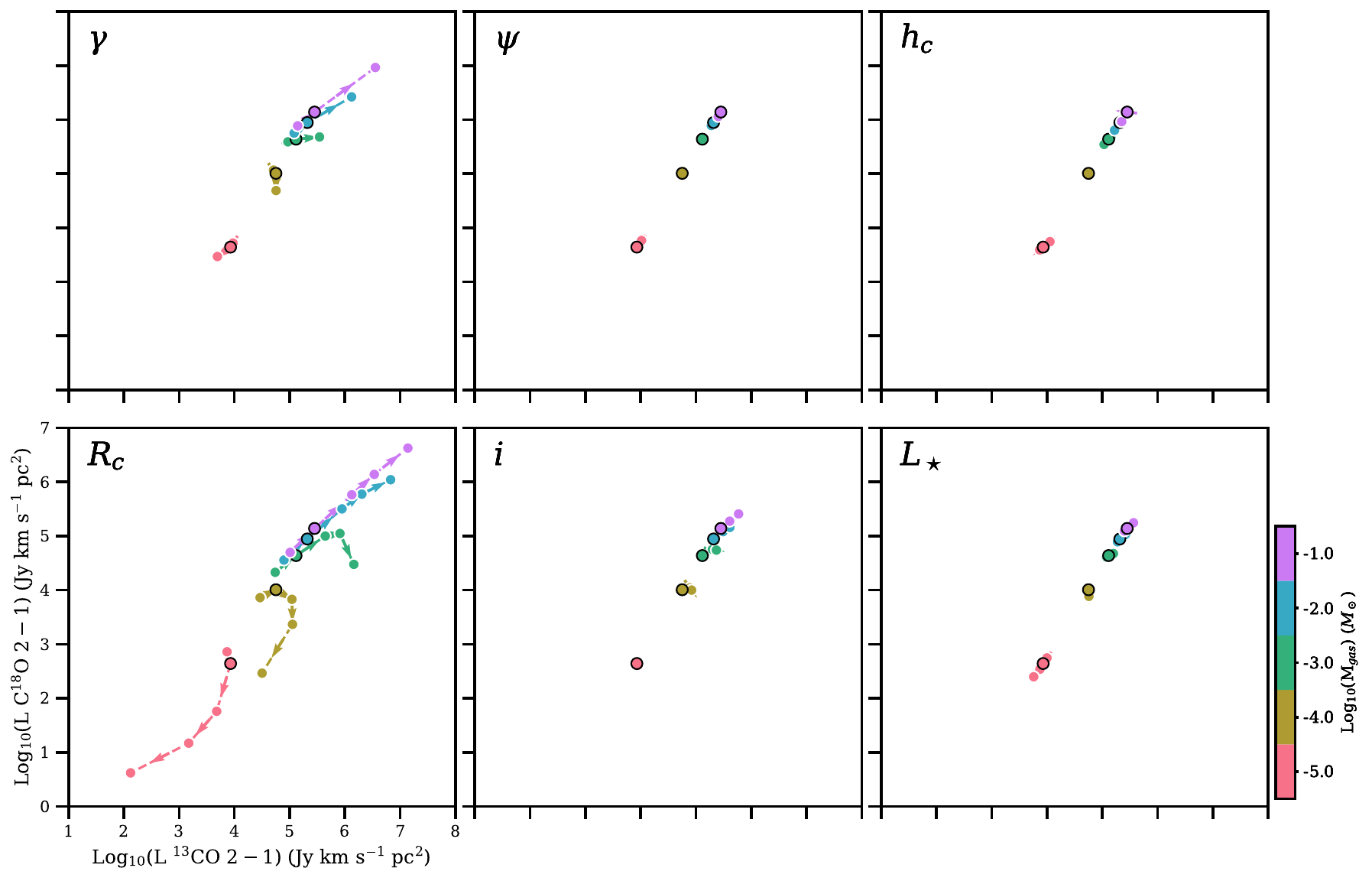}
    \caption{Same as Fig.~\ref{fig:parameter_overview}, but with $R_c=10$~au}
    \label{fig:parameter_overview_Rc_10}
\end{figure*}

\begin{figure*}
    \centering
    \includegraphics[width=0.9\textwidth]{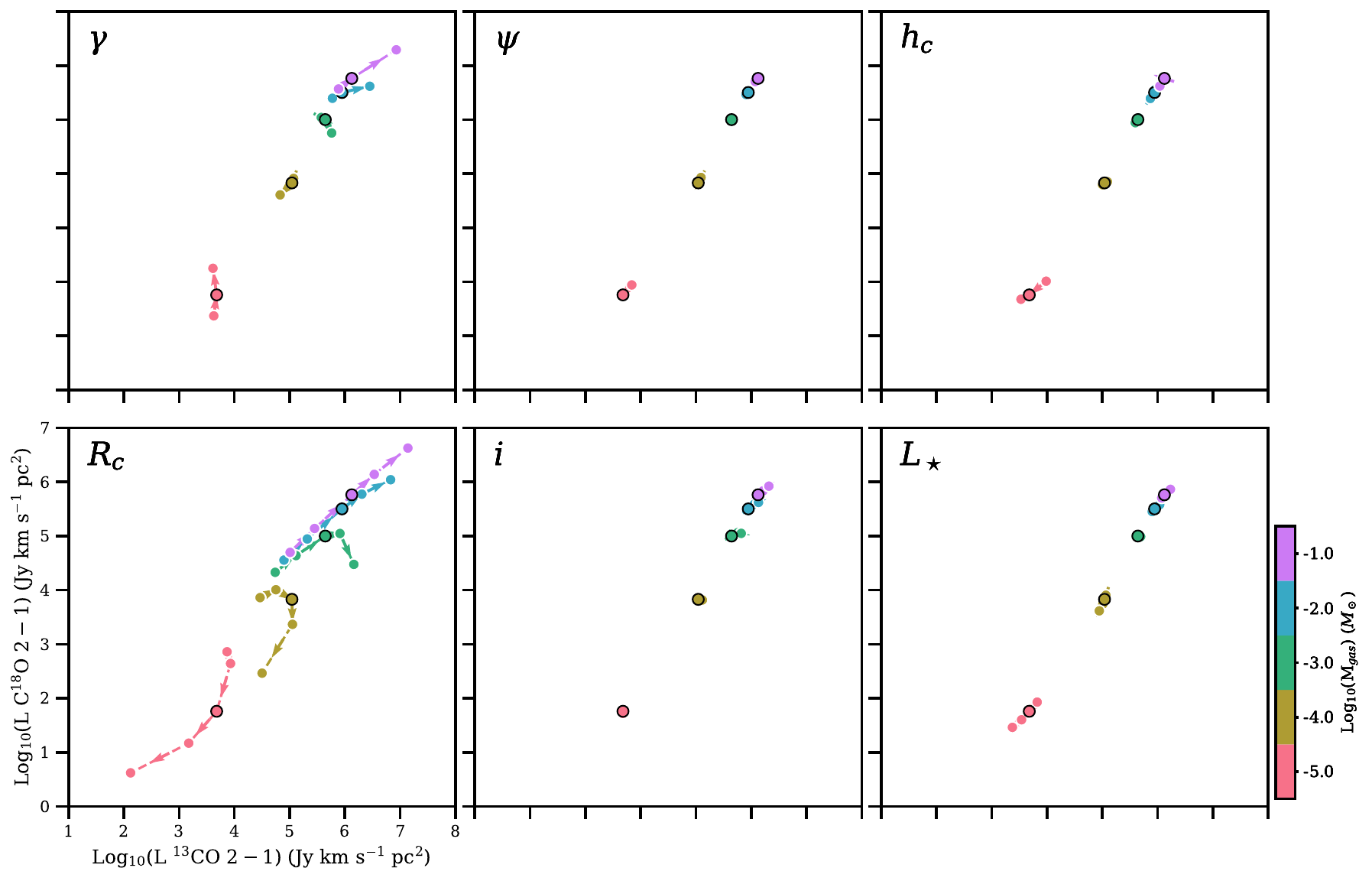}
    \caption{Same as Fig.~\ref{fig:parameter_overview}, but with $R_c=30$~au}
    \label{fig:parameter_overview_Rc_30}
\end{figure*}

\begin{figure*}
    \centering
    \includegraphics[width=0.9\textwidth]{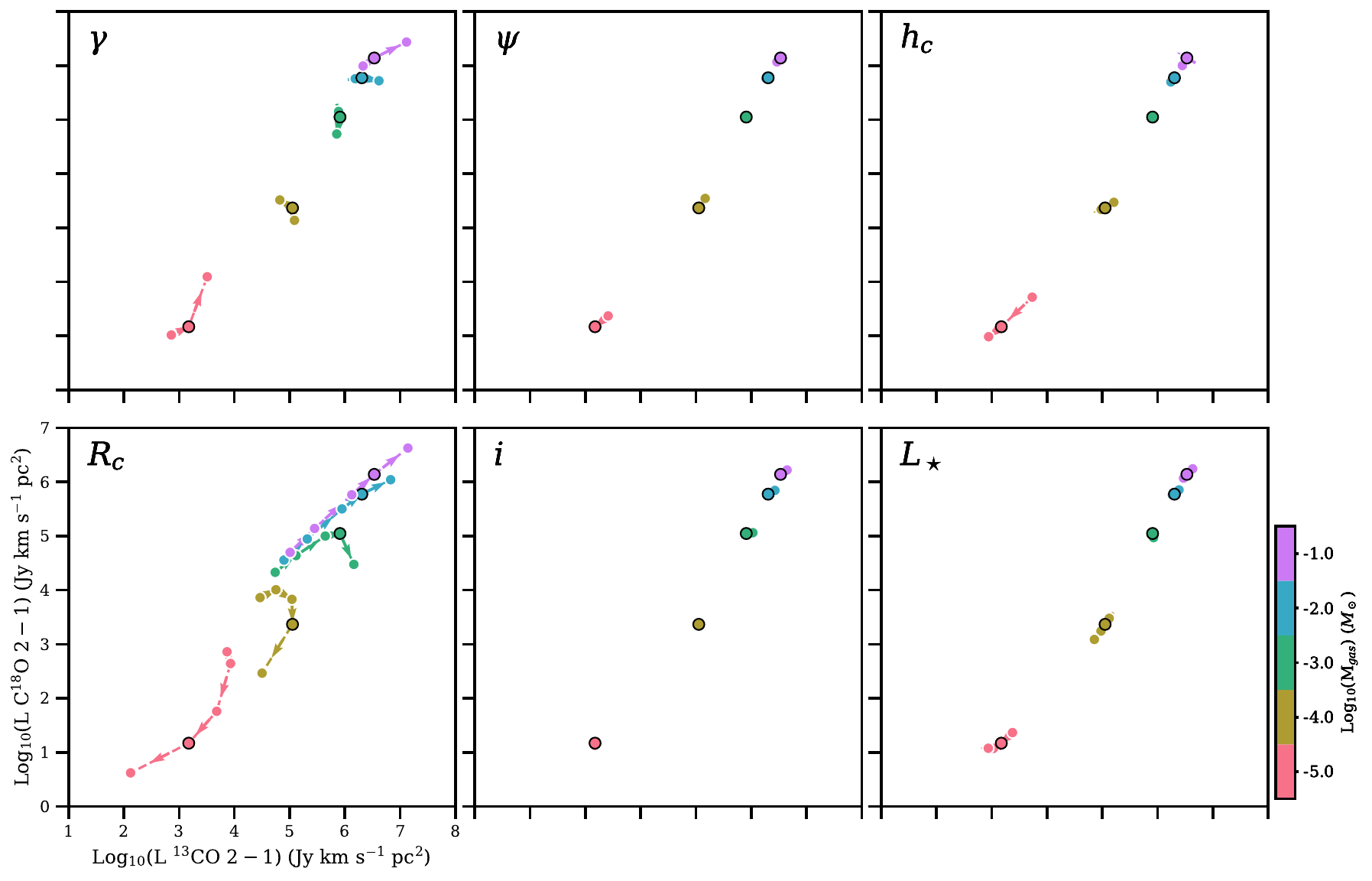}
    \caption{Same as Fig.~\ref{fig:parameter_overview}, but with $R_c=60$~au}
    \label{fig:parameter_overview_Rc_60}
\end{figure*}

\begin{figure*}
    \centering
    \includegraphics[width=0.9\textwidth]{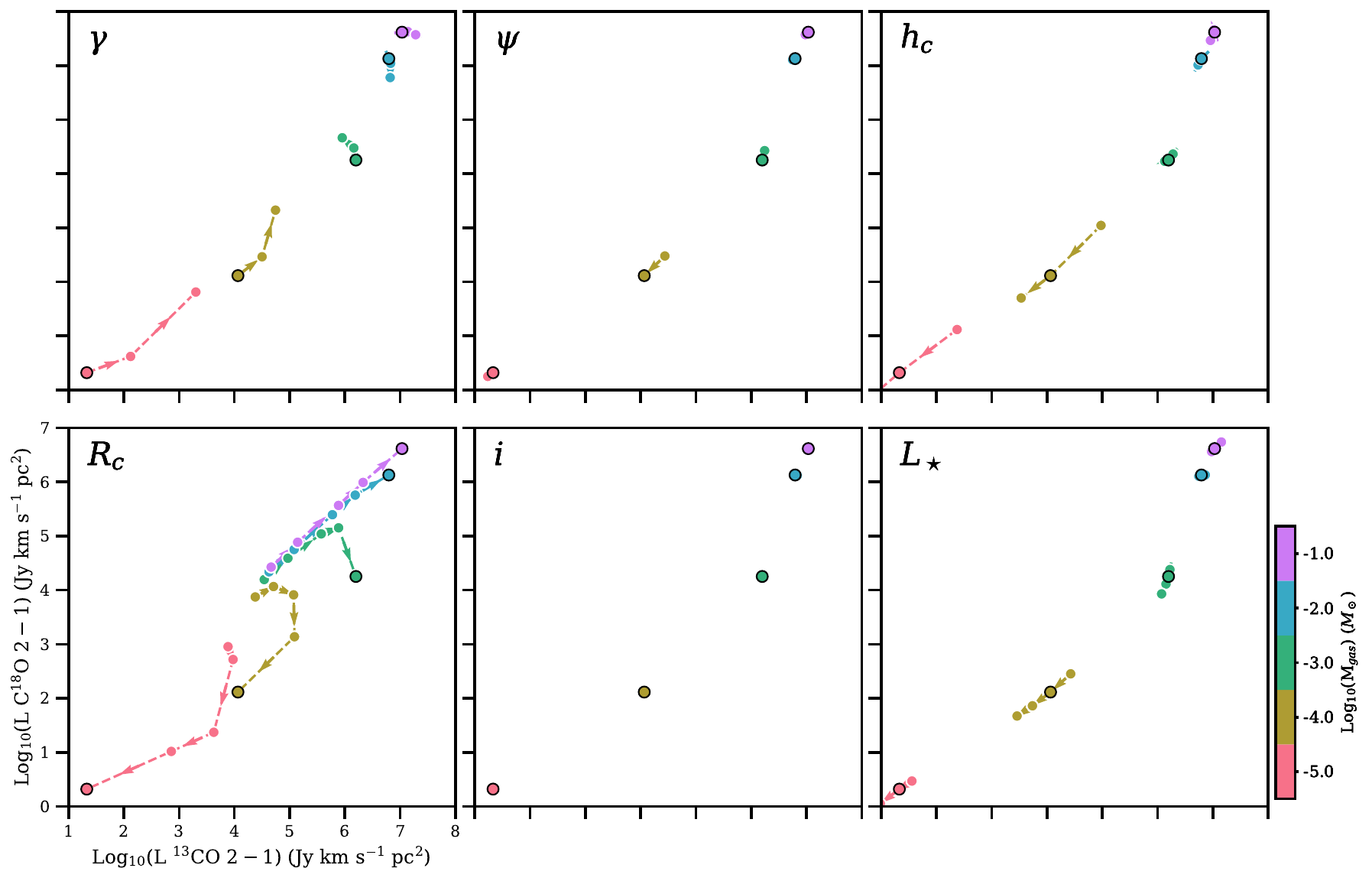}
    \caption{Same as Fig.~\ref{fig:parameter_overview}, but with $\gamma=0.4$}
    \label{fig:parameter_overview_gamma_0.4}
\end{figure*}

\begin{figure*}
    \centering
    \includegraphics[width=0.9\textwidth]{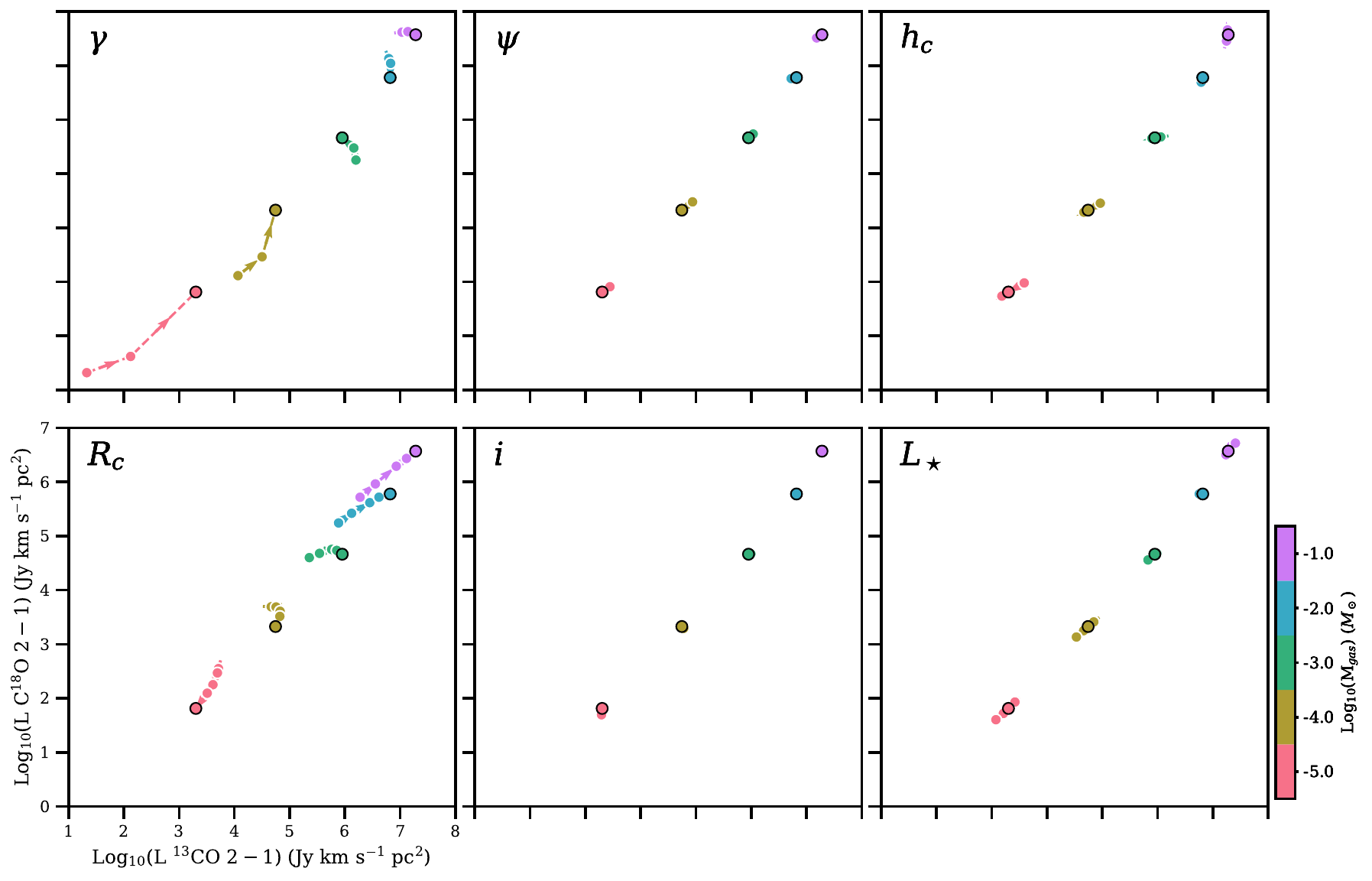}
    \caption{Same as Fig.~\ref{fig:parameter_overview}, but with $\gamma=1.5$}
    \label{fig:parameter_overview_gamma_1.5}
\end{figure*}

\begin{figure*}
    \centering
    \includegraphics[width=0.9\textwidth]{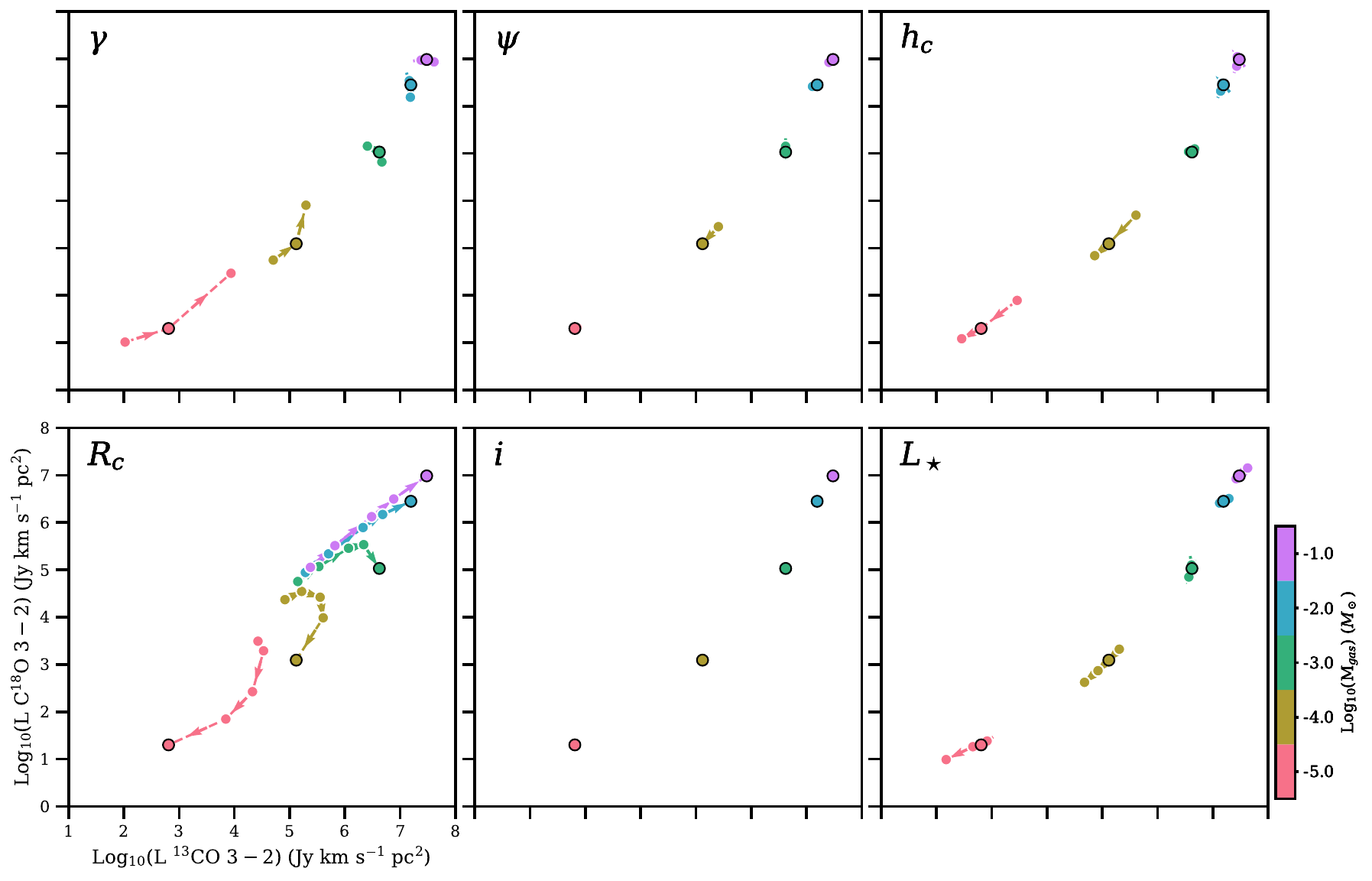}
    \caption{Same as Fig.~\ref{fig:parameter_overview}, but with the $J=3-2$ transition.}
    \label{fig:parameter_overview_J32}
\end{figure*}

\begin{figure*}
    \centering
    \includegraphics[width=0.9\textwidth]{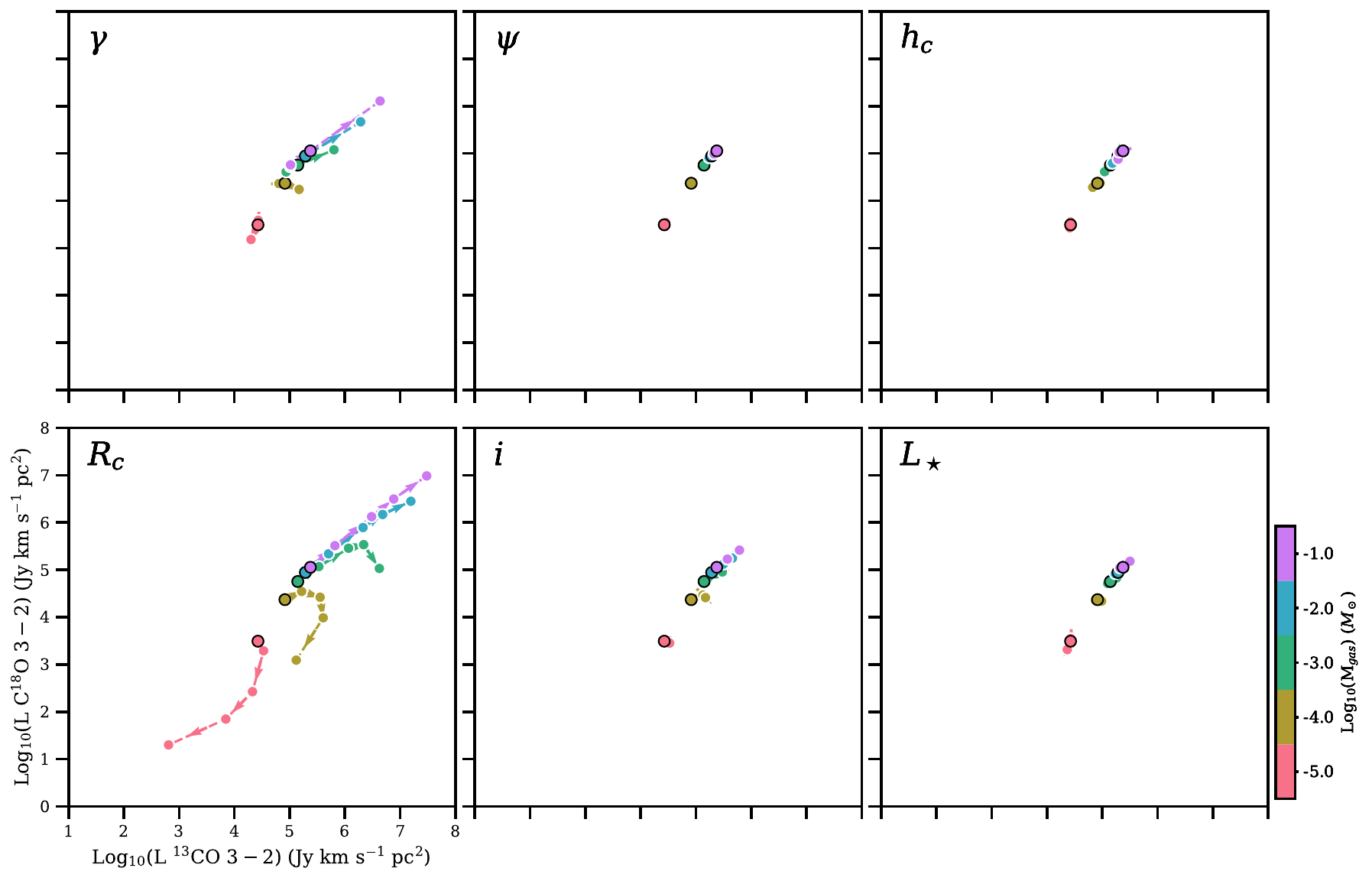}
    \caption{Same as Fig.~\ref{fig:parameter_overview_J32}, but with $R_c=5$~au}
    \label{fig:parameter_overview_Rc_5_J32}
\end{figure*}

\begin{figure*}
    \centering
    \includegraphics[width=0.9\textwidth]{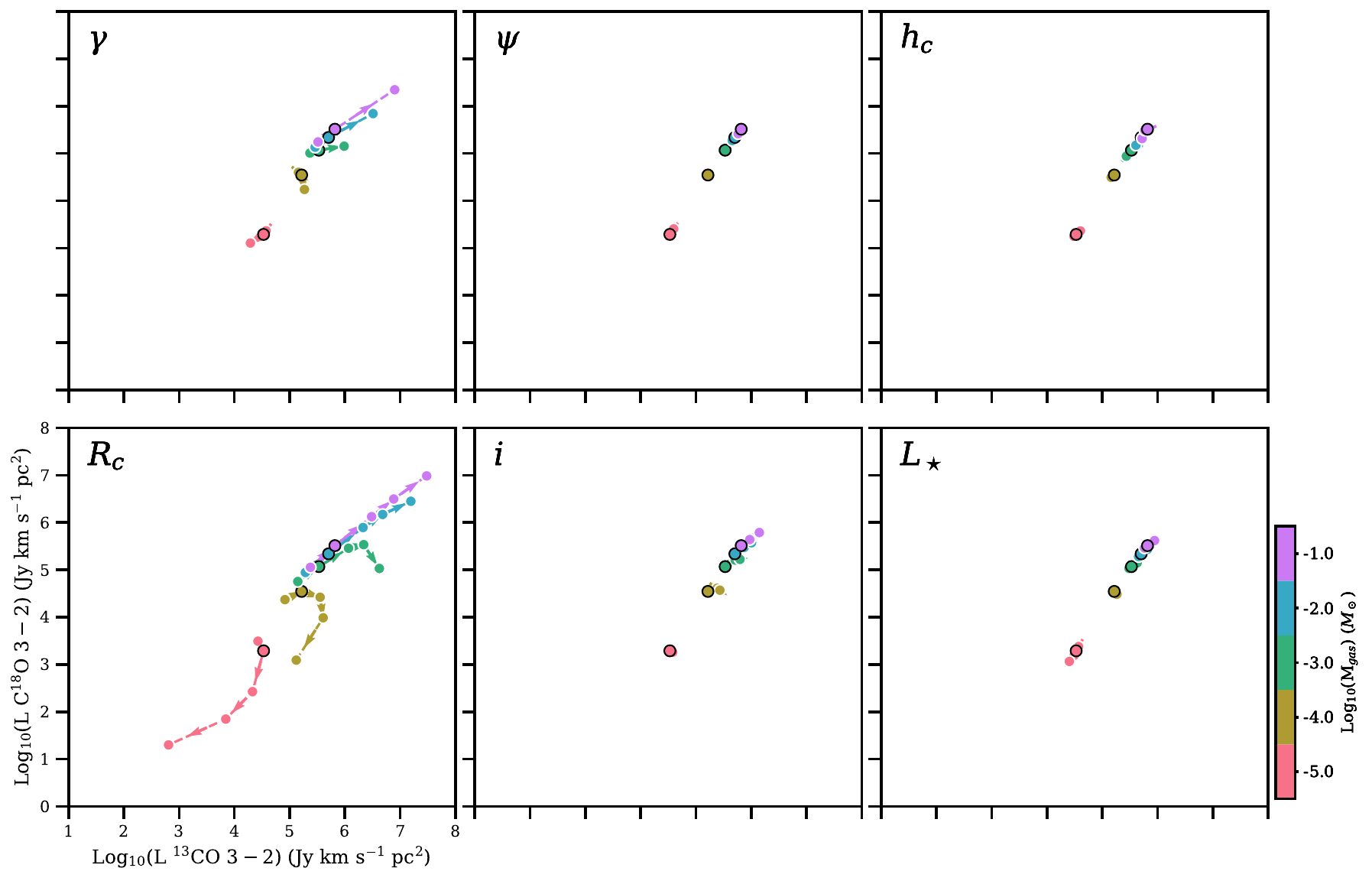}
    \caption{Same as Fig.~\ref{fig:parameter_overview_J32}, but with $R_c=10$~au}
    \label{fig:parameter_overview_Rc_10_J32}
\end{figure*}

\begin{figure*}
    \centering
    \includegraphics[width=0.9\textwidth]{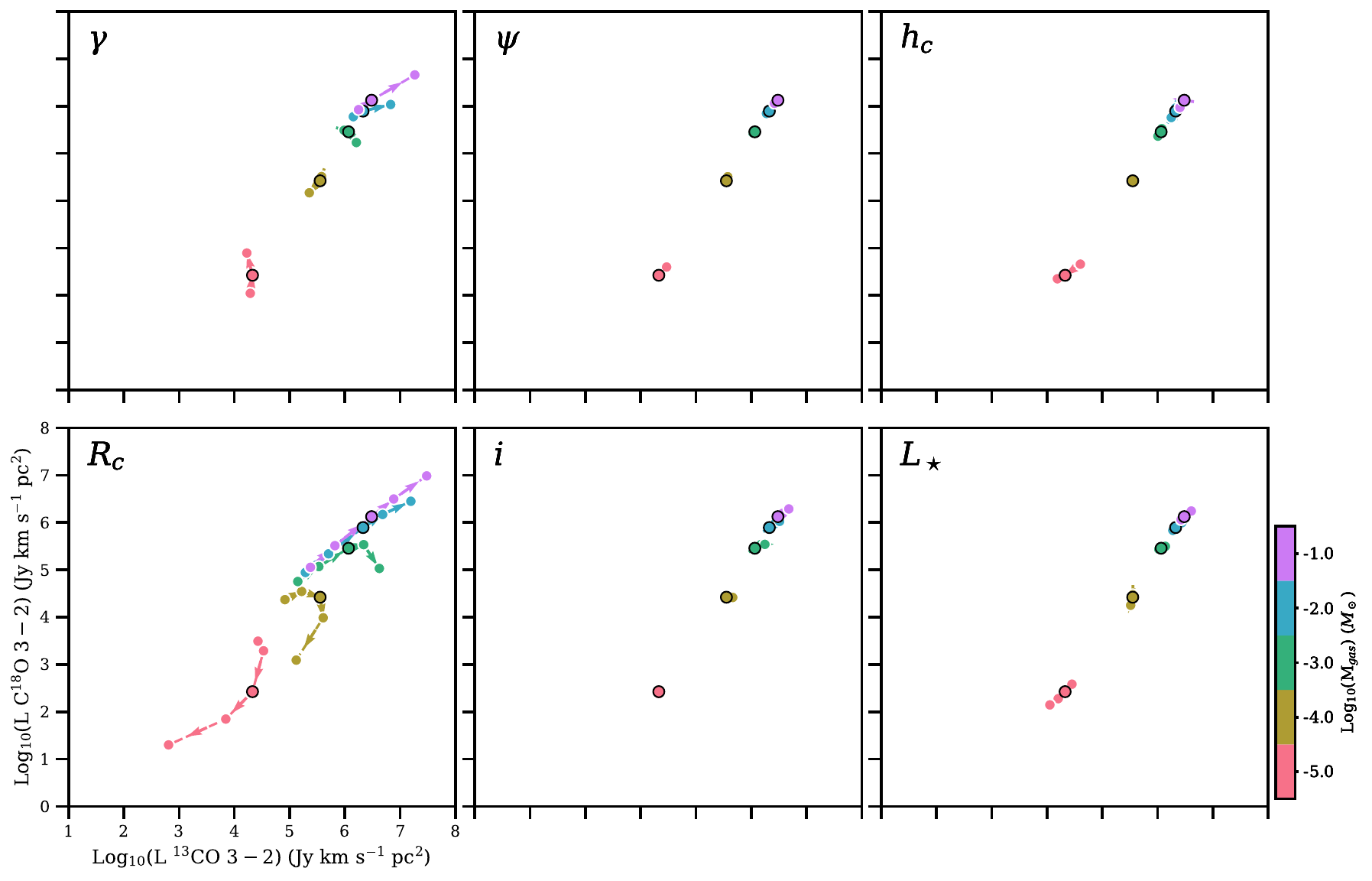}
    \caption{Same as Fig.~\ref{fig:parameter_overview_J32}, but with $R_c=30$~au}
    \label{fig:parameter_overview_Rc_30_J32}
\end{figure*}

\begin{figure*}
    \centering
    \includegraphics[width=0.9\textwidth]{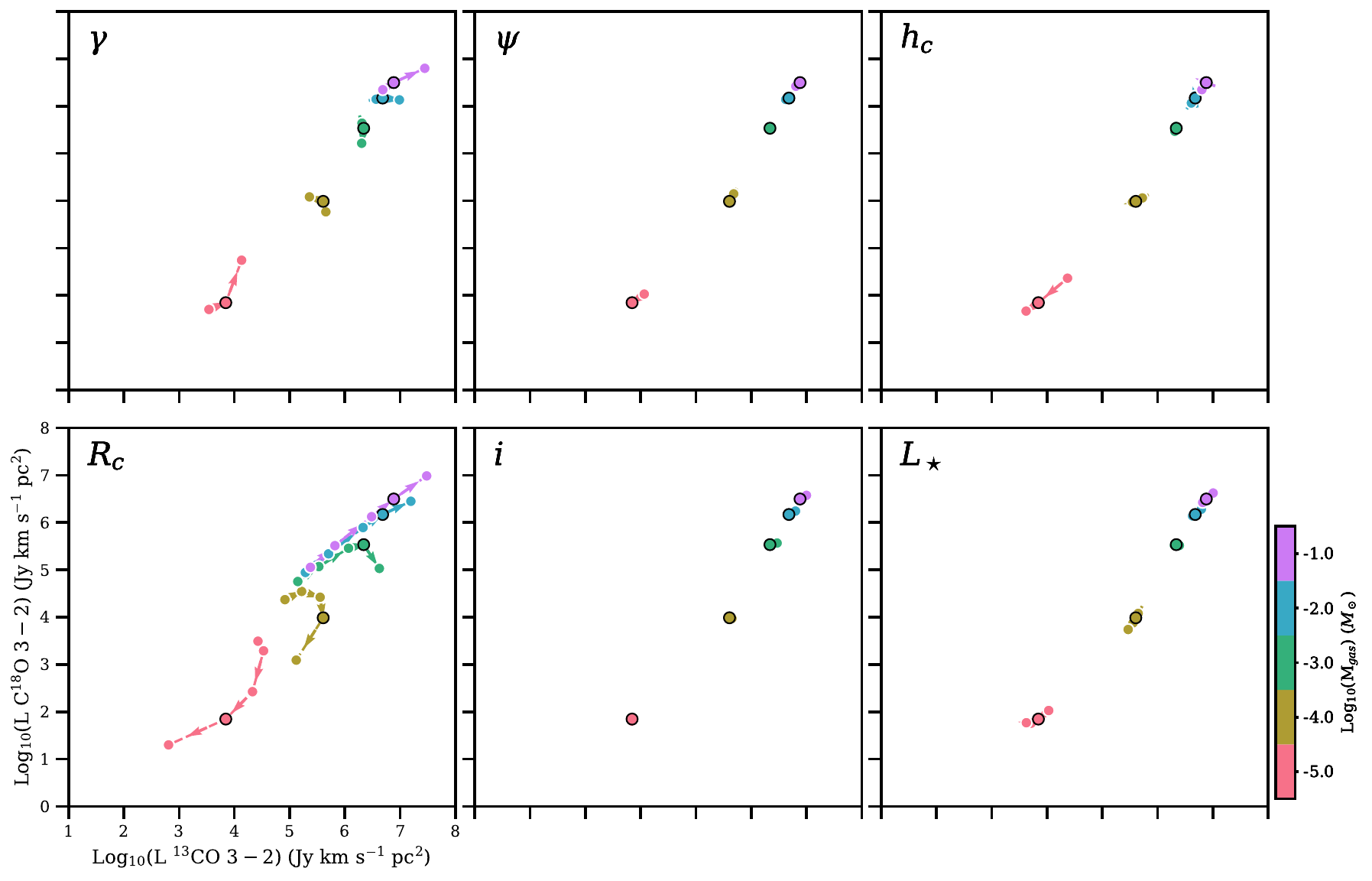}
    \caption{Same as Fig.~\ref{fig:parameter_overview_J32}, but with $R_c=60$~au}
    \label{fig:parameter_overview_Rc_60_J32}
\end{figure*}

\begin{figure*}
    \centering
    \includegraphics[width=0.9\textwidth]{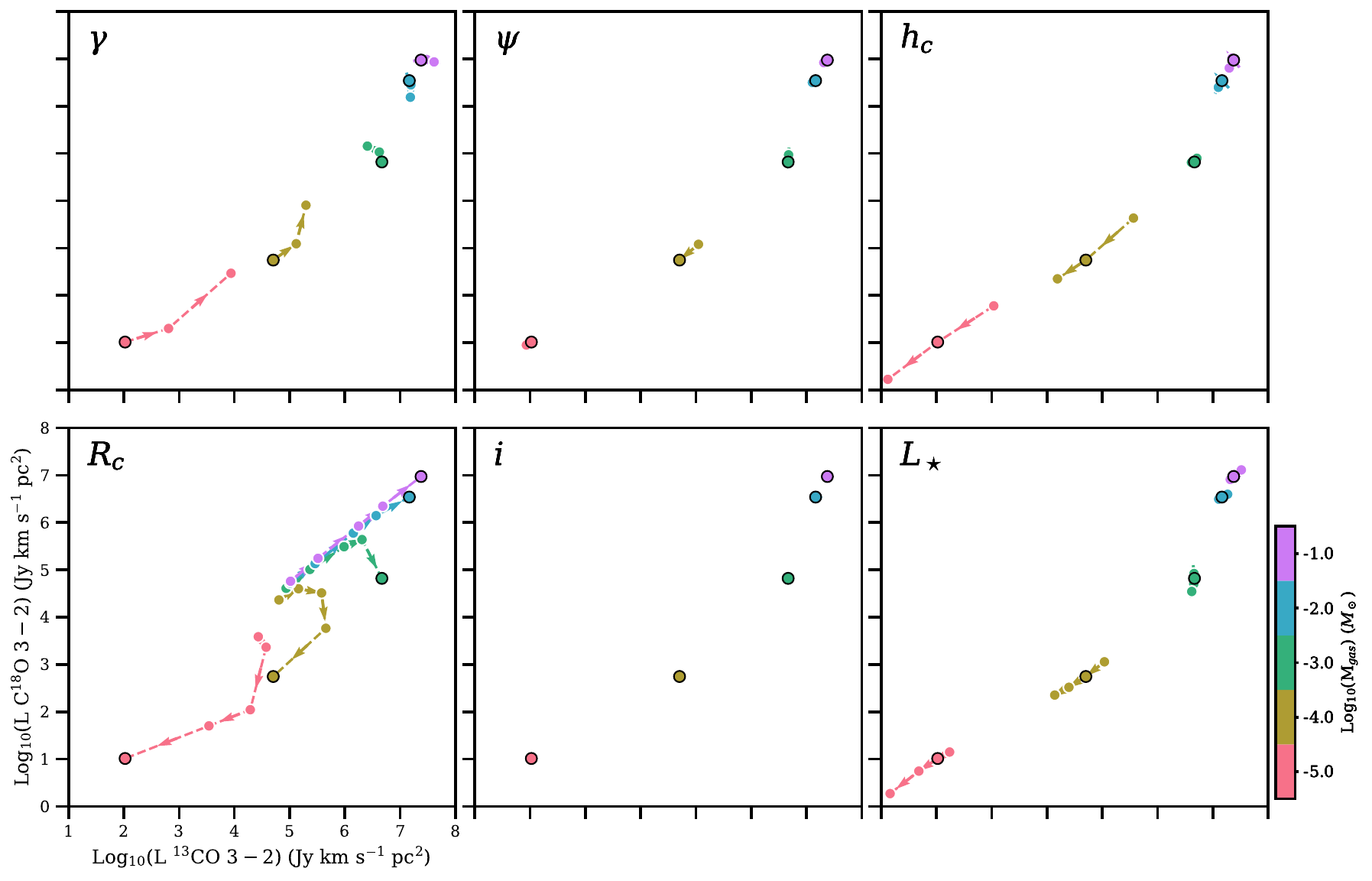}
    \caption{Same as Fig.~\ref{fig:parameter_overview_J32}, but with $\gamma=0.4$}
    \label{fig:parameter_overview_gamma_0.4_J32}
\end{figure*}

\begin{figure*}
    \centering
    \includegraphics[width=0.9\textwidth]{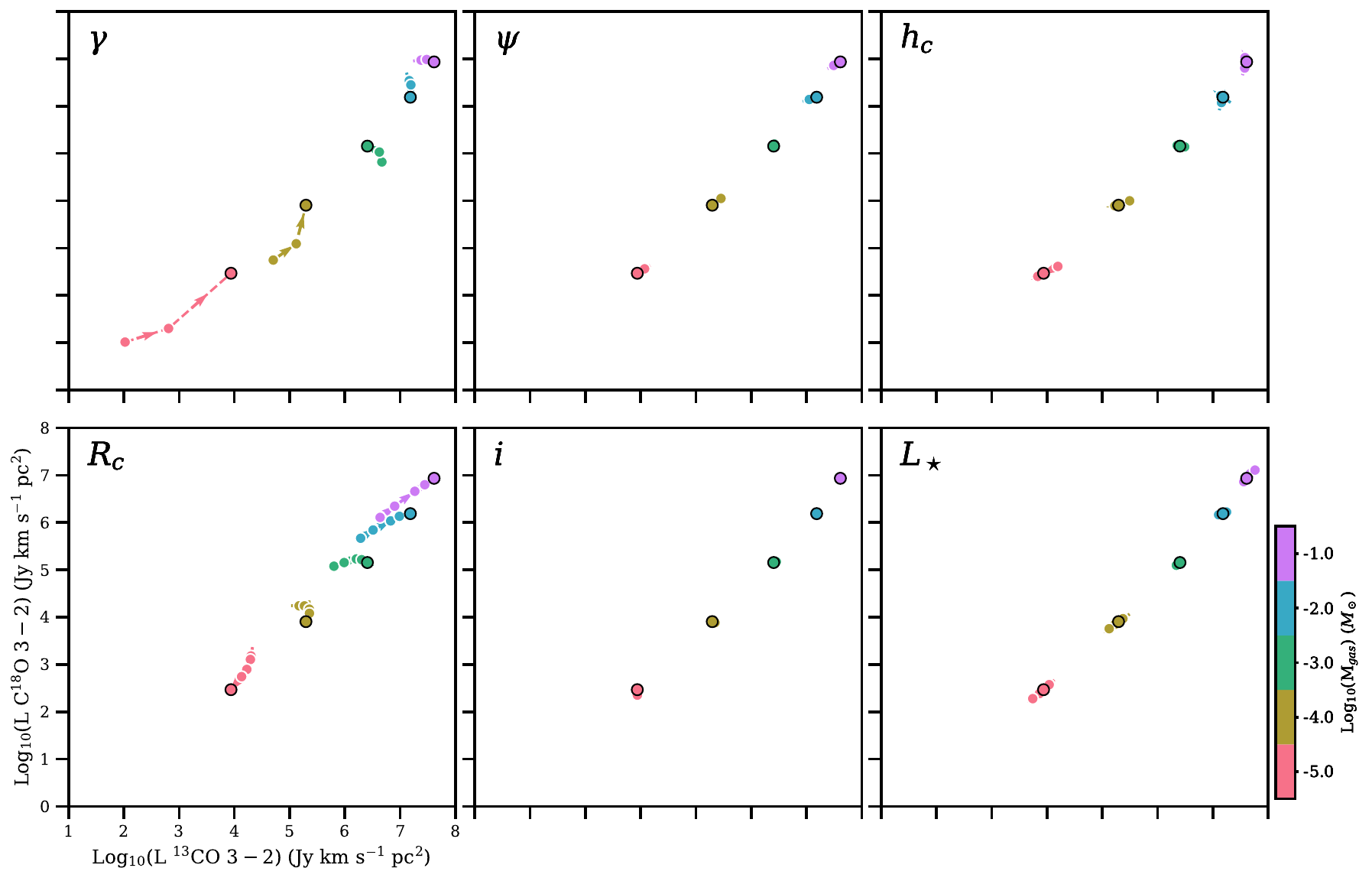}
    \caption{Same as Fig.~\ref{fig:parameter_overview_J32}, but with $\gamma=1.5$}
    \label{fig:parameter_overview_gamma_1.5_J32}
\end{figure*}

This appendix presents plots showing in which direction the parameters from Table~\ref{tab:model_params} change the \ce{^13CO} and \ce{C^18O} luminosities, similar to Fig.~\ref{fig:parameter_overview}. Figures \ref{fig:parameter_overview_Rc_5} to \ref{fig:parameter_overview_Rc_60} are the same as Fig.~\ref{fig:parameter_overview}, but for the other $R_c$ values used. Similarly, Figures \ref{fig:parameter_overview_gamma_0.4} and \ref{fig:parameter_overview_gamma_1.5} are the same as Fig.~\ref{fig:parameter_overview}, but for the other $\gamma$ values. Lastly, Figures \ref{fig:parameter_overview_J32} to \ref{fig:parameter_overview_gamma_1.5_J32} show the same figures but for the $J=3-2$ transition.

% J=3-2 transitions

\section{Disk masses}
\label{app:disk_masses}
Table~\ref{tab:gas_masses} presents the mean disk masses and range in possible disk masses based on our models for all Herbig disks presented in this work.

\begin{table}[]
\caption{Gas mass range and average values of the Herbig disk gas masses together with the resulting gas-to-dust ratio when combined with the dust masses from \citet{stapper2022} and the dust masses from Table~\ref{tab:NOEMA_dust_masses}.}
\tiny\centering
\begin{tabular}{lccr}\hline\hline\\
[-0.7em]
\makecell{Name \\ \hspace{1mm}} & \makecell{Log$_{10}$($\Delta$M$_\text{g}$) \\ ($M_\odot$)} & \makecell{Log$_{10}$($\bar{M}_\text{g}$) \\ (M$_\odot$)} & \makecell{$\Delta_{g/d}$ \\ \hspace{1mm}}  \\ \hline
AB~Aur & >-0.5 & -0.75 &5010\\
AK~Sco & -3.5 -- -0.5 & -2.64 &124\\
BH~Cep & <-0.5 & -3.98 &>23\\
BO~Cep & -3.0 -- -0.5 & -1.67 &122\\
CQ~Tau & -2.5 -- -1.0 & -1.90 &95\\
HD~9672 & -4.5 -- -3.5 & -3.99 &271\\
HD~31648 & -1.25 -- -0.75 & -1.00 &470\\
HD~34282 & -1.5 -- -0.5 & -0.91 &474\\
HD~36112 & -2.25 -- -1.75 & -2.00 &177\\
HD~58647 & <-4.0 & -4.73 &6\\
HD~97048 & -1.5 -- -0.5 & -1.00 &214\\
HD~100453 & -2.5 -- -0.5 & -1.50 &603\\
HD~100546 & -2.0 -- -0.5 & -1.50 &277\\
HD~104237 & -3.0 -- -0.5 & -1.62 &745\\
HD~135344B & -2.5 -- -1.0 & -1.68 &199\\
HD~139614 & -2.5 -- -0.5 & -1.59 &205\\
HD~141569 & -3.25 -- -2.75 & -3.00 &933\\
HD~142527 & >-0.5 & -0.83 &228\\
HD~142666 & -2.5 -- -0.5 & -1.50 &420\\
HD~163296 & -1.25 -- -0.75 & -1.00 &714\\
HD~169142 & -2.0 -- -0.5 & -1.25 &818\\
HD~176386 & <-3.5 & -4.70 &>112\\
HD~200775 & <-0.5 & -3.92 &10\\
HD~290764 & -1.5 -- -0.5 & -1.00 &369\\
HR~5999 & <-4.0 & -4.73 &2\\
KK~Oph & <-0.5 & -3.95 &2\\
MWC~297 & <-0.5 & -4.14 &<1\\
SV~Cep & <-0.5 & -3.96 &5\\
TY~CrA & <-3.5 & -4.68 &68\\
V718~Sco & -2.5 -- -0.5 & -1.44 &1022\\
V892~Tau & -3.0 -- -0.5 & -2.40 &17\\
XY~Per & <-0.5 & -3.84 &3\\
Z~CMa & <-2.5 & -4.61 &<1\\\hline
\end{tabular}
\label{tab:gas_masses}\\
\end{table}

\end{document}